\documentclass[10pt]{article}
\usepackage{hyperref}
\usepackage{amsmath}
\usepackage{amsfonts}
\usepackage{amsthm}
\usepackage{amssymb}
\usepackage{graphicx}
\usepackage{bm}
\usepackage{setspace}
\usepackage{fullpage}
\usepackage{color}

\newcommand{\nc}{\newcommand}
\nc{\R}{\mathbb{R}}
\nc{\bx}{{\bf x}}
\nc{\bU}{{\bf U}}
\nc{\bJ}{{\bf J}}
\nc{\bn}{{\bf n}}
\nc{\by}{{\bf y}}
\nc{\Bs}{{\bf s}}
\nc{\bd}{{\bf d}}
\nc{\bF}{{\bf F}}
\nc{\bFc}{\bm{\mathcal F}}
\nc{\bG}{{\bf G}}
\nc{\bg}{{\bm{\gamma}}}
\nc{\bss}{{\bm{\sigma}}}
\nc{\be}{{\bm{\epsilon}}}
\nc{\bka}{{\bm{\kappa}}}
\nc{\bET}{{\bm{\eta}}}
\nc{\LG}{{\bm{\Lambda}_\gamma}}
\nc{\LGr}{{\bm{\Lambda}_{\gamma^{(o)}}}}
\nc{\LGd}{{\bm{\Lambda}^\dagger_\gamma}}
\nc{\bPhi}{{\bm{\phi}}}
\nc{\bmP}{{\bf{\Phi}}}
\nc{\bW}{{\bf W}}
\nc{\bY}{{\bf Y}}
\nc{\bV}{{\bf V}}
\nc{\Yr}{\bY^{(o)}}
\nc{\bA}{{\bf A}}
\nc{\Ar}{\bA^{(o)}}
\nc{\bB}{{\bf B}}
\nc{\bGG}{{\bf G}}
\nc{\bE}{{\bf E}}
\nc{\bFF}{{\bf F}}
\nc{\bH}{{\bf H}}
\nc{\bq}{\overline{q}}
\nc{\bAq}{\bA^{(\bq)}}
\nc{\bYq}{\bY^{(\bq)}}
\nc{\Dq}{{\bm{\Delta}}^{(\bq)}}
\nc{\Dr}{{\bm{\Delta}}^{(o)}}
\nc{\s}{\sigma}
\nc{\g}{\gamma}
\nc{\G}{\Gamma}
\nc{\T}{\theta}
\nc{\bTe}{\bm{\theta}}
\nc{\bTTe}{\bm{\mathcal Y}}
\nc{\al}{\alpha}
\nc{\om}{\omega}
\nc{\la}{\lambda}
\nc{\ep}{\epsilon}
\nc{\de}{\delta}
\nc{\der}{\de^{(o)}}
\nc{\xir}{\xi^{(o)}}
\nc{\B}{\mathcal B} 
\nc{\I}{\mathcal I}
\nc{\Pc}{\mathcal P}
\nc{\E}{\mathcal E}
\nc{\U}{\mathcal U}
\nc{\Rm}{\mathcal R}
\nc{\M}{\mathcal M}
\nc{\J}{\mathcal J}
\nc{\Q}{\mathcal Q}
\nc{\Sc}{\mathcal S}
\nc{\D}{\mathcal D}
\nc{\K}{\mathcal K}
\nc{\F}{\mathcal F}
\nc{\Gc}{\mathcal G}
\nc{\C}{\mathcal C}
\nc{\Qc}{\mathcal{Q}}
\nc{\ppr}{\bpi_{\mbox{\tiny pr}}}
\nc{\sM}{\s_{\mbox{\tiny MAP}}}
\nc{\BsM}{\Bs_{\mbox{\tiny MAP}}}
\nc{\uB}{{u_{_\B}}} 
\nc{\JB}{{J}_{_\B}}
\nc{\bUI}{{\bU_{_\I}}}
\nc{\bUB}{{\bU_{_\B}}}
\nc{\bJB}{{\bJ_{_\B}}}
\nc{\KII}{{{\bf K}_{_{\I \I}}}}
\nc{\KIB}{{{\bf K}_{_{\I \B}}}}
\nc{\KBI}{{{\bf K}_{_{\B \I}}}}
\nc{\KBB}{{{\bf K}_{_{\B \B}}}}
\nc{\hT}{\widehat{\theta}}
\nc{\hr}{\widehat{r}}
\nc{\hg}{\widehat{\g}}
\nc{\hz}{\widehat{z}}
\nc{\hal}{\widehat{\al}}
\nc{\hs}{\widehat{\s}}
\nc{\hW}{\widehat{W}}
\nc{\tA}{\widetilde{{\bf A}}} 
\nc{\tY}{\widetilde{{\bf Y}}}
\nc{\mhf}{m_{1/2}}
\nc{\fd}{f^\dagger}
\nc{\Fd}{F^\dagger}
\nc{\sr}{\sigma^{(o)}}
\nc{\ar}{\al^{(o)}}
\nc{\har}{\hal^{(o)}}
\nc{\fdr}{f^{\dagger (o)}}
\nc{\Fdr}{F^{\dagger (o)}}
\nc{\rr}{r^{(o)}}
\nc{\hrr}{\hr^{(o)}}
\nc{\zr}{z^{(o)}}
\nc{\hzr}{\hz^{(o)}}
\nc{\zer}{\zeta^{(o)}}
\nc{\hzeq}{\widehat{\zeta}^{(\bq)}}
\nc{\zeq}{\zeta^{(\bq)}}
\nc{\hzer}{\widehat{\zeta}^{(o)}}
\nc{\Gr}{\mathcal{G}^{(o)}}
\nc{\aq}{\al^{(\bq)}}
\nc{\haq}{\hal^{(\bq)}}
\nc{\sq}{\s^{(\bq)}}
\nc{\hsq}{\hs^{(\bq)}}
\nc{\deq}{\de^{(\bq)}}
\nc{\xiq}{\xi^{(\bq)}}

\newcommand{\bS}{{\bf S}}
\newcommand{\bDe}{{\bf D}}

\newcommand{\bone}{\mathbf{1}}
\newcommand{\bzero}{\mathbf{0}}
\newcommand{\BM}{\mathbf{M}}

\newcommand{\diag}{\mbox{diag}\,}

\newtheorem{alg}{Algorithm}
\newtheorem{remark}{Remark}
\newtheorem{lemma}{Lemma}
\newtheorem{theorem}{Theorem}
\newtheorem{corollary}{Corollary}
\begin{document}

\title{Resistor network approaches to electrical impedance tomography}
\author{L. Borcea\footnotemark[1] \and V. Druskin \footnotemark[2]
\and F. Guevara Vasquez\footnotemark[3] \and
A.V. Mamonov\footnotemark[4]}

\renewcommand{\thefootnote}{\fnsymbol{footnote}}
\footnotetext[1]{Computational and Applied Mathematics, Rice
University, MS 134, Houston, TX 77005-1892. (borcea@caam.rice.edu)}
\footnotetext[2]{Schlumberger Doll Research Center, One Hampshire St.,
Cambridge, MA 02139-1578. (druskin1@slb.com)}
\footnotetext[3]{Mathematics, University of Utah, 155 S
1400 E RM 233, Salt Lake City, UT
84112-0090. (fguevara@math.utah.edu)} \footnotetext[4]{Institute for
Computational Engineering and Sciences, University of Texas at Austin,
1 University Station C0200, Austin, TX 78712. (mamonov@ices.utexas.edu)} 
\date{}
\maketitle 
\begin{abstract}
  We review a resistor network approach to the  numerical solution
  of the inverse problem of electrical impedance tomography (EIT).
  The networks arise in the context of finite volume discretizations
  of the elliptic equation for the electric potential, on sparse and
  adaptively refined grids that we call optimal. The name refers to
  the fact that the grids give spectrally accurate approximations of
  the Dirichlet to Neumann map, the data in EIT.  The fundamental
  feature of the optimal grids in inversion is that they connect the
  discrete inverse problem for resistor networks to the continuum EIT
  problem.
\end{abstract}
\section{Introduction}
\label{sect:intro}
We consider the inverse problem of electrical impedance tomography
(EIT) in two dimensions \cite{borcea2002electrical}.  It seeks the
scalar valued positive and bounded conductivity $\s(\bx)$, the
coefficient in the elliptic partial differential equation for the
potential $u \in H^1(\Omega)$,
\begin{equation}
\label{eq:CONDeq}
\nabla \cdot \left[ \s(\bx) \nabla u(\bx) \right] = 0, \qquad \bx \in
\Omega.
\end{equation}
The domain $\Omega$ is a bounded and simply connected set in
$\mathbb{R}^2$ with smooth boundary $\B$.  Because all such domains
are conformally equivalent by the Riemann mapping theorem, we assume
throughout that $\Omega$ is the unit disk,
\begin{equation}
\label{eq:Om}
\Omega = \left\{ \bx = (r \cos \T, r \sin \T), \quad r \in [0,1], 
~ ~ \T \in [0,2 \pi) \right\}.
\end{equation}
 The EIT problem is to
determine $\s(\bx)$ from measurements of the Dirichlet to Neumann
(DtN) map $\Lambda_\s$ or equivalently, the Neumann to Dirichlet map
$\Lambda_\s^\dagger$. We consider the \emph{full boundary setup}, with
access to the entire boundary, and the \emph{partial measurement
  setup}, where the measurements are confined to an accessible subset
$\B_A$ of $\B$, and the remainder $\B_I = \B \setminus \B_A$ of the
boundary is grounded ($u|_{\B_I} = 0$).

The DtN map $\Lambda_\s: H^{1/2}(\B) \to H^{-1/2}(\B)$ takes arbitrary
boundary potentials $\uB$ in the trace space $ H^{1/2}(\B)$ to normal
boundary currents
\begin{equation}
\Lambda_\s \uB (\bx) = \s(\bx) \bn(\bx) \cdot \nabla u(\bx), \qquad
\bx \in \B,
\label{eq:DtN}
\end{equation}
where $\bn(\bx)$ is the outer normal at $\bx \in \B$ and $u(\bx)$
solves (\ref{eq:CONDeq}) with Dirichlet boundary conditions
\begin{equation}
u(\bx) = \uB(\bx), \quad \bx \in \B.
\label{eq:DirBC}
\end{equation}
Note that $\Lambda_\s$ has a null space consisting of constant
potentials and thus, it is invertible only on a subset $\J$ of
$H^{-1/2}(\B)$, defined by
\begin{equation}
\J = \left\{ J \in H^{-1/2}(\B) ~ ~ \mbox{such that} ~ ~ \int_\B J(\bx)
ds(\bx) = 0  \right\}.
\label{eq:RestrJ}
\end{equation}
Its generalized inverse is the NtD map $\Lambda_\s^\dagger:\J \to
H^{1/2}(\B)$, which takes boundary currents $\JB \in \J$ to boundary
potentials
\begin{equation}
\Lambda_\s^\dagger \JB(\bx) = u(\bx), \qquad \bx \in \B.
\label{eq:NtD}
\end{equation}
Here $u$ solves (\ref{eq:CONDeq}) with Neumann boundary conditions
\begin{equation}
\s(\bx) \bn(\bx) \cdot \nabla u(\bx) = \JB(\bx), \qquad \bx \in \B,
\label{eq:NeumBC}
\end{equation}
and it is defined up to an additive constant, that can be fixed for
example by setting the potential to zero at one boundary point, as if
it were connected to the ground.

It is known that $\Lambda_\s$ determines uniquely $\s$ in the full
boundary setup \cite{astala2005cip}. See also the earlier uniqueness
results \cite{nachman1996gut,brown1997uniqueness} under some
smoothness assumptions on $\sigma$.  Uniqueness holds for the partial
boundary setup as well, at least for $\s \in C^{3 + \epsilon} ( \bar
\Omega )$ and $\epsilon > 0$, \cite{imanuvilov2008gup}. The 
case of real-analytic or piecewise real-analytic $\sigma$ is resolved in  
\cite{druskin1982usi, druskin1985udt, kohn1984dcb,
  kohn1985dcb}.

However, the
problem is exponentially unstable, as shown in
\cite{alessandrini1988sdc,barcelo2001sic,mandache2001eii}. Given two
sufficiently regular conductivities $\s_1$ and $\s_2$, the best
possible stability estimate is of logarithmic type
\begin{equation}
\label{stab}
\|\sigma_1-\sigma_2 \|_{L^\infty(\Omega)} \le c \left| \log \|
  \Lambda_{\s_1} - \Lambda_{\s_2}\|_{ H^{1/2}(\B), H^{-1/2}(\B)}
  \right|^{-\alpha},
\end{equation}
with some positive constants $c$ and $\alpha$.  This means that if we
have noisy measurements, we cannot expect the conductivity to be close
to the true one uniformly in $\Omega$, unless the noise is
exponentially small.

In practice the noise plays a role and the inversion can be carried
out only by imposing some regularization constraints on $\sigma$.
Moreover, we have finitely many measurements of the DtN map and we
seek numerical approximations of $\s$ with finitely many degrees of
freedom (parameters). The stability of these approximations depends on
the number of parameters and their distribution in the domain
$\Omega$. 

It is shown in \cite{alessandrini2005lipschitz} that if $\s$ is
piecewise constant, with a bounded number of unknown values, then the
stability estimates on $\s$ are no longer of the form (\ref{stab}),
but they become of Lipschitz type. However, it is not really
understood how the Lipschitz constant depends on the distribution of
the unknowns in $\Omega$. Surely, it must be easier to determine the
features of the conductivity near the boundary than deep inside
$\Omega$.

Then, the question is how to parametrize the unknown conductivity in
numerical inversion so that we can control its stability and we do not
need excessive regularization with artificial penalties that introduce
artifacts in the results. Adaptive parametrizations for EIT have been
considered for example in \cite{isaacson,macmillan2004first} and
\cite{ameur2002refinement,ameur2002regularization}. Here we review our
inversion approach that is based on resistor networks that arise in
finite volume discretizations of (\ref{eq:CONDeq}) on sparse and
adaptively refined grids which we call \emph{optimal}. The name refers
to the fact that they give spectral accuracy of approximations of
$\Lambda_\s$ on finite volume grids. One of their important features
is that they are refined near the boundary, where we make the
measurements, and coarse away from it. Thus they capture the expected
loss of resolution of the numerical approximations of $\s$.

Optimal grids were introduced in \cite{DruKni,druskin2000gsr,
  IngDruKni,asvadurov2000adg,asvadurov2004ofd} for accurate
approximations of the DtN map in forward problems.  Having such
approximations is important for example in domain decomposition
approaches to solving second order partial differential equations and
systems, because the action of a sub-domain can be replaced by the DtN
map on its boundary \cite{quarteroni1999domain}. In addition, accurate
approximations of DtN maps allow truncations of the computational
domain for solving hyperbolic problems. The studies in
\cite{DruKni,druskin2000gsr,
  IngDruKni,asvadurov2000adg,asvadurov2004ofd} work with spectral
decompositions of the DtN map, and show that by just placing grid
points optimally in the domain, one can obtain exponential convergence
rates of approximations of the DtN map with second order finite
difference schemes. That is to say, although the solution of the
forward problem is second order accurate inside the computational
domain, the DtN map is approximated with spectral accuracy. Problems
with piecewise constant and anisotropic coefficients are considered in
\cite{DruMos,asvadurov2007optimal}.

The optimal grids are useful in the context of numerical inversion,
because they resolve the inconsistency that arises from the
exponential ill posedness of the problem and the second order
convergence of typical discretization schemes applied to  equation
(\ref{eq:CONDeq}), on ad-hoc grids that are usually uniform.  The
forward problem for the approximation of the DtN map is the inverse of
the EIT problem, so it should converge exponentially. This can be
achieved by discretizing on the optimal grids.  

In this article we review the use of optimal grids in inversion, as it
was developed over the last few years in
\cite{BorDru,BorDruKni,BorDruGue,
  GuevaraPhD,BDM-10,BDMG-10,MamonovPhD}.  We present first, in section
\ref{sect:Layered}, the case of layered conductivity $\s = \s(r)$ and
full boundary measurements, where the DtN map has eigenfunctions $e^{i
  k \T}$ and eigenvalues denoted by $f(k^2)$, with integer $k$. Then,
the forward problem can be stated as one of rational approximation of
$f(\la)$, for $\la$ in the complex plane, away from the negative real
axis. We explain in section \ref{sect:Layered} how to compute the
optimal grid from such rational approximants and also how to use it in
inversion. The optimal grid depends on the type of discrete
measurements that we make of $\Lambda_\s$ (i.e., $f(\la)$) and so does
the accuracy and stability of the resulting approximations of $\s$.

The two dimensional problem $\s = \s(r,\T)$ is reviewed in sections
\ref{sect:2DFB} and \ref{sect:2DPB}. The easier case of full access to
the boundary, and discrete measurements at $n$ equally distributed
points on $\B$ is in section \ref{sect:2DFB}. There, the grids are
essentially the same as in the layered case and the finite volumes
discretization leads to circular networks with topology determined by
the grids. We show how to use the discrete inverse problem theory for
circular networks developed in
\cite{CurtMooMor,CurtIngMor,IngerLayer,deverdiere1994rep,
  deverdiere1996rep} for the numerical solution of the EIT problem.
Section \ref{sect:2DPB} considers the more difficult, partial boundary
measurement setup, where the accessible boundary consists of either
one connected subset of $\B$ or two disjoint subsets. There, the
optimal grids are truly two dimensional and cannot be computed
directly from the layered case. 

The theoretical review of our results in
\cite{BorDru,BorDruKni,BorDruGue,
  GuevaraPhD,BDM-10,BDMG-10,MamonovPhD} is complemented by some
numerical results. For brevity, all the results are in the noiseless
case. We refer the reader to \cite{BGM-11} for an extensive study of
noise effects on our inversion approach.

\section{Resistor networks as discrete models for EIT}
\label{sect:Rnets}
\setcounter{equation}{0} 
\begin{figure}[t!]
 \begin{center}
   \includegraphics[width=0.35\textwidth]{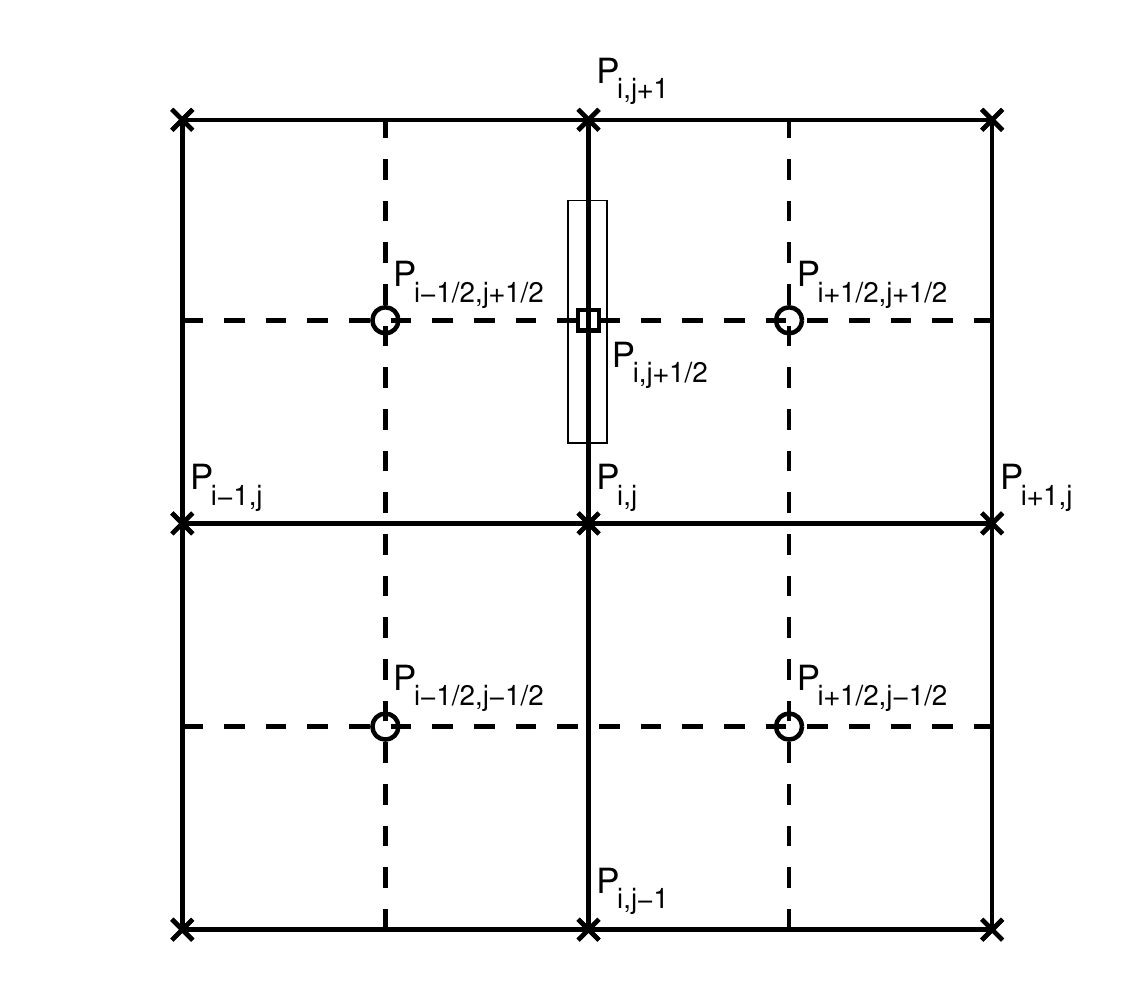}
 \end{center}
 \caption[Finite Volumes]{
   \label{fig:FinVol}
   \renewcommand{\baselinestretch}{1} \small\normalsize Finite volume
   discretization on a staggered grid. The primary grid lines are
   solid and the dual ones are dashed.  The primary grid nodes are
   indicated with $\times$ and the dual nodes with $\circ$.  The dual
   cell $C_{i,j}$, with vertices (dual nodes) $P_{i\pm\frac{1}{2},j\pm
     \frac{1}{2}}$ surrounds the primary node $P_{i,j}$. A resistor is
   shown as a rectangle with axis along a primary line, that
   intersects a dual line at the point indicated with $\Box$. }
\end{figure}
Resistor networks arise naturally in the context of finite volume
discretizations of the elliptic equation (\ref{eq:CONDeq}) on
staggered grids with interlacing primary and dual lines that may be
curvilinear, as explained in section \ref{sect:FVol}. Standard finite
volume discretizations use arbitrary, usually equidistant tensor
product grids. We consider \emph{optimal grids} that are designed to
obtain very accurate approximations of the measurements of the DtN
map, the data in the inverse problem. The geometry of these grids
depends on the measurement setup. We describe in section
\ref{sect:FBgrids} the type of grids used for the full measurement case, where
we have access to the entire boundary $\B$. The grids for the partial
boundary measurement setup are discussed later, in section
\ref{sect:2DPB}.

\subsection{Finite volume discretization and resistor networks}
\label{sect:FVol}

See Figure \ref{fig:FinVol} for an illustration of a staggered
grid.  The potential
$u(\bx)$ in equation (\ref{eq:CONDeq}) is discretized at the primary
nodes $P_{i,j}$, the intersection of the primary grid lines, and the
finite volumes method balances the fluxes across the boundary of the
dual cells $C_{ij}$,
\begin{equation}
\int_{C_{i,j}} \nabla \cdot \left[ \s(\bx) \nabla u(\bx) \right] d \bx
= \int_{\partial C_{i,j}} \s(\bx) \bn(\bx) \cdot \nabla u(\bx) ds(\bx)
= 0.
\label{eq:BalF}
\end{equation}
A dual cell $C_{i,j}$ contains a primary point $P_{i,j}$, it has
vertices (dual nodes) $P_{i\pm\frac{1}{2},j\pm \frac{1}{2}}$, and
boundary
\begin{equation}
\partial C_{i,j} =
\Sigma_{i, j+\frac{1}{2}} \cup \Sigma_{i+\frac{1}{2}, j} \cup
\Sigma_{i, j-\frac{1}{2}} \cup \Sigma_{i-\frac{1}{2}, j}, 
\label{eq:PartC}
\end{equation}
the union of the dual line segments $\Sigma_{i, j \pm \frac{1}{2}}=(
P_{i-\frac{1}{2}, j \pm \frac{1}{2}},P_{i+\frac{1}{2},j \pm
  \frac{1}{2}})$ and $\Sigma_{i \pm \frac{1}{2}, j}=( P_{i \pm
  \frac{1}{2}, j - \frac{1}{2}},P_{i \pm \frac{1}{2},j +
  \frac{1}{2}}).$ Let us denote by $\Pc = \{P_{i,j}\}$ the set of
primary nodes, and define the potential function $U:\Pc \to
\mathbb{R}$ as the finite volume approximation of $u(\bx)$ at the
points in $\Pc$,
\begin{equation}
U_{i,j} \approx u(P_{i,j}), \qquad P_{i,j} \in \Pc.
\end{equation}
The set $\Pc$ is the union of two disjoint sets $\Pc_\I$ and $\Pc_\B$
of interior and boundary nodes, respectively. Adjacent nodes in $\Pc$
are connected by edges in the set $\E \subset \Pc \times \Pc$. We
denote the edges by $E_{i,j\pm\frac{1}{2}} = (P_{i,j},P_{i,j\pm 1})$
and $E_{i\pm \frac{1}{2},j} = (P_{i\pm 1,j},P_{i,j})$.  

The finite volume discretization results in a system of linear
equations for the potential
\begin{eqnarray}
\gamma_{i+\frac{1}{2},j}(U_{i+1,j}-U_{i,j}) +
\gamma_{i-\frac{1}{2},j}(U_{i-1,j}-U_{i,j}) +
\gamma_{i,j+\frac{1}{2}}(U_{i,j+1}-U_{i,j}) +
\gamma_{i,j-\frac{1}{2}}(U_{i,j-1}-U_{i,j}) = 0,
\label{eq:FVeq}
\end{eqnarray}
with terms given by approximations of the fluxes
\begin{eqnarray}
 \int_{\Sigma_{i, j \pm \frac{1}{2}}} \hspace{-0.2in}
\sigma(\bx)\bn(\bx) \cdot \nabla u(\bx) d s(\bx) \approx
\gamma_{i,j \pm \frac{1}{2}}(U_{i,j \pm 1}-U_{i,j}), \nonumber \\
\int_{\Sigma_{i \pm \frac{1}{2}, j}} \hspace{-0.2in}
\sigma(\bx)\bn(\bx) \cdot \nabla u(\bx) d s (\bx) \approx
\gamma_{i \pm \frac{1}{2},j}(U_{i \pm 1,j}-U_{i,j}).
\label{eq:RAlex}
\end{eqnarray}
Equations \eqref{eq:FVeq} are Kirchhoff's law for the interior nodes
in a resistor network $(\Gamma,\gamma)$ with graph $\Gamma = (\Pc,\E)$
and conductance function $\gamma : \E \to \mathbb{R}^+$, that assigns
to an edge like $E_{i\pm \frac{1}{2},j}$ in $\E$ a positive
conductance $\gamma_{i\pm \frac{1}{2},j}$.  At the boundary nodes we
discretize either the Dirichlet conditions (\ref{eq:DirBC}), or the
Neumann conditions (\ref{eq:NeumBC}), depending on what we wish to
approximate, the DtN or the NtD map.

To write the network equations in compact (matrix) form, let us number
the primary nodes in some fashion, starting with the interior ones and
ending with the boundary ones. Then we can write $\Pc = \left\{ {\bf p}_q
\right\}$, where ${\bf p}_q$ are the numbered nodes.  They correspond to
points like $P_{i,j}$ in Figure \ref{fig:FinVol}.  Let also $\bUI$ and
$\bUB$ be the vectors with entries given by the potential at the
interior nodes and boundary nodes, respectively.  The vector of
boundary fluxes is denoted by $\bJB$. We assume throughout that
there are $n$ boundary nodes, so $\bUB, \bJB \in \mathbb{R}^n$.
The network equations are
\begin{equation}
\label{eq:Kirc}
{\bf K} \bU 
 = \left( \begin{array}{c} {\bf 0} \\ \bJ_\B
\end{array} \right), \qquad  \bU = \left( \begin{array}{c} \bU_\I \\ \bU_\B
\end{array} \right), \qquad 
{\bf K} = \left( \begin{array}{cc} \KII & \KIB 
    \\ \KIB & \KBB \end{array} \right),
\end{equation}
where ${\bf K} = \left( K_{ij}\right)$ is the Kirchhoff matrix with entries
\begin{equation}
K_{i,j} = \left\{ \begin{array}{cl} - \gamma(E), \quad & \mbox{if} ~ i
  \ne j \mbox{ and } E = \left( {\bf p}_i,{\bf p}_j\right) \in \E,
  \\ 0, & \mbox{if} ~ i \ne j \mbox{ and } \left( {\bf p}_i,{\bf
    p}_j\right) \notin \E, \\ \displaystyle \sum_{k: ~ E = \left( {\bf
      p}_i,{\bf p}_k\right) \in \E} \hspace{-0.05in} \gamma(E), &
  \mbox{if} ~ i = j.
\end{array} \right. 
\label{eq:KircM}
\end{equation}
In (\ref{eq:Kirc}) we write it in block form, with $\KII$ the block
with row and column indices restricted to the interior nodes, $\KIB$
the block with row indices restricted to the interior nodes and column
indices restricted to the boundary nodes, and so on. Note that
${\bf K}$ is symmetric, and its rows and columns sum to zero, which
is just the condition of conservation of currents.

It is shown in \cite{CurtMooMor} that the potential $\bU$ satisfies a
discrete maximum principle. Its minimum and maximum entries are
located on the boundary. This implies that the network equations with
Dirichlet boundary conditions
\begin{equation}
\KII \bUI = -\KIB \bUB 
\end{equation}
have a unique solution if $\KIB$ has full rank. That is to
say, $\KII$ is invertible and we can eliminate $\bUI$ from
(\ref{eq:Kirc}) to obtain
\begin{equation}
\bJB = \left( \KBB - \KBI {\bf K}^{-1}_{_{\I \I}} \KIB \right) \bUB = \LG
\bUB.
\label{eq:RDtN}
\end{equation} 
The matrix $\LG \in \mathbb{R}^{n \times n}$ is
the Dirichlet to Neumann map of the network.  It takes the boundary
potential $\bUB$ to the vector $\bJB$ of boundary fluxes, and is given
by the Schur complement of the block $\KBB$
\begin{equation}
\LG = \KBB - \KBI {\bf K}^{-1}_{_{\I \I}}\KIB.
\label{eq:R1DtN}
\end{equation}
The DtN map is symmetric, with nontrivial null space spanned by the
vector ${\bf 1}_\B \in \mathbb{R}^n$ of all ones.  The symmetry
follows directly from the symmetry of ${\bf K}$. Since the columns
of ${\bf K}$ sum to zero, ${\bf K} {\bf 1} = {\bf 0}$, where
${\bf 1}$ is the vector of all ones. Then, (\ref{eq:RDtN}) gives
$\bJ_\B = {\bf 0} = \LG {\bf 1}_\B$, which means that ${\bf
  1}_\B$ is in the null space of $\LG$.

The inverse problem for a network $(\Gamma,\gamma)$ is to determine
the conductance function $\gamma$ from the DtN map $\LG$. The graph
$\Gamma$ is assumed known, and it plays a key role in the solvability
of the inverse problem
\cite{CurtMooMor,CurtIngMor,IngerLayer,deverdiere1994rep,
  deverdiere1996rep}. More precisely, $\Gamma$ must satisfy a certain
criticality condition for the network to be uniquely recoverable from
$\LG$, and its topology should be adapted to the type of measurements
that we have. We review these facts in detail in sections
\ref{sect:Layered}-\ref{sect:2DPB}. We also show there how to relate
the continuum DtN map $\Lambda_\s$ to the discrete DtN map $\LG$. The
inversion algorithms in this paper use the solution of the discrete
inverse problem for networks to determine approximately 
the solution $\s(\bx)$ of the continuum EIT problem.

\subsection{Tensor product grids for the full boundary measurements setup}
\label{sect:FBgrids}

\begin{figure}[t]
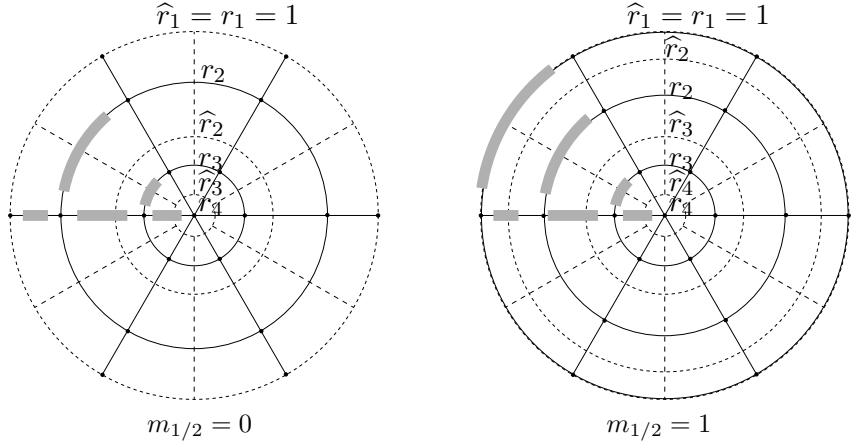

   \centering
   \begin{tabular}{cc}
     \input{Figures/c2_0_6} &
   \hspace{-2em}\input{Figures/c2_1_6}\\
   \hspace{-4em}$\mhf = 0$ & \hspace{-6em}$\mhf = 1$
   \end{tabular}

  \caption[Examples of the two kinds of grids
    used.]{\label{fig:CircGrids} \renewcommand{\baselinestretch}{1}
    \small\normalsize Examples of grids. The primary grid lines are
    solid and the dual ones are dotted. Both grids have $n = 6$
    primary boundary points, and index of the layers $\ell = 3$.  We
    have the type of grid indexed by $\mhf = 0$ on the left and by
    $\mhf = 1$ on the right. }
 \end{figure}

In the full boundary measurement setup, we have access to the entire
boundary $\B$, and it is natural to discretize the domain
(\ref{eq:Om}) with tensor product grids that are uniform in angle, as
shown in Figure \ref{fig:CircGrids}.  Let
\begin{equation}
\T_j = \frac{2 \pi(j-1)}{n}, \qquad \hT_j = \frac{2 \pi
  \left(j-1/2\right)}{n}, \qquad j = 1, \ldots, n,
\label{eq:angles}
\end{equation}
be the angular locations of the primary and dual nodes.  The radii of
the primary and dual layers are denoted by $r_i$ and $\hr_i$, and we
count them starting from the boundary. We can have two types of grids,
so we introduce the parameter $\mhf \in \{0,1\}$ to distinguish
between them. We have
\begin{equation}
1 = r_1 = \hr_1 > r_2 > \hr_2 > \ldots > r_{\ell} > \hr_{\ell} >
r_{\ell+1} \ge 0
\label{eq:Grid1}
\end{equation}
when $\mhf = 0$, and 
\begin{equation}
1 = \hr_1 = r_1 > \hr_2 > r_2 > \ldots > r_{\ell} > \hr_{\ell+1} >
r_{\ell+1} \ge 0
\label{eq:Grid2}
\end{equation}
for $\mhf = 1$. In either case there are $\ell+1$ primary layers and
$\ell+ \mhf$ dual ones, as illustrated in Figure \ref{fig:CircGrids}.
We explain in sections \ref{sect:Layered} and \ref{sect:2DFB} how to
place optimally in the interval $[0,1]$ the primary and dual radii, so
that the finite volume discretization gives an accurate approximation
of the DtN map $\Lambda_\s$.

The graph of the network is given by the primary grid. We follow
\cite{CurtMooMor,CurtIngMor} and call it a circular network. It has
$n$ boundary nodes and $n(2\ell +\mhf-1)$ edges. Each edge is
associated with an unknown conductance that is to be determined from
the discrete DtN map $\LG$, defined by measurements of $\Lambda_\s$,
as explained in sections \ref{sect:Layered} and \ref{sect:2DFB}.
Since $\LG$ is symmetric, with
columns summing to zero, it contains $n(n-1)/2$ measurements. Thus, we
have the same number of unknowns as data points when
\begin{equation}
2 \ell + \mhf - 1 = \frac{n-1}{2}, \qquad n = \mbox{odd integer}.
\label{eq:GridCOND}
\end{equation}
This condition turns out to be necessary and sufficient for the DtN
map to determine uniquely a circular network, as shown in
\cite{deverdiere1996rep,CurtIngMor,BorDruGue}. We assume henceforth
that it holds.
 
\section{Layered media}
\label{sect:Layered}
\setcounter{equation}{0} 

In this section we assume a layered conductivity function $\s(r)$ in
$\Omega$, the unit disk, and access to the entire boundary $\B$. Then,
the problem is rotation invariant and can be simplified by writing the
potential as a Fourier series in the angle $\T$. We begin in section
\ref{sect:Lay_eig} with the spectral decomposition of the continuum
and discrete DtN maps and define their eigenvalues, which contain all
the information about the layered conductivity. Then, we explain in
section \ref{sect:Lay_rat} how to construct finite volume grids that
give discrete DtN maps with eigenvalues that are accurate, rational
approximations of the eigenvalues of the continuum DtN map.  One such
approximation brings an interesting connection between a classic
Sturm-Liouville inverse spectral problem
\cite{gel1951determination,chadan1997introduction,hochstadt1973inverse,marchenko2011sturm,mclaughlin1987uniqueness}
and an inverse eigenvalue problem for Jacobi matrices
\cite{chu2002structured}, as described in sections
\ref{sect:TRUNC_meas} and \ref{sect:Lay_ISP}. This connection allows
us to solve the continuum inverse spectral problem with efficient,
linear algebra tools. The resulting algorithm is the first example of
resistor network inversion on optimal grids proposed and analyzed in
\cite{BorDruKni}, and we review its convergence study in section
\ref{sect:Lay_ISP}.

\subsection{Spectral decomposition of the continuum and discrete 
DtN maps}
\label{sect:Lay_eig}
Because equation (\ref{eq:CONDeq}) is separable in layered media, we 
write the potential $u(r,\theta)$ as a Fourier series
\begin{equation}
u(r,\theta) = v_\B(0) + \sum_{k \in \mathbb{Z}, k \ne 0} v(r,k) e^{i k \theta},
\label{eq:L1}
\end{equation}
with coefficients $v(r,k)$ satisfying the differential
equation
\begin{equation}
\frac{r}{\s(r)} \frac{d}{dr} \left[ r \s(r) \frac{d v(r,k)}{dr}
  \right] - k^2 v(r,k) = 0, \qquad r \in (0,1), 
\label{eq:L2}
\end{equation}
and the condition
\begin{equation}
v(0,k) = 0.
\label{eq:L3}
\end{equation}
The first term $v_\B(0)$ in (\ref{eq:L1}) is the average boundary potential 
\begin{equation}
v_\B(0) = \frac{1}{2 \pi} \int_0^{2 \pi} u(1,\theta) \, d \theta.
\end{equation}
The boundary conditions at $r = 1$ are Dirichlet or Neumann, depending
on which map we consider, the DtN or the NtD map.  

\subsubsection{The DtN map}
\label{sect:DtN}
The DtN map is determined by the potential $v$ satisfying
(\ref{eq:L2}-\ref{eq:L3}), with Dirichlet boundary condition 
\begin{equation}
v(1,k) = v_\B(k),
\label{eq:L3p}
\end{equation}
where $v_\B(k)$ are the Fourier coefficients of the boundary potential
$u_\B(\T)$. The normal boundary flux has the Fourier series expansion
\begin{equation}
\s(1) \frac{\partial u(1,\theta)}{\partial r} = \Lambda_\s \uB(\theta)
= \s(1) \sum_{k \in \mathbb{Z}, k \ne 0} \frac{d v(1,k)}{dr} e^{i k
  \theta},
\label{eq:L4}
\end{equation} 
and we assume for simplicity that $\s(1) = 1$. Then, we deduce
formally from (\ref{eq:L4}) that $e^{i k \theta}$ are the
eigenfunctions of the DtN map $\Lambda_\s$, with eigenvalues
\begin{equation}
f(k^2) = {\frac{d v(1,k)}{dr}}/{v(1,k)}.
\label{eq:L5}
\end{equation}
Note that $f(0) = 0$.

A similar diagonalization applies to the DtN map $\LG$ of networks
arising in the finite volume discretization of (\ref{eq:CONDeq}) if
the grids are equidistant in angle, as described in section
\ref{sect:FBgrids}. Then, the resulting network is layered in the
sense that the conductance function is rotation invariant.  We can
define various quadrature rules in (\ref{eq:RAlex}), with minor
changes in the results \cite[Section 2.4]{BDM-10}. In this section we
use the definitions
\begin{equation}
\label{eq:alphas}
\gamma_{j+\frac{1}{2},q} = \frac{h_\T}{z(r_{j+1}) - z(r_{j})} =
\frac{h_\T}{ \al_{j}}, \qquad \gamma_{j,q + \frac{1}{2}} =
\frac{\hz(\hr_{j+1}) -\hz(\hr_{j})}{h_\T} = \frac{\hal_j}{h_\T},
\end{equation}
derived in appendix \ref{ap:Quadrature}, where $h_\T = 2\pi/n$ and
\begin{equation}
z(r) = \int_r^1 \frac{dt}{t \s(t)}, \qquad \hz(r) = \int_{r}^1
\frac{\s(t)}{t} dt.
\label{eq:L8}
\end{equation}
The network equations (\ref{eq:FVeq}) become
\begin{equation}
\frac{1}{\hal_j} \left(\frac{U_{j+1,q}-U_{j,q}}{\al_j} - \frac{U_{j,q}
  - U_{j-1,q}}{\al_{j-1}}\right) - \frac{2
  U_{j,q}-U_{j,q+1}-U_{j,q-1}}{h_\T^2} = 0,
\label{eq:L6}
\end{equation}
and we can write them in block form as 
\begin{equation}
  \frac{1}{\hal_j} \left(\frac{\bU_{j+1}-\bU_{j}}{\al_j} - \frac{\bU_{j}
      - \bU_{j-1}}{\al_{j-1}}\right) - \left[-\partial^2_\T \right] 
  \bU_j = {\bf 0},
\label{eq:L6B}
\end{equation}
where 
\begin{equation}
{\bf U}_j = \left(U_{j,1}, \ldots, U_{j,n}\right)^T,
\label{eq:L6BB}
\end{equation}
and $\left[-\partial^2_\T \right]$ is the circulant matrix
\begin{equation}
\left[-\partial^2_\T \right] = \frac{1}{h^2_\T} \left(
\begin{array}{ccccccc}
2 & -1 & 0 & \ldots & \ldots & 0 & -1 \\
-1 & 2 & 1 & 0 & \ldots & 0 & 0 \\
\ddots & \ddots & \ddots & \ddots & \ddots & \ddots & \ddots \\
-1 & 0 & \ldots & \ldots & 0 & -1 & 2 \end{array} 
\right),
\label{eq:CIRCULANT}
\end{equation}
the discretization of the operator $-\partial^2_\T$ with periodic
boundary conditions. It has the eigenvectors 
\begin{equation}
\left[ e^{i k \T}
  \right] = \left( e^{i k \T_1}, \ldots, e^{i k \T_n} \right)^T, 
\label{eq:L6BBB}
\end{equation}
with entries given by the restriction of the continuum eigenfunctions
$e^{i k \T}$ at the primary grid angles. Here $k$ is integer,
satisfying $|k| \le (n-1)/2$, and the eigenvalues are $\om_k^2$, where
\begin{equation}
  \om_k = |k| \left| \mbox{sinc} \left( \frac{k h_\T}{2}\right)
  \right|,
\label{eq:eig}
\end{equation}
and $\mbox{sinc}(x) = \sin(x)/x$. Note that $\om_k^2 \approx k^2$ only
for $|k| \ll n$.

To determine the spectral decomposition of the discrete DtN map $\LG$
we proceed as in the continuum and write the potential $\bU_{j}$ as a
Fourier sum
\begin{equation}
\bU_{j} = v_\B(0){\bf 1}_\B + \sum_{|k| \le \frac{n-1}{2}, k \ne 0}
V_j(k) \left[e^{i k \T}\right],
\end{equation}
where we recall that ${\bf 1}_\B\in \R^{n}$ is a vector of all ones.
We obtain the finite difference equation for the coefficients
$V_j(k)$,
\begin{equation}
\frac{1}{\hal_j}\left(\frac{V_{j+1}(k)-V_{j}(k)}{\al_j} -
\frac{V_{j}(k) - V_{j-1}(k)}{\al_{j-1}}\right) - \om_k^2 V_j(k) = 0,
\label{eq:L9}
\end{equation}
where $j = 2, 3, \ldots, \ell$.  It is the discretization of
(\ref{eq:L2}) that takes the form
\begin{equation}
\frac{d}{d \hz} \left( \frac{dv(z,k)}{dz} \right) - k^2 v(z,k) = 0,
\label{eq:L11}
\end{equation} 
in the coordinates (\ref{eq:L8}), where we let in an abuse of notation
$v(r,k) \leadsto v(z,k)$.  The boundary condition at $r = 0$  is mapped to 
\begin{equation}
\lim_{z \to \infty} v(z,k) = 0,
\label{eq:CentC}
\end{equation}
and it is implemented in the discretization as $ V_{\ell+1}(k) = 0.  $
At the boundary $r = 1$, where $z= 0$, we specify $V_1(k)$ as some
approximation of $v_\B(k)$.

The discrete DtN map $\LG$ is diagonalized in the basis $\{[e^{i k
    \T}]\}_{|k| \le \frac{n-1}{2}}$, and we denote its eigenvalues by
    $F(\om^2_k)$. Its definition depends on the type of grid that we
    use, indexed by $\mhf$, as explained in section
    \ref{sect:FBgrids}.  In the case $\mhf = 0$, the first radius next
    to the boundary is $r_2$, and we define the boundary flux at
    $\hr_1 = 1$ as $(V_1(k)-V_2(k))/\al_1$.  When $\mhf = 1$, the
    first radius next to the boundary is $\hr_2$, so to compute the
    flux at $\hr_1$ we introduce a ghost layer at $r_0 > 1$ and use
    equation (\ref{eq:L9}) for $j = 1$ to define the boundary flux as
\[
\frac{V_0(k)-V_1(k)}{\alpha_o} = \hal_1\om_k^2 V_1(k) +
\frac{V_1(k)-V_2(k)}{\al_1}.
\]
Therefore, the eigenvalues of the discrete DtN map are
\begin{equation}
F(\om_k^2) = \mhf \hal_1 \om_k^2 + \frac{V_1(k)-V_2(k)}{\alpha_1 V_1(k)}.
\label{eq:DtNMHf0}
\end{equation}

\subsubsection{The NtD map}
\label{sect:NtD}
The NtD map $\Lambda_\s^\dagger$ has eigenfunctions $e^{i k \T}$ for
$k \ne 0$ and eigenvalues $\fd(k^2) = 1/f(k^2)$. Equivalently, in
terms of the solution $v(z,k)$ of equation (\ref{eq:L11}) with
boundary conditions (\ref{eq:CentC}) and 
\begin{equation}
-\frac{d v(0,k)}{dz} = \frac{1}{2 \pi}\int_0^{2 \pi} J_\B(\theta)
 e^{-i k \theta} d \T = \varphi_{_\B}(k),
\label{eq:L12}
\end{equation}
we have 
\begin{equation}
f^\dagger(k^2) = \frac{v(0,k)}{\varphi_{_\B}(k)}.
\label{eq:L13}
\end{equation}
In the discrete case, let us use the grids with $\mhf = 1$. We obtain
that the potential $V_j(k)$ satisfies (\ref{eq:L9}) for $j = 1, 2,
\ldots, \ell$, with boundary conditions
\begin{equation}
-\frac{V_1(k)-V_0(k)}{\al_0} = \Phi_{_\B}(k), \qquad V_{\ell+1} = 0.
\label{eq:L14}
\end{equation}
Here $\Phi_\B(k)$ is some approximation of $\varphi_\B(k)$.  The
eigenvalues of $\LGd$ are
\begin{equation}
F^\dagger(\om_k^2) = \frac{V_1(k)}{\Phi_{_\B}(k)}.
\label{eq:L15}
\end{equation}

\subsection{Rational approximations, optimal grids and reconstruction
  mappings}
\label{sect:Lay_rat}
Let us define by analogy to (\ref{eq:L13}) and (\ref{eq:L15}) the
functions
\begin{equation}
\fd(\la) = \frac{v(0)}{\varphi_{_\B}}, \qquad \Fd(\la)
 = \frac{V_1}{\Phi_{_\B}},
\label{eq:R1}
\end{equation}
where $v$ solves equation (\ref{eq:L11}) with $k^2$ replaced by $\la$
and $V_j$ solves equation (\ref{eq:L9}) with $\om_k^2$ replaced by
$\la$. The spectral parameter $\la$ may be complex, satisfying $\la
\in \mathbb{C}\setminus(-\infty, 0]$.  For simplicity, we suppress in
the notation the dependence of $v$ and $V_j$ on $\la$. We consider in
detail the discretizations on grids indexed by $\mhf = 1$, but the
results can be extended to the other type of grids, indexed by $\mhf =
0$.

\begin{lemma}
\label{lem.1}
The function $\fd(\la)$ is of form
\begin{equation}
\fd(\la) = \int_{-\infty}^0 \frac{d \mu(t)}{\la-t},
\label{eq:Stieljf}
\end{equation}
where $\mu(t)$ is the positive spectral measure on $(-\infty,0]$ of
  the differential operator $d_{\hz} d_z$, with homogeneous Neumann
  condition at $z = 0$ and limit condition (\ref{eq:CentC}). The
  function $\Fd(\la)$ has a similar form
\begin{equation}
\Fd(\la) = \int_{-\infty}^0 \frac{d \mu^F(t)}{\la-t},
\label{eq:StieljF}
\end{equation}
where $\mu^F(t)$ is the spectral measure of the difference operator in
(\ref{eq:L9}) with boundary conditions (\ref{eq:L14}).
\end{lemma}
\emph{Proof:} The result (\ref{eq:Stieljf}) is shown in
\cite{kac1974spectral} and it says that $\fd(\la)$ is essentially a
Stieltjes function. To derive the representation (\ref{eq:StieljF}), we
write our difference equations in matrix form for ${\bf V} = (V_1,
\ldots, V_{\ell})^T$,
\begin{equation}
\left( {\bf A} - \la {\bf I} \right){\bf V} =
-\frac{\Phi_{_\B}(\la)}{\hal_1} {\bf e}_1.
\label{eq:L9Mat}
\end{equation}
Here ${\bf I}$ is the $\ell \times \ell$ identity matrix, ${\bf e}_1 =
(1, \ldots, 0)^T \in \mathbb{R}^\ell$ and ${\bf A}$ is the tridiagonal
matrix with entries
\begin{equation}
A_{ij} = \left\{ \begin{array}{lc} -\frac{1}{\hal_i} \left(
  \frac{1}{\al_i} + \frac{1}{\al_{i-1}} \right)\delta_{i,j} +
  \frac{1}{\hal_i \al_{i-1}} \delta_{i-1,j} +
  \frac{1}{\hal_i \al_i}
  \delta_{i+1,j} ~ ~ & \mbox{if} ~ 1< i \le \ell, ~ 1 \le j \le \ell, \\
-\frac{1}{\hal_1 \al_1}\delta_{1,j} +
  \frac{1}{\hal_1 \al_1} \delta_{2,j} & \mbox{if} ~ i = 1, ~ 1 \le j \le \ell.
\end{array} \right. 
\label{eq:JacA}
\end{equation}
The Kronecker delta symbol $\delta_{i,j}$ is one when $i = j$ and zero
otherwise. Note that ${\bf A}$ is a Jacobi matrix when it is defined
on the vector space $\mathbb{R}^\ell$ with weighted inner product
\begin{equation}
\left< {\bf a},{\bf b} \right> = \sum_{j=1}^\ell \hal_j a_j b_j , \qquad 
{\bf a} = (a_1, \ldots, a_\ell)^T, ~ ~ 
{\bf b} = (b_1, \ldots, b_\ell)^T.
\label{eq:L16}
\end{equation}
That is to say, 
\begin{equation}
\label{eq:tA}
\widetilde{\bf A} = \mbox{diag} \left(\hal_1^{1/2}, \ldots,
\hal_\ell^{1/2} \right) {\bf A} \, \mbox{diag} \left(\hal_1^{-1/2},
\ldots, \hal_\ell^{-1/2} \right)
\end{equation}  
is a symmetric, tridiagonal matrix, with negative entries on its
diagonal and positive entries on its upper/lower diagonal. It follows
from \cite{chu2002structured} that ${\bf A}$ has simple, negative
eigenvalues $-\delta^2_j$ and eigenvectors ${\bf Y}_j = \left(Y_{1,j},
\ldots, Y_{\ell,j}\right)^T$ that are orthogonal with respect to the
inner product (\ref{eq:L16}). We order the eigenvalues as
\begin{equation}
\delta_1 < \delta_2 < \ldots < \delta_\ell,
\label{eq:L17}
\end{equation}
and normalize the eigenvectors 
\begin{equation}
\|{\bf Y}_j\|^2 = \left< {\bf Y}_j,{\bf Y}_j\right> = \sum_{p=1}^\ell 
\hal_p^2 Y_{p,j}^2 = 1.
\label{eq:NORM}
\end{equation}
Then, we obtain from (\ref{eq:R1}) and
(\ref{eq:L9Mat}), after expanding ${\bf V}$ in the basis of the
eigenvectors, that
\begin{equation}
\Fd(\la) = \sum_{j=1}^\ell \frac{Y_{1,j}^2}{\la + \delta_j^2}.
\label{eq:L18}
\end{equation}
This is precisely (\ref{eq:StieljF}), for the discrete spectral measure 
\begin{equation}
\mu^F(t) = - \sum_{j=1}^\ell \xi_j H\left(-t - \delta_j^2\right), \qquad 
\xi_j = Y_{1,j}^2,
\label{eq:L19}
\end{equation}
where $H$ is the Heaviside step function. $\Box$.

Note that any function of the form (\ref{eq:L18}) defines the
eigenvalues $\Fd(\om_k^2)$ of the NtD map $\LGd$ of a finite volumes
scheme with $\ell+1$ primary radii and uniform discretization in
angle.  This follows from the decomposition in section
\ref{sect:Lay_eig} and the results in \cite{kac1974spectral}. Note
also that there is an explicit, continued fraction representation of
$\Fd(\la)$, in terms of the network conductances, i.e., the parameters
$\al_j$ and $\hal_j$, 
\begin{equation}
\label{eq:CFr}
\Fd(\la) = \cfrac{1}{\hal_1 \la+ \cfrac{1}{\al_1+\dots
 \cfrac{1}{\hal_{\ell} \la + \cfrac{1}{\al_\ell}}}}.
\end{equation}
This representation is known in the theory of rational function
approximations \cite{nikishin1991rational,kac1974spectral} and its
derivation is given in appendix \ref{ap:CF}.

Since both $\fd(\la)$ and $\Fd(\la)$ are Stieltjes functions, we can
design finite volume schemes (i.e., layered networks) with accurate,
rational approximations $\Fd(\la)$ of $\fd(\la)$. There are various
approximants $\Fd(\la)$, with different rates of convergence to
$\fd(\la)$, as $\ell \to \infty$. We discuss two choices below, in
sections \ref{sect:RAT_interp} and \ref{sect:TRUNC_meas}, but refer
the reader to \cite{druskin2000gsr,DruKni,
druskin2002three} for details on various Pad\'{e} approximants and the
resulting discretization schemes.  No matter which approximant we
choose, we can compute the network conductances, i.e., the parameters
$\al_j$ and $\hal_j$ for $j = 1, \ldots, \ell$, from $2 \ell$
measurements of $\fd(\la)$. The type of measurements dictates  the type
of approximant, and only some of them are directly accessible in the
EIT problem. For example, the spectral measure $\mu(\la)$ cannot be
determined in a stable manner in EIT. However, we can measure the
eigenvalues $\fd(k^2)$ for integer $k$, and thus we can design a
rational, multi-point Pad\'{e} approximant. 

\begin{remark}
\label{rem.1}
We describe in detail in appendix \ref{ap:Euclid} how to determine the
parameters $\{\al_j, \hal_j\}_{j = 1, \ldots, \ell}$ from $2 \ell$
point measurements of $\fd(\la)$, such as $\fd(k^2)$, for $k = 1,
\ldots, \frac{n-1}{2} = 2 \ell$. The are two steps. The first is to
write $\Fd(\la)$ as the ratio of two polynomials of $\la$, and
determine the $2 \ell$ coefficients of these polynomials from the
measurements $\Fd(\om_k^2)$ of $\fd(k^2)$, for $1 \le k \le
\frac{n-1}{2}$. See section \ref{sect:RAT_interp} for examples of such
measurements.  The exponential instability of EIT comes into play in
this step, because it involves the inversion of a Vandermonde
matrix. It is known \cite{gautschi1987lower} that such matrices have
condition numbers that grow exponentially with the dimension
$\ell$. The second step is to determine the parameters $\{\al_j,
\hal_j\}_{j = 1, \ldots, \ell}$ from the coefficients of the
polynomials. This can be done in a stable manner with the Euclidean
division algorithm \cite{lang2005undergraduate}.
\end{remark}

The approximation problem can also be formulated in terms of the DtN
map, with $F(\la) = 1/\Fd(\la)$.  Moreover, the representation
(\ref{eq:CFr}) generalizes to both types of grids, by replacing
$\hal_1 \la$ with $\hal_1 \mhf \la$. Recall equation
(\ref{eq:DtNMHf0}) and note the parameter $\hal_1$ does not play any
role when $\mhf = 0$.

\subsubsection{Optimal grids and reconstruction mappings}
\label{sect:Lay_OPT}
Once we have determined the network conductances, that is the
coefficients
\begin{equation}
\al_j = \int_{r_{j+1}}^{r_j} \frac{dr}{r \s(r)}, \qquad 
\hal_j = \int_{\hr_{j+1}}^{\hr_{j}} \frac{\s(r)}{r} d r,\qquad 
j = 1, \ldots, \ell,
\label{eq:O1}
\end{equation}
we could determine the optimal placement of the radii $r_j$ and
$\hr_j$, if we knew the conductivity $\s(r)$. But $\s(r)$ is the
unknown in the inverse problem. The key idea behind the resistor
network approach to inversion is that the grid depends only weakly on
$\s$, and we can compute it approximately for the reference
conductivity $\sr \equiv 1$.

Let us denote by $\fdr(\la)$ the analog of (\ref{eq:R1}) for
conductivity $\sr$, and let $\Fdr(\la)$ be its rational approximant
defined by (\ref{eq:CFr}), with coefficients $\ar_j$ and $\har_j$
given by
\begin{equation}
  \ar_j = \int_{\rr_{j+1}}^{\rr_j} \frac{dr}{r} = 
  \log \frac{\rr_j}{\rr_{j+1}}, \qquad 
  \har_j = \int_{\hr_{j+1}}^{\hr_{j}} \frac{dr}{r} = 
  \log \frac{\hrr_{j}}{\hrr_{j+1}}, 
  \qquad j = 1, \ldots,\ell.
\label{eq:O2}
\end{equation}
Since $\rr_1 = \hrr_1 = 1$, we obtain 
\begin{equation}
\rr_{j+1} = \exp \left( - \sum_{q=1}^j \ar_q \right), \qquad 
\hrr_{j+1} =  \exp \left( - \sum_{q=1}^j \har_q \right), \qquad 
j = 1, \ldots, \ell. 
\label{eq:O3}
\end{equation}
We call the radii (\ref{eq:O3}) {\em optimal}. The name refers to the
fact that finite volume discretizations on grids with such radii give
an NtD map that matches the measurements of the continuum map
$\Lambda_{\sr}^\dagger$ for the reference conductivity $\sr$. 
\begin{remark}
\label{rem:OGRID}
It is essential that the parameters $\{\al_j,\hal_j\}$ and
$\{\ar_j,\har_j\}$ are computed from the same type of measurements.
For example, if we measure $\fd(k^2)$, we compute $\{\al_j,\hal_j\}$
so that
\[
\Fd(\om_k^2) = \fd(k^2), 
\]
and $\{\ar_j,\har_j\}$ so that 
\[
\Fdr(\om_k^2) = \fdr(k^2), 
\]
where $k = 1, \ldots, (n-1)/2$.  This is because the distribution of
the radii (\ref{eq:O3}) in the interval $[0,1]$ depends on what
measurements we make, as illustrated with examples in sections
\ref{sect:RAT_interp} and \ref{sect:TRUNC_meas}.
\end{remark}

Now let us denote by $\D_{n}$ the set in $\mathbb{R}^{\frac{n-1}{2}}$
of measurements of $\fd(\la)$, and introduce the reconstruction
mapping $\Qc_n$ defined on $\D_{n}$, with values in
$\mathbb{R}_+^{\frac{n-1}{2}}$.  It takes the measurements of
$\fd(\la)$ and returns the $(n-1)/2$ positive numbers
\begin{eqnarray}
  \s_{j+1-\mhf} &=& \frac{\hal_j}{\har_j}, 
  \quad j = 2-\mhf, \ldots \ell, \nonumber \\
  \hs_{j+\mhf} &=& \frac{\ar_j}{\al_j}, \quad j = 1, 2, \ldots, \ell.,
\label{eq:LQn}
\end{eqnarray}
where we recall the relation (\ref{eq:GridCOND}) between $\ell$ and
$n$.  We call $\Qc_n$ a reconstruction mapping because if we take
$\s_j$ and $\hs_{j}$ as point values of a conductivity at nodes
$\rr_j$ and $\hrr_{j}$, and interpolate them on the optimal grid, we
expect to get a conductivity that is close to the interpolation of the
true $\s(r)$. This is assuming that the grid does not depend strongly
on $\s(r)$.  The proof that the resulting sequence of conductivity
functions indexed by $\ell$ converges to the true $\s(r)$ as $\ell \to
\infty$ is carried out in \cite{BorDruKni}, given the spectral measure
of $\fd(\la)$. We review it in section \ref{sect:Lay_ISP}, and discuss
the measurements in section \ref{sect:TRUNC_meas}. The convergence
proof for other measurements remains an open question, but the
numerical results indicate that the result should hold. Moreover, the
ideas extend to the two dimensional case, as explained in detail in
sections \ref{sect:2DFB} and \ref{sect:2DPB}.

\subsubsection{Examples of rational interpolation grids}
\label{sect:RAT_interp}
Let us begin with an example that arises in the discretization of the
problem with lumped current measurements
\begin{equation*}
J_q = \frac{1}{h_\T} \int_{\hT_{q}}^{\hT_{q+1}} \Lambda_\s u_\B(\T)
d \T,
\end{equation*}
for $h_\T = \frac{2\pi}{n}$, and vector $ \bU_\B = \left(u_\B(\T_1),
  \ldots, u_\B(\T_n)\right)^T $ of boundary potentials.  If we take
harmonic boundary excitations $ u_\B(\theta) = e^{i k \T}$, the
eigenfunction of $\Lambda_\s$ for eigenvalue $f(k^2)$, we obtain
\begin{equation}
  J_q = \frac{1}{h_\T} \int_{\hT_{q}}^{\hT_{q+1}} \Lambda_\s e^{i k
    \T} d \T = f(k^2)\left| \mbox{sinc}\left(\frac{k
        h_\T}{2}\right) \right| e^{i k \T_q} = \frac{f(k^2)}{|k|} \om_k e^{i
    k \T_q}, \qquad q = 1, \ldots, n.
\label{eq:Rat.2}
\end{equation}
These measurements, for all integers $k$ satisfying $|k| \le
\frac{n-1}{2}$, define a discrete DtN map ${\bf M}_n(\Lambda_\s)$. It
is a symmetric matrix with eigenvectors $[e^{i k \T}] = \left(e^{i k
  \T_1}, \ldots, e^{i k \T_n} \right)^T$, and eigenvalues
$\frac{f(k^2)}{|k|} \om_k$.

The approximation problem is to find the finite volume discretization
with DtN map $\LG = {\bf M}_n(\Lambda_\s)$. Since both $\LG$ and ${\bf
  M}_n$ have the same eigenvectors, this is equivalent to the rational
approximation problem of finding the network conductances
(\ref{eq:alphas}) (i.e., $\al_j$ and $\hal_j$), so that
\begin{equation}
F(\om_k^2) = \frac{f(k^2)}{|k|} \om_k, \qquad k = 1, \ldots,
\frac{n-1}{2}.
\label{eq:Rat.3}
\end{equation}
The eigenvalues depend only on $|k|$, and the case $k = 0$ gives no
information, because it corresponds to constant boundary potentials
that lie in the null space of the DtN map. This is why we take in
(\ref{eq:Rat.3}) only the positive values of $k$, and obtain the same
number $(n-1)/2$ of measurements as unknowns: $\{\al_j\}_{j = 1,
\ldots, \ell}$ and $\{\hal_j\}_{j = 2-\mhf, \ldots, \ell}$.

\begin{figure}[t!]
 \begin{center}
   \includegraphics[width=0.45\textwidth]{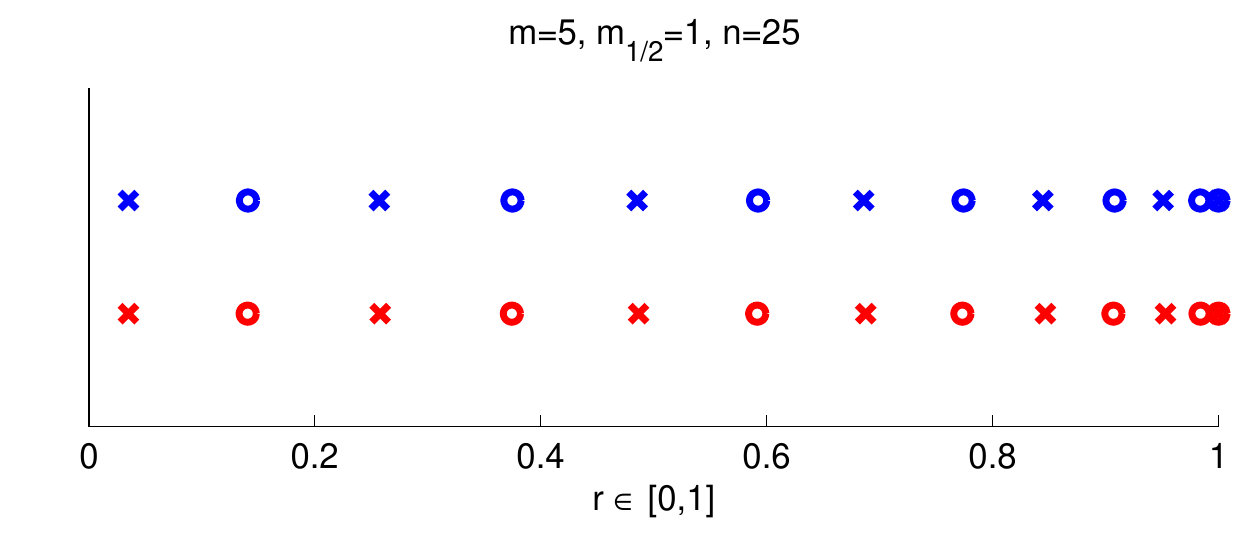}
   \includegraphics[width=0.45\textwidth]{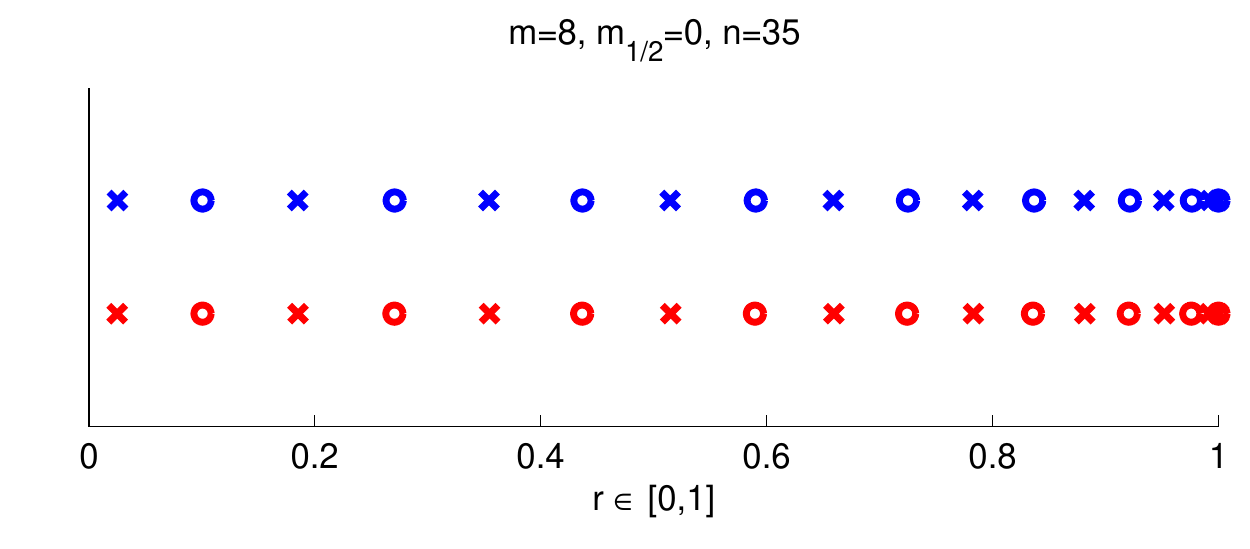}
 \end{center}
 \caption[Interpolation Grids]{
   \label{fig:InterpGrids}
   \renewcommand{\baselinestretch}{1} \small\normalsize Examples of
   optimal grids with $n$ equidistant boundary points and primary and
   dual radii shown with $\times$ and $\circ$. On the left we have $n
   = 25$ and a grid indexed by $\mhf = 1$, with $\ell = m + 1 = 6$.
   On the right we have $n = 35$ and a grid indexed by $\mhf = 0$,
   with $\ell = m+1 = 8.$ The grid shown in red is computed with
   formulas (\ref{eq:Rat.6}). The grid shown in blue is obtained from
   the rational approximation (\ref{eq:RR3}).}
\end{figure}

When we compute the optimal grid, we take the reference $\sr \equiv
1$, in which case $f^{(o)}(k^2) = |k|$. Thus, the optimal grid
computation reduces to that of rational interpolation of $f(\la)$,
\begin{equation}
F^{(o)}(\om_k^2) = \om_k = f^{(o)}(\om_k^2), \qquad k = 1, \ldots,
\frac{n-1}{2}.
\label{eq:Rat.5}
\end{equation}
This is solved explicitly in \cite{biesel}. For example, when
$\mhf = 1$, the coefficients $\ar_j$ and $\har_j$ are given by
\begin{eqnarray}
  \ar_j &=& h_\T \cot \left[\frac{h_\T}{2} (2 \ell -2 j + 1)
  \right], \qquad  \har_j =
  h_\T \cot \left[ \frac{h_\T}{2} (2 \ell -2 j + 2)\right] ,  
  \qquad j = 1,2 \ldots, \ell,
\label{eq:Rat.6}
\end{eqnarray}
and the radii follow from (\ref{eq:O3}). They satisfy the interlacing
relations
\begin{equation}
  1 = \hrr_1 = \rr_1 > \hrr_2 > \rr_2 > \ldots >  \hrr_{\ell+1} > \rr_{\ell+1} \ge 0,
\label{eq:Rat.7}
\end{equation}
as can be shown easily using the monotonicity of the cotangent and
exponential functions. We show an illustration of the resulting grids
in red, in Figure \ref{fig:InterpGrids}. Note the refinement toward
the boundary $r = 1$ and the coarsening toward the center $r = 0$ of
the disk. Note also that the dual points shown with $\circ$ are almost
half way between the primary points shown with $\times$. The last
primary radii $\rr_{\ell+1}$ are small, but the points do not reach
the center of the domain at $r = 0$.

In sections \ref{sect:2DFB} and \ref{sect:2DPB} we work with slightly
different measurements of the DtN map $\LG = {\bf
  M}_n(\Lambda_{\s})$, with entries defined by
\begin{equation}
\left( \LG\right)_{p,q} = \int_{0}^{2 \pi} \chi_p(\T)
\Lambda_{\s} \chi_q(\T) d \T, \qquad p \ne q, \qquad 
\left( \LG \right)_{p,p} = -\sum_{q \ne p} \left( \LG\right)_{p,q},
\label{eq:RAT_MEAS2}
\end{equation} 
using the non-negative measurement (electrode) functions $\chi_q(\T)$
that are compactly supported in $(\hT_{q},\hT_{q+1})$, and are
normalized by
\[
\int_0^{2 \pi} \chi_q(\T) d\T = 1.
\]
For example, we can take 
\[
\chi_q(\T) = \left\{ \begin{array}{ll} \frac{1}{h_\T}, \quad &
  \mbox{if} ~ ~ \hT_q < \T < \hT_{q+1}, \\ 0, & \mbox{otherwise}.
\end{array} \right., 
\]
and obtain after a calculation given in appendix \ref{ap:DRR} that the
entries of $\LG$ are given by
\begin{equation}
\left( \LG\right)_{p,q} = \frac{1}{2 \pi} \sum_{k \in \mathbb{Z}} e^{i
  k ( \T_p-\T_q)} f(k^2) \left[ \mbox{sinc}\left(\frac{k
  h_\T}{2}\right)\right]^2, \qquad p, q, = 1, \ldots, n.
\label{eq:RR}
\end{equation}
We also show in appendix \ref{ap:DRR} that
\begin{equation}
\LG \left[ e^{i k \T} \right] = \frac{1}{h_\T}\widetilde{F}(\om_k^2)
\left[ e^{i k \T} \right], \qquad |k| \le \frac{n-1}{2},
\label{eq:RR1}
\end{equation}
with eigenvectors $\left[ e^{i k \T} \right]$ defined in
(\ref{eq:L6BBB}) and scaled eigenvalues 
\begin{equation}
\widetilde{F}(\om_k^2) = f(k^2) \left[\mbox{sinc}\left(\frac{k
  h_\T}{2}\right)\right]^2 = F(\om_k^2) 
\left|\mbox{sinc}\left(\frac{k h_\T}{2}\right)\right|.
\label{eq:RR2}
\end{equation}
Here we recalled (\ref{eq:Rat.3}) and (\ref{eq:eig}).  

There is no explicit formula for the optimal grid satisfying
\begin{equation}
\label{eq:RR3}
\widetilde{F}^{(o)}(\om_k^2) = F^{(o)}(\om_k^2)
\left|\mbox{sinc}\left(\frac{k h_\T}{2}\right)\right| =
\om_k\left|\mbox{sinc}\left(\frac{k h_\T}{2}\right)\right|,
\end{equation}
but we can compute it as explained in Remark \ref{rem.1} and
appendix \ref{ap:Euclid}.  We show in Figure \ref{fig:InterpGrids} two
examples of the grids, and note that they are very close to those
obtained from the rational interpolation (\ref{eq:Rat.5}).  This is
not surprising because the sinc factor in (\ref{eq:RR3}) is not
significantly different from $1$ over the range $|k| \le
\frac{n-1}{2}$,
\[
\frac{2}{\pi} < \frac{\sin \left[ \frac{\pi}{2} \left( 1-\frac{1}{n}
    \right) \right]}{ \frac{\pi}{2} \left(1-\frac{1}{n}\right)} \le
\left|\mbox{sinc}\left(\frac{k h_\T}{2}\right)\right| \le 1.
\]
Thus, many eigenvalues $\widetilde{F}^{(o)}(\om_k^2)$ are
approximately equal to $\om_k$, and this is why the grids are similar.

\subsubsection{Truncated measure and optimal grids}
\label{sect:TRUNC_meas}
Another example of rational approximation arises in a modified
problem, where the positive spectral measure $\mu$ in Lemma
\ref{lem.1} is discrete
\begin{equation}
\mu(t) = - \sum_{j=1}^\infty \xi_j H\left(-t - \delta_j^2\right).
\label{eq:DiscrMeas}
\end{equation}
This does not hold for equation (\ref{eq:L2}) or equivalently
(\ref{eq:L11}), where the origin of the disc $r = 0$ is mapped to
$\infty$ in the logarithmic coordinates $z(r)$, and the measure
$\mu(t)$ is continuous. To obtain a measure like (\ref{eq:DiscrMeas}),
we change the problem here and in the next section to
\begin{equation}
\frac{r}{\s(r)} \frac{d}{dr} \left[ r \s(r) \frac{d v(r)}{dr}
  \right] - \la v(r) = 0, \qquad r \in (\ep,1), 
\label{eq:T1}
\end{equation}
with $\ep \in (0,1)$ and boundary conditions
\begin{equation}
\frac{\partial v(o)}{\partial r} = \varphi_\B, \qquad v(\ep) = 0.
\label{eq:T2}
\end{equation}
The Dirichlet boundary condition at $r = \ep$ may be realized if we
have a perfectly conducting medium in the disk concentric with
$\Omega$ and of radius $\ep$. Otherwise, $v(\ep) = 0$ gives an
approximation of our problem, for small but finite $\ep$.

\subsubsection*{Coordinate change and scaling}

\noindent It is convenient here and in the next section to introduce
the scaled logarithmic coordinate
\begin{equation}
\zeta(r) = \frac{\zr(r)}{Z} =  \frac{1}{Z} \int_r^1
\frac{dt}{t}, \qquad Z = - \log(\ep) = \zr(\ep),  
\label{eq:SC}
\end{equation}
and write (\ref{eq:L8}) in the scaled form
\begin{eqnarray}
\frac{z(r)}{Z} = \int_0^\zeta \frac{dt}{\s(r(t))} = z'(\zeta),
\label{eq:zSC} \qquad 
\frac{\hz(r)}{Z} = \int_0^\zeta \s(r(t))dt = \hz{\,'}(\zeta).
\end{eqnarray}
The conductivity function in the transformed coordinates 
is 
\begin{equation}
\sigma'(\zeta) = \sigma(r(\zeta)), \qquad r(\zeta) = e^{-Z \zeta},
\label{eq:ScSIG}
\end{equation}
and the potential 
\begin{equation}
v'(z') = \frac{v(r(z'))}{\varphi_\B}
\label{eq:ScPot}
\end{equation}
satisfies the scaled equations 
\begin{eqnarray}
\frac{d}{d\hz^{\,'}}\left(\frac{dv'}{dz'}\right) - \la' v' &=& 0,
\quad z' \in (0,L'), \nonumber \\ \frac{dv(0)}{dz'} &=& -1,
\label{eq:TT} \qquad v(L') = 0,
\end{eqnarray}
where we let  $\la'= \la/Z^2$ and 
\begin{equation}
L' = z'(1) = \int_{0}^1 \frac{dt}{\s'(t)}.
\label{eq:TT1}
\end{equation}

\begin{remark}
\label{rem.2}
We assume in the remainder of this section and in section
\ref{sect:Lay_ISP} that we work with the scaled equations
(\ref{eq:TT}) and drop the primes for simplicity of notation.
\end{remark}

\subsubsection*{The inverse spectral problem}

\noindent The differential operator $\frac{d}{d\hz} \frac{d}{dz}$
acting on the vector space of functions with homogeneous Neumann
conditions at $z = 0$ and Dirichlet conditions at $z = L$ is symmetric
with respect to the weighted inner product
\begin{equation}
\left(a,b\right) = \int_{0}^{\widehat L} a(z) b(z) d\hz = \int_0^1
a(z(\zeta)) b(z(\zeta)) \s(\zeta) d \zeta, \qquad \widehat{L} =
\hz(1).
\label{eq:T3}
\end{equation}
It has negative eigenvalues $\{-\delta_j^2\}_{j = 1, 2, \ldots}$, the
points of increase of the measure (\ref{eq:DiscrMeas}), and
eigenfunctions $y_j(z)$. They are orthogonal with respect to the inner
product (\ref{eq:T3}), and we normalize them by
\begin{equation}
\|y_j\|^2 = \left(y_j,y_j\right) = \int_{0}^{\widehat{L}} y_j^2(z) d\hz = 1.
\label{eq:T4}
\end{equation}
The weights $\xi_j$ in (\ref{eq:DiscrMeas}) are defined by 
\begin{equation}
\xi_j = y_j^2(0).
\label{eq:T5}
\end{equation}

For the discrete problem we assume in the remainder of the section
that $\mhf = 1$, and work with the NtD map, that is with $\Fd(\la)$
represented in Lemma \ref{lem.1} in terms of the discrete measure
$\mu^F(t)$. Comparing (\ref{eq:DiscrMeas}) and (\ref{eq:L19}), we note
that we ask that $\mu^F(t)$ be the truncated version of $\mu(t)$,
given the first $\ell$ weights $\xi_j$ and eigenvalues $-\delta_j^2$,
for $j = 1, \ldots, \ell$. We arrived at the classic \emph{inverse
spectral problem}
\cite{gel1951determination,chadan1997introduction,hochstadt1973inverse,marchenko2011sturm,mclaughlin1987uniqueness},
that seeks an approximation of the conductivity $\s$ from the
truncated measure. We can solve it using the theory of resistor
networks, via an \emph{inverse eigenvalue problem}
\cite{chu2002structured} for the Jacobi like matrix ${\bf A}$ defined
in (\ref{eq:JacA}). The key ingredient in the connection between the
continuous and discrete eigenvalue problems is the optimal grid, as
was first noted in \cite{BorDru} and proved in \cite{BorDruKni}. We
review this result in section \ref{sect:Lay_ISP}.

\subsubsection*{The truncated measure optimal grid}

\noindent The optimal grid is obtained by solving the discrete inverse
problem with spectral data for the reference conductivity
$\sr(\zeta)$,
\begin{equation}
\D^{(o)}_n = \left\{ \xi^{(o)}_j = 2, \delta^{(o)}_j=
\pi\left(j-\frac{1}{2}\right), \quad j = 1, \ldots, \ell \right\}.
\label{eq:T6r}
\end{equation}
The parameters $\{ \ar_j,\har_j \}_{j = 1, \ldots, \ell}$ can be
determined from $\D^{(o)}_n$ with the Lanczos algorithm
\cite{trefethen1997numerical,chu2002structured} reviewed briefly in
appendix \ref{ap:Lanczos}. The grid points are given by 
\begin{equation}
\zer_{j+1} = \ar_j + \zer_j = \sum_{q=1}^j \ar_q, \qquad \hzer_{j+1} =
\har_j + \hzer_j = \sum_{q=1}^j \har_q, \qquad j = 1, \ldots, \ell,
\label{eq:T9}
\end{equation}
where $\zer_1 = \hzer_1 = 0$. This is in the logarithmic coordinates
that are related to the optimal radii as in (\ref{eq:ScSIG}). The grid
is calculated explicitly in \cite[Appendix A]{BorDruKni}. We summarize
its properties in the next lemma, for large $\ell$.
\begin{lemma}
\label{lem.2}
The steps $\{ \ar_j,\har_j \}_{j = 1, \ldots, \ell}$ of the truncated
measure optimal grid satisfy the monotone relation
\begin{equation}
\har_1 < \ar_1 < \har_2 < \ar_2 < \ldots < \har_k < \ar_k.
\label{eq:MonGrid}
\end{equation}
Moreover, for large $\ell$, the primary grid steps are 
\begin{equation}
\ar_j = \left\{ \begin{array}{ll} 
\frac{2 + O \left[ (\ell-j)^{-1} + j^{-2}\right]}{\pi \sqrt{\ell^2-j^2}}, 
\quad & \mbox{if} ~1 \le j \le \ell -1 , \vspace{0.05in}\\
\frac{\sqrt{2} +O(\ell^{-1})}{\sqrt{\pi \ell}}, & \mbox{if} ~ j = \ell,
\end{array} \right.
\label{eq:PrimGrid}
\end{equation}
and the dual grid steps are
\begin{equation}
\har_j = \frac{2 + O \left[ (\ell+1-j)^{-1} + j^{-2}\right]}{\pi
\sqrt{\ell^2-(j-1/2)^2}}, \quad 1 \le j \le \ell.
\label{eq:DualGrid}
\end{equation}
\end{lemma}

\begin{figure}[t!]
 \begin{center}
   \includegraphics[width=0.5\textwidth]{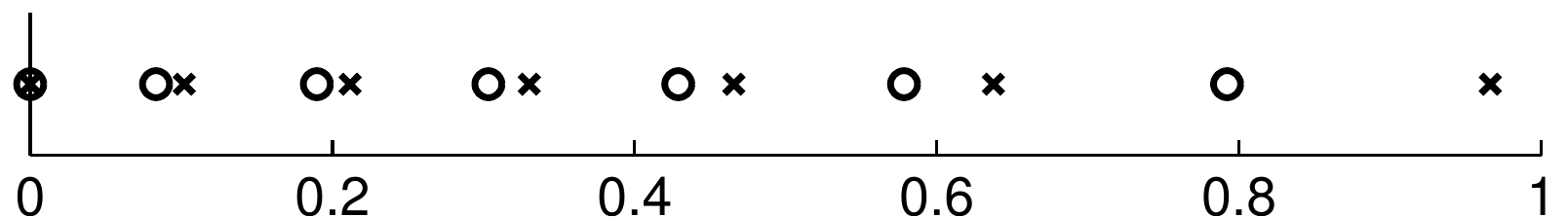}
  \end{center}
 \caption[Truncated Measure Grid]{
   \label{fig:Tmeas}
   \renewcommand{\baselinestretch}{1} \small\normalsize Example of
   truncated measure optimal grid with $\ell = 6$. This is in the logarithmic 
   scaled coordinates $\zeta \in [0,1]$. The primary points are denoted with 
   $\times$ and the dual ones with $\circ$.}
\end{figure}

\begin{figure}[t!]
 \begin{center}
   \includegraphics[width=0.5\textwidth]{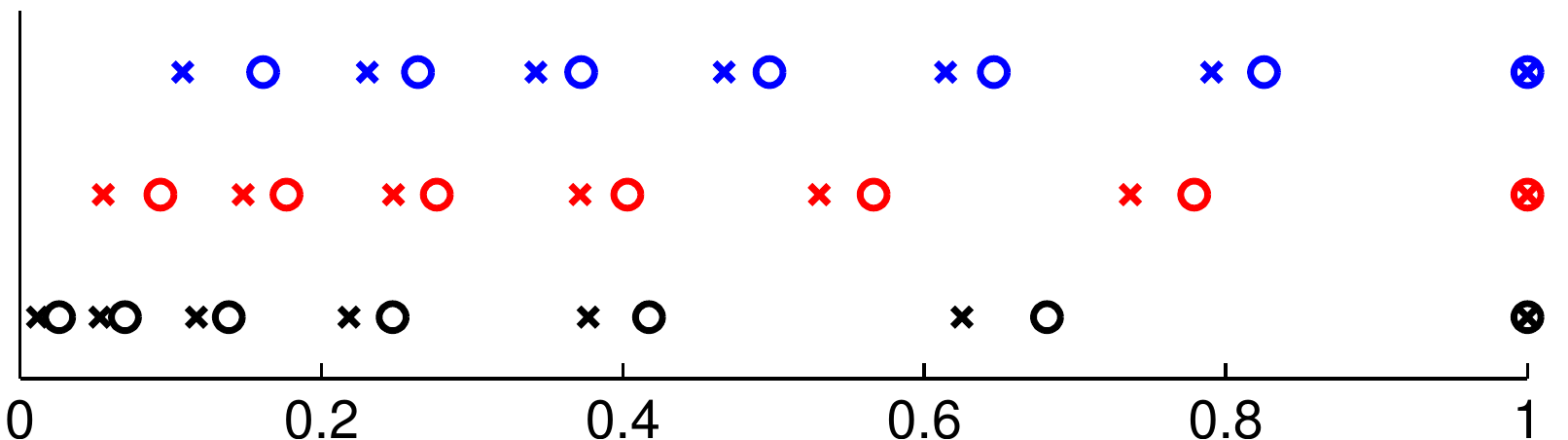}
  \end{center}
 \caption[Truncated Measure radial Grid]{
   \label{fig:Tmeas1}
   \renewcommand{\baselinestretch}{1} \small\normalsize The radial
   grid obtained with the coordinate change $r = e^{- Z \zeta}$. The
   scale $Z = -\log \epsilon$ affects the distribution of the
   radii. The choice $\ep = 0.1$ is in blue, $\ep = 0.05$ is in red
   and $\ep = 0.01$ is in black. The primary radii are indicated with
   $\times$ and the dual ones with $\circ$. }
\end{figure}

We show in Figure \ref{fig:Tmeas} an example for the case $\ell =
6$. To compare it with the grid in Figure \ref{fig:InterpGrids}, we
plot in Figure \ref{fig:Tmeas1} the radii given by the coordinate
transformation (\ref{eq:ScSIG}), for three different parameters $\ep$.
Note that the primary and dual points are interlaced, but the dual
points are not half way between the primary points, as was the case in
Figure \ref{fig:InterpGrids}. Moreover, the grid is not refined near
the boundary at $r = 1$. In fact, there is accumulation of the grid
points near the center of the disk, where we truncate the domain. The
smaller the truncation radius $\ep$, the larger the scale $Z = -\log
\epsilon$, and the more accumulation near the center.

Intuitively, we can say that the grids in Figure \ref{fig:InterpGrids}
are much superior to the ones computed from the truncated measure, for
both the forward and inverse EIT problem. Indeed, for the forward
problem, the rate of convergence of $\Fd(\la)$ to $\fd(\la)$ on the
truncated measure grids is algebraic \cite{BorDruKni}
\[
\left| \fd(\la)-\Fd(\la)\right| = \left|\sum_{j = \ell+1}^\infty
\frac{\xi_j}{\la + \de_j^2}\right| = O \left(\sum_{j = \ell+1}^\infty
\frac{1}{j^2} \right) = O\left(\frac{1}{\ell} \right).
\]
The rational interpolation grids described in section
\ref{sect:RAT_interp} give exponential convergence of $\Fd(\la)$ to
$\fd(\la)$ \cite{MamonovMasters}. For the inverse problem, we expect
that the resolution of reconstructions of $\s$ decreases rapidly away
from the boundary where we make the measurements, so it makes sense to
invert on grids like those in Figure \ref{fig:InterpGrids}, that are
refined near $r = 1$.

The examples in Figures \ref{fig:InterpGrids} and \ref{fig:Tmeas1}
show the strong dependence of the grids on the measurement setup.
Although the grids in Figure \ref{fig:Tmeas1} are not good for the EIT
problem, they are optimal for the inverse spectral problem. The
optimality is in the sense that the grids give an exact match of the
spectral measurements (\ref{eq:T6r}) of the NtD map for conductivity
$\sr$. Furthermore, they give a very good match of the spectral
measurements (\ref{eq:T6}) for the unknown $\s$, and the reconstructed
conductivity on them converges to the true $\s$, as we show next.

\subsection{\textbf{Continuum limit of the discrete inverse spectral 
problem on optimal grids}} \label{sect:Lay_ISP}

 Let $\Qc_n:\D_n \to \R_+^{2 \ell}$ be the reconstruction mapping that
 takes the data
\begin{equation}
\D_n = \left\{ \xi_j, \delta_j, \quad j = 1, \ldots, \ell \right\}
\label{eq:T6}
\end{equation}
to the $2 \ell = \frac{n-1}{2}$ positive values $\{\s_j,\hs_j\}_{j =
  1, \ldots, \ell}$ given by
\begin{eqnarray}
\s_{j} = \frac{\hal_j}{\har_j}, \qquad \hs_{j+1} =
\frac{\ar_j}{\al_j}, \quad j = 1, 2, \ldots, \ell. 
\label{eq:CL1}
\end{eqnarray}
The computation of $\{ \al_j,\hal_j \}_{j = 1, \ldots, \ell }$
requires solving the discrete inverse spectral problem with data
$\D_n$, using for example the Lanczos algorithm reviewed in appendix
\ref{ap:Lanczos}.  We define the \emph{reconstruction}
$\s^\ell(\zeta)$ of the conductivity as the piecewise constant
interpolation of the point values (\ref{eq:CL1}) on the optimal grid
(\ref{eq:T9}). We have
\begin{equation}
\s^\ell(\zeta) = \left\{ \begin{array}{ll} \s_j, \qquad & \mbox{if} ~
\zeta \in [\zer_j,\hzer_{j+1}), ~ ~ j = 1, \ldots, \ell,
\vspace{0.07in}\\ \vspace{0.07in}\hs_{j}, & \mbox{if} ~ \zeta \in
[\hzer_{j},\zer_{j}), ~ ~ j = 2, \ldots, \ell+1, \\ \hs_{\ell+1}, &
\mbox{if} ~ \zeta \in [\zer_{l+1},1]
\end{array} \right.  
\label{eq:CL2} 
\end{equation}
and we discuss here its convergence to the true conductivity function
$\s(\zeta)$, as $\ell \to \infty$.

To state the convergence result, we need some assumptions on the decay
with $j$ of the perturbations of the spectral data
\begin{equation}
\Delta \de_j = \de_j - \der_j, \qquad \Delta \xi_j = \xi_j-\xir_j.
\label{eq:CL3}
\end{equation}
The asymptotic behavior of $\de_j$ and $\xi_j$ is well known, under
various smoothness requirements on $\s(z)$
\cite{mclaughlin1987uniqueness,Trub,ColM}.  For example, if $\s(\zeta)
\in H^3[0,1]$, we have
\begin{equation}
\Delta \de_j = \de_j - \der_j = \frac{ \int_0^1 q(\zeta) d \zeta}{(2 j
  - 1)\pi} + O\left(j^{-2}\right) ~ \mbox{ and } ~ \Delta \xi_j =
  \xi_j - \xir_j= O\left(j^{-2}\right), \label{eq:CL4}
\end{equation}
where $q(\zeta)$ is the Schr\"{o}dinger potential 
\begin{equation} 
q(\zeta) = \s(\zeta)^{-\frac{1}{2}} \frac{d^2 \s(\zeta)^{\frac{1}{2}}}{d
\zeta^2}.
\label{eq:CL5}
\end{equation}
We have the following convergence result proved in \cite{BorDruKni}.
\begin{theorem}
\label{thm.1}
Suppose that $\s(\zeta)$ is a positive and bounded scalar
conductivity function, with spectral data satisfying the asymptotic
behavior 
\begin{equation}
\Delta \de_j = O \left(\frac{1}{j^s \log(j)}\right), \qquad \Delta
\xi_j =O \left(\frac{1}{j^s}\right), \quad \mbox{for some } s > 1,
\mbox{ as } j \to \infty.  
\label{eq:CL6} 
\end{equation}
Then $\s^\ell(\zeta)$ converges to $\s(\zeta)$ as $\ell \to \infty$, 
pointwise and in $L^1[0,1]$.
\end{theorem}

Before we describe the outline of the proof in \cite{BorDruKni}, let
us note that it appears from (\ref{eq:CL4}) and (\ref{eq:CL6}) that
the convergence result applies only to the class of conductivities
with zero mean potential. However, if 
\begin{equation}
\bq = \int_0^1 q(\zeta) d \zeta \ne 0,
\label{eq:CL7}
\end{equation}
we can modify the point values (\ref{eq:CL1}) of the reconstruction
$\s^\ell(\zeta)$ by replacing $\ar_j$ and $\har_j$ with $\aq_j$ and
$\haq_j$, for $ j = 1,\ldots, \ell$. These are computed by solving the
discrete inverse spectral problem with data $ \D^{(\bq)}_n = \left\{
\xiq_j, \deq_j, \quad j = 1, \ldots, \ell \right\}, $ for conductivity
function
\begin{equation}
\sq(\zeta) = \frac{1}{4} \left( e^{\sqrt{\bq}\, \zeta} + e^{-\sqrt{\bq} \,
\zeta} \right)^2. \label{eq:CL9}
\end{equation}
This conductivity satisfies the initial value problem
\begin{equation}
\frac{d^2 \sqrt{\sq(\zeta)}}{d \zeta^2} = \bq \sqrt{\sq(\zeta)}
\quad\mbox{for}\quad 0 < \zeta \le 1, \qquad\frac{d\sq(0)}{d \zeta} = 0
\quad\mbox{and}\quad \sq(0) = 1, \label{eq:CL10}
\end{equation}
and we assume that
\begin{equation}
\bq > -\frac{\pi^2}{4},
\label{eq:CL10p}
\end{equation}
so that (\ref{eq:CL9}) stays positive for $\zeta \in [0,1]$.

As seen from (\ref{eq:CL4}), the perturbations $\de_j-\deq_j$ and
$\xi_j-\xiq_j$ satisfy the assumptions (\ref{eq:CL6}), so Theorem
\ref{thm.1} applies to reconstructions on the grid given by $\sq$. We
show below in Corollary \ref{cor.1} that this grid is asymptotically
the same as the {\em optimal grid}, calculated for $\sr$. Thus, the
convergence result in Theorem \ref{thm.1} applies after all, without
changing the definition of the reconstruction (\ref{eq:CL2}).

\subsubsection{The case of constant Schr\"{o}dinger potential} 
\label{sect:CONSTq}
The equation (\ref{eq:TT}) for $\s \leadsto \sq$ can be transformed to
Schr\"{o}dinger form with constant potential $\bq$
\begin{eqnarray}
\frac{d^2 w(\zeta)}{d \zeta^2} - (\la+\bq) w(\zeta) &=& 0, \qquad
\zeta \in (0,1) ,
\label{eq:CL13} \\
\frac{d w(0)}{d \zeta} &=& -1, \qquad w(1) = 0,
\nonumber
\end{eqnarray}
by letting $w(\zeta) = v(\zeta) \sqrt{\sq(\zeta)}$. Thus, the
eigenfunctions $y_j^{(\bq)}(\zeta)$ of the differential operator
associated with $\sq(\zeta)$ are related to $y_j^{(o)}(\zeta)$, the
eigenfunctions for $\sr \equiv 1$, by
\begin{equation}
y_j^{(\bq)}(\zeta) = \frac{y_j^{(o)}(\zeta)}{\sqrt{\sq(\zeta)}}.
\end{equation}
They satisfy the orthonormality condition
\begin{equation}
\int_0^1 y_j^{(\bq)}(\zeta) y_p^{(\bq)}(\zeta) \sq(\zeta) d \zeta = 
\int_0^1 y_j^{(o)}(\zeta) y_p^{(o)}(\zeta) d \zeta  = \delta_{jp},
\end{equation}
and since $\sq(0) = 1$,
\begin{equation}
  \xiq_j = \left[ y_j^{(\bq)}(0)\right]^2 = \left[ y_j^{(o)}(0)\right]^2 = 
  \xir_j,
  \qquad j = 1, 2, \ldots
\label{eq:CL12}
\end{equation}
The eigenvalues are shifted by $\bq$,
\begin{equation}
-\left(\deq_j\right)^2 = -\left(\der_j \right)^2 - \bq, \qquad j = 1, 2
, \ldots
\label{eq:CL14}
\end{equation}

Let $\{\aq_j,\haq_j\}_{j=1, \ldots, \ell}$ be the parameters obtained
by solving the discrete inverse spectral problem with data
$\D_n^{(\bq)}$.  The reconstruction mapping $\Qc_n:\D_n^{(\bq)} \to
\mathbb{R}^{2 \ell}$ gives the sequence of $2 \ell = \frac{n-1}{2}$
pointwise values
\begin{equation}
\sq_j = \frac{\haq_j}{\har_j}, \qquad \hsq_{j+1} =
\frac{\ar_j}{\aq_j}, \quad j = 1,\ldots, \ell.
\label{eq:Q0}
\end{equation}
We have the following result stated and proved in
\cite{BorDruKni}. See the review of the proof in appendix
\ref{ap:ProofLem.3}.
\begin{lemma}
\label{lem.3} 
The point values $\sq_j$ satisfy the finite difference discretization
of initial value problem (\ref{eq:CL10}), on the optimal grid,
\begin{eqnarray}
\frac{1}{\har_j}\left[
\left(\frac{\sqrt{\sq_{j+1}}-\sqrt{\sq_j}}{\ar_j}\right) -
\left(\frac{\sqrt{\sq_{j}}-\sqrt{\sq_{j-1}}}{\ar_{j-1}}\right)
\right]
 -\bq\sqrt{\sq_j}&=& 0, \quad j = 2, 3, \ldots, \ell, \nonumber \\
\frac{1}{\har_1}
\left(\frac{\sqrt{\sq_{2}}-\sqrt{\sq_1}}{\ar_1}\right)
 -\bq\sqrt{\sq_1}&=& 0, \qquad \sq_1=1.
 \label{eq:Q1}
\end{eqnarray}
Moreover, $ \hsq_{j+1} = \sqrt{\sq_j \sq_{j+1}}$, for $j = 1, \ldots,
\ell. $
\end{lemma}

The convergence of the reconstruction $\sigma^{(\bq),\ell}(\zeta)$
follows from this lemma and a standard finite-difference error
analysis \cite{Godunov} on the optimal grid satisfying Lemma
\ref{lem.2}. The reconstruction is defined as in (\ref{eq:CL2}), by
the piecewise constant interpolation of the point values (\ref{eq:Q0})
on the optimal grid.
\begin{theorem}
\label{thm.2}
As $\ell \to \infty$ we have 
\begin{equation}
\max_{1\le j\le \ell}\left|\sq_j-\sq(\zer_j)\right| \rightarrow 0
\qquad\mbox{and}\qquad\max_{1\le j\le
\ell}\left|\hsq_{j+1}-\sq(\hzer_{j+1}) \right| \rightarrow
0, \label{eq:Q3}
\end{equation}
and the reconstruction $\sigma^{(\bq),\ell}(\zeta)$ converges to
$\sq(\zeta)$ in $L^{\infty}[0,1]$.
\end{theorem}

As a corollary to this theorem, we can now obtain that the grid
induced by $\sq(\zeta)$, with primary nodes $\zeq_j$ and dual nodes
$\hzeq_j$, is asymptotically close to the optimal grid. The proof is
in appendix \ref{ap:ProofLem.3}.
\begin{corollary}
\label{cor.1}
The grid induced by $\sq(\zeta)$ is defined by equations
\begin{equation}
\int_0^{\zeq_{j+1}} \frac{d \zeta}{\sq(\zeta)}= \sum_{p=1}^j \aq_p,
\qquad \int_0^{\hzeq_{j+1}} \sq(\zeta) d \zeta = \sum_{p=1}^j\haq_p,
\qquad j = 1, \ldots, \ell, \qquad \zeq_1 = \hzeq_1 = 0,
\label{eq:GQ}
\end{equation}
and satisfies
\begin{equation}
\max_{1 \le j \le \ell+1} \left|\zeq_j-\zer_j\right| \to 0, 
\qquad \max_{1 \le j \le \ell+1} \left|\hzeq_j-\hzer_j\right| \to 0, \quad 
\mbox{as} ~ \ell \to \infty.
\label{eq:GQ1}
\end{equation}
\end{corollary}

\subsubsection{Outline of the proof of Theorem \ref{thm.1}}
\label{sect:ProofThm1}
The proof given in detail in \cite{BorDruKni} has two main steps. The
first step is to establish the compactness of the set of reconstructed
conductivities. The second step uses the established compactness and
the uniqueness of solution of the continuum inverse spectral problem
to get the convergence result.

\subsubsection*{Step 1: Compactness}

\noindent We show here that the sequence $\{\s^\ell(\zeta)\}_{\ell \ge 1}$
of reconstructions (\ref{eq:CL2}) has bounded variation.

\begin{lemma}
\label{lem.4}
The sequence $\{\s_j,\hs_{j+1}\}_{j=1,\ldots, \ell}$ (\ref{eq:CL1})
returned by the reconstruction mapping $\Qc_n$ satisfies
\begin{equation}
\sum_{j=1}^{\ell} \left| \log \hs_{j+1} - \log \s_j \right| +
\sum_{j=1}^{\ell} \left| \log \hs_{j+1} - \log \s_{j+1} \right| \le C,
\label{eq:lem4.1}
\end{equation}
where $C$ is independent of $\ell$. Therefore the sequence of
reconstructions $\{\s^\ell(\zeta)\}_{\ell \ge 1}$ has uniformly
bounded variation.
\end{lemma}

Our original formulation is not convenient for proving
(\ref{eq:lem4.1}), because when written in Schr\"{o}dinger form, it
involves the second derivative of the conductivity as seen from
(\ref{eq:CL5}).  Thus, we rewrite the problem in first order system
form, which involves only the first derivative of $\s(\zeta)$, which
is all we need to show (\ref{eq:lem4.1}). At the discrete level, the
linear system of $\ell$ equations
\begin{equation}
\bA \bV - \la \bV = -\frac{{\bf e}_1}{\hal_1}
\end{equation}
for the potential $\bV = \left(V_1, \ldots, V_\ell\right)^T$  
is transformed to the system of $2 \ell$ equations
\begin{equation}
\bB \bH^{\frac{1}{2}} \bW - \sqrt{\la} \bH^{\frac{1}{2}}\bW =
-\frac{{\bf e}_1}{\sqrt{\la \hal_1}}
\label{eq:PC1}
\end{equation}
for the vector $\bW = \left(W_1,\hW_2,\ldots,
W_{\ell},\hW_{\ell+1}\right)^T$ with components
\begin{equation}
 W_j = \sqrt{\s_j}V_j, \quad \hW_{j+1} = \frac{\hs_{j+1}}{\sqrt{\la
     \s_j}} \left( \frac{V_{j+1}-V_j}{\ar_j}\right), \quad j = 1,
 \ldots, \ell.
\label{eq:PC4}
\end{equation}
Here $\bH = \mbox{diag}\left( \har_1,\ar_1,\ldots, \har_\ell,\ar_\ell\right)$ and 
$\bB$ is the tridiagonal, skew-symmetric matrix
\begin{equation}
\bB = \begin{pmatrix} 0 & \beta_1 & 0 &0  &\ldots \\ -\beta_1 & 0 &
\beta_2 & 0 & \ldots  \\ 
0 &-\beta_2 & 0 &\ddots &\vdots\\
 \vdots \\
0& \ldots&  & 
-\beta_{2\ell-1} & 0
\end{pmatrix}
\label{eq:PC2}
\end{equation}
with entries
\begin{eqnarray}
\beta_{2 p} &=& \frac{1}{\sqrt{\al_p \hal_{p+1}}} =
\frac{1}{\sqrt{\ar_p \har_{p+1}}} \sqrt{\frac{\hs_{p+1}}{\s_p}} =
\beta_{2p}^{(o)} \sqrt{\frac{\hs_{p+1}}{\s_{p+1}}},
\label{eq:PC3.1} \\ 
\beta_{2p-1} &=& \frac{1}{\sqrt{\al_p \hal_p}} = \frac{1}{\sqrt{\ar_p
    \har_{p}}} \sqrt{\frac{\hs_{p+1}}{\s_p}} = \beta_{2p-1}^{(o)}
\sqrt{\frac{\hs_{p+1}}{\s_{p}}}.
\label{eq:PC3.2}
\end{eqnarray}

Note that we have
\begin{equation}
\sum_{p=1}^{2 \ell -1} \left| \log \frac{\beta_p}{\beta_p^{(o)}} \right| = 
\frac{1}{2} \sum_{p=1}^{\ell} \left| \log \hs_{p+1} - \log \s_p \right| + 
\frac{1}{2} \sum_{p=1}^{\ell} \left| \log \hs_{p+1} - \log \s_{p+1} \right|,
\label{eq:PC5}
\end{equation}
and we can prove (\ref{eq:lem4.1}) by using a method of small
perturbations. Recall definitions (\ref{eq:CL3}) and let
\begin{equation}
\Delta \de_j^r = r \Delta \de_j, \qquad \Delta \xi_j^r = r \Delta \xi_j, \quad 
j = 1, \ldots, \ell, 
\label{eq:PC6}
\end{equation}
where $r \in [0,1]$ is an arbitrary continuation parameter.  Let also
$\beta_j^r$ be the entries of the tridiagonal, skew-symmetric matrix
$\bB^r$ determined by the spectral data $\de_j^r = \de_j^{(o)} +
\Delta \de_j^r$ and $\xi_j^r = \xi_j^{(o)} + \Delta \xi_j^r$, for
$j=1, \ldots, \ell$. We explain in appendix \ref{ap:perturbation} how
to obtain explicit formulae for the perturbations $d \log \beta_j^r$
in terms of the eigenvalues and eigenvectors of matrix $\bB^r$ and
perturbations $d \de_j^r = \Delta \de_j dr$ and $d \xi_j^r = \Delta
\xi_j dr$. These perturbations satisfy the uniform bound
\begin{equation}
\sum_{j=1}^{2 \ell -1} \left| d \log \beta_j^r \right| \le C_1 | dr|,
\label{eq:PC7}
\end{equation}
with constant $C_1$ independent of $\ell$ and $r$.  Then,
\begin{equation}
\log \frac{\beta_j}{\beta_j^{(o)}} = \int_0^1 d \log \beta_j^r
\end{equation}
satisfies the uniform bound $ \displaystyle \sum_{j=1}^{2 \ell -1}
\left| \log \frac{\beta_j}{\beta_j^{(o)}} \right| \le C_1 $ and
(\ref{eq:lem4.1}) follows from (\ref{eq:PC5}).

\subsubsection*{Step 2: Convergence}

\noindent Recall section \ref{sect:Lay_rat} where we state that the
eigenvectors $\bY_j$ of $\bA$ are orthonormal with respect to the
weighted inner product (\ref{eq:L16}). Then, the matrix $\tY$ with
columns $\mbox{diag}\left(\hal_1^{\frac{1}{2}}, \ldots,
\hal_\ell^{\frac{1}{2}}\right)\bY_j$ is orthogonal and we have 
\begin{equation}
\left(\tY \tY^T\right)_{11} = \hal_1 \sum_{j=1}^\ell \xi_j = 1.
\label{eq:PC8}
\end{equation}
Similarly
\begin{equation}
\har_1 \sum_{j=1}^\ell \xir_j = 2 \ell \har_1 = 1,
\label{eq:PC8}
\end{equation}
where we used (\ref{eq:T6r}), and since $\Delta \xi_j$ are summable by
assumption (\ref{eq:CL6}),
\begin{equation}
\s_1 = \frac{\hal_1}{\har_1} = \left(1 + \har_1 \sum_{j=1}^\ell \Delta
\xi_j\right)^{-1} = 1 + O(\har_1) = 1 + O\left(\frac{1}{\ell}\right).
\end{equation}
But $\s^\ell(0) = \s_1$, and since $\s^\ell(\zeta)$ has bounded
variation by Lemma \ref{lem.4}, we conclude that $\s^\ell(\zeta)$ is
uniformly bounded in $\zeta \in [0,1]$. 

Now, to show that $\s^\ell(\zeta) \to \s(\zeta)$ in $L^1[0,1]$,
suppose for contradiction that it does not. Then, there exists
$\varepsilon > 0$ and a subsequence $\s^{\ell_k}$ such that 
\[
\|\s^{\ell_k} - \s\|_{L^1[0,1]} \ge \varepsilon.
\]
But since this subsequence is bounded and has bounded variation, we
conclude from Helly's selection principle and the compactness of the
embedding of the space of functions of bounded variation in $L^1[0,1]$
\cite{Natanson} that it has a convergent subsequence pointwise and in
$L^1[0,1]$. Call again this subsequence $\s^{\ell_k}$ and denote its
limit by $\s^\star \ne \s$. Since the limit is in $L^1[0,1]$, we have
by definitions (\ref{eq:zSC}) and Remark \ref{rem.2},
\begin{equation}
z(\zeta; \s^{\ell_k}) = \int_0^\zeta \frac{dt}{\s^{\ell_k}(t)} \to
z(\zeta;\s) = \int_0^\zeta \frac{dt}{\s^\star(t)}, \qquad \hz(\zeta;
\s^{\ell_k}) = \int_0^\zeta \s^{\ell_k}(t) dt \to \hz(\zeta;\s^\star)
= \int_0^\zeta \s(t) dt.
\end{equation}
Furthermore, the continuity of $\fd$ with respect to the conductivity
gives $\fd(\la;\s^{\ell_k}) \to \fd(\la; \s^\star)$. However, Lemma
\ref{lem.1} and (\ref{eq:DiscrMeas}) show that $\fd(\la;\s^{\ell}) \to
\fd(\la;\s)$ by construction, and since the inverse spectral problem
has a unique solution \cite{gelfand1951dde,levitan1987isl,ColM,Trub},
we must have $\s^\star = \s$. We have reached a contradiction, 
so $\s^\ell(\zeta) \to \s(\zeta)$ in $L^1[0,1]$. The pointwise convergence 
can be proved analogously.

\begin{remark}
\label{rem.3}
All the elements of the proof presented here, except for establishing
the bound (\ref{eq:PC7}), apply to any measurement setup. The
challenge in proving convergence of inversion on optimal grids for
general measurements lies entirely in obtaining sharp stability
estimates of the reconstructed sequence with respect to perturbations
in the data. The inverse spectral problem is stable, and this is why
we could establish the bound (\ref{eq:PC7}). The EIT problem is
exponentially unstable, and it remains an open problem to show the
compactness of the function space of reconstruction sequences
$\s^\ell$ from measurements such as (\ref{eq:RR2}).
\end{remark}

%%%%%%%%%%%%%%%%%%%%%%%%%%%%%%%%%%%%%%%%%%%%%%%%%%%%%%%%%%%%%%%%%%%%%%%
\section{Two dimensional media and full boundary measurements}
\label{sect:2DFB}
\setcounter{equation}{0} We now consider the two dimensional EIT
problem, where $\s = \s(r,\theta)$ and we cannot use separation of
variables as in section~\ref{sect:Layered}. More explicitly, we cannot
reduce the inverse problem for resistor networks to one of rational
approximation of the eigenvalues of the DtN map.  We start by
reviewing in section~\ref{sect:rnetip} the conditions of unique
recovery of a network $(\Gamma,\gamma)$ from its DtN map $\LG$,
defined by measurements of the continuum $\Lambda_\s$. The
approximation of the conductivity $\s$ from the network conductance
function $\gamma$ is described in section~\ref{sect:fmog}.

%%%%%%%%%%%%%%%%%%%%%%%%%%%%%%%%%%%%%%%%%%%%%%%%%%%%%%%%%%%%%%%%%%%%%%%
\subsection{The inverse problem for planar resistor networks}
\label{sect:rnetip}

The unique recoverability from $\LG$ of a network $(\Gamma,\gamma)$
with known circular planar graph $\Gamma$ is established in
\cite{deverdiere1994rep,deverdiere1996rep,CurtMooMor,CurtIngMor}.  A
graph $\Gamma = (\Pc,\E)$ is called circular and planar if it can be
embedded in the plane with no edges crossing and with the boundary
nodes  lying on a circle.  We call by association the
networks with such graphs circular planar.  The recoverability result
states that if the data is {\em consistent} and the graph $\Gamma$ is
{\em critical} then the DtN map $\LG$ determines uniquely the
conductance function $\gamma$.  By consistent data we mean that the
measured matrix $\LG$ belongs to the set of DtN maps of circular
planar resistor networks.

A graph is critical if and only if it is {\em well-connected} and the
removal of any edge breaks the well-connectedness. A graph is
well-connected if all its {\em circular pairs} $(P,Q)$ are {\em
  connected}. Let $P$ and $Q$ be two sets of boundary nodes with the
same cardinality $|P| = |Q|$. We say that $(P,Q)$ is a circular pair
when the nodes in $P$ and $Q$ lie on disjoint segments of the boundary
$\B$. The pair is {\em connected} if there are $|P|$ disjoint paths
joining the nodes of $P$ to the nodes of $Q$.

A symmetric $n\times n$ real matrix $\LG$ is the DtN map of a circular
planar resistor network with $n$ boundary nodes if its rows sum to
zero $\LG \bone = \bzero$ (conservation of currents) and all its {\em
  circular minors} $(\LG)_{P,Q}$ have non-positive determinant.  A
circular minor $(\LG)_{P,Q}$ is a square submatrix of $\LG$ defined
for a circular pair $(P,Q)$, with row and column indices corresponding
to the nodes in $P$ and $Q$, ordered according to a predetermined
orientation of the circle $\B$. Since subsets of $P$ and $Q$ with the
same cardinality also form circular pairs, the determinantal
inequalities are equivalent to requiring that all circular minors be
totally non-positive. A matrix is totally non-positive if all its
minors have non-positive determinant.

Examples of critical networks were given in
section~\ref{sect:FBgrids}, with graphs $\Gamma$ determined by tensor
product grids. Criticality of such networks is proved in
\cite{CurtMooMor} for an odd number $n$ of boundary points. As
explained in section~\ref{sect:FBgrids} (see in particular equation
\eqref{eq:GridCOND}), criticality holds when the number of edges in
$\E$ is equal to the number $n(n-1)/2$ of independent entries of the
DtN map $\LG$.

The discussion in this section is limited to the tensor product
topology, which is natural for the full boundary measurement setup.
Two other topologies admitting critical networks (pyramidal and
two-sided), are discussed in more detail in
sections~\ref{sect:piramidal} and \ref{sect:twosided}.  They are
better suited for partial boundary measurements setups
\cite{BDMG-10,BGM-11}.

\begin{figure}
\begin{center}
\input{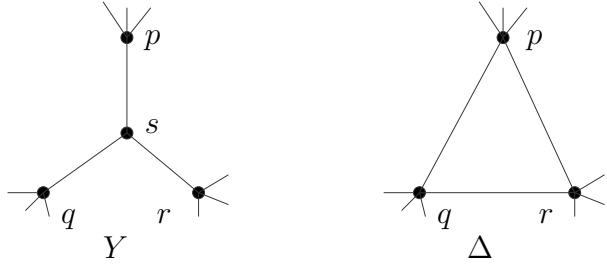}
\end{center}
\caption{Given some conductances in the $Y$ network, there is a choice
  of conductances in the $\Delta$ network for which the two networks
  are indistinguishable from electrical measurements at the nodes $p$,
  $q$ and $r$.}
\label{fig:ydelta}
\end{figure}

\begin{remark}
\label{rem:YD}
It is impossible to recover both the
topology and the conductances from the DtN map of a network.  An
example of this indetermination is the so-called $Y-\Delta$
transformation given in figure~\ref{fig:ydelta}.  A critical network
can be transformed into another by a sequence of $Y-\Delta$
transformations without affecting the DtN map \cite{CurtIngMor}.
\end{remark}

%%%%%%%%%%%%%%%%%%%%%%%%%%%%%%%%%%%%%%%%%%%%%%%%%%%%%%%%%%%%%%%%%%%%%%%
\subsubsection{From the continuum to the discrete DtN map}
\label{sect:consistency}
Ingerman and Morrow \cite{MorInger} showed that pointwise values of
the kernel of $\Lambda_\s$ at any $n$ distinct nodes on $\B$ define a
matrix that is consistent with the DtN map of a circular planar
resistor network, as defined above. We consider a generalization of
these measurements, taken with electrode functions $\chi_q(\theta)$,
as given in equation (\ref{eq:RAT_MEAS2}). It is shown in
\cite{BorDruGue} that the measurement operator ${\bf M}_n$ in
(\ref{eq:RAT_MEAS2}) gives a matrix $\BM_n (\Lambda_\sigma)$ that
belongs to the set of DtN maps of circular planar resistor networks.
We can equate therefore
\begin{equation}
 \BM_n(\Lambda_\s) = \LG,
 \label{eq:exact:match}
\end{equation}
and solve the inverse problem for the network $(\Gamma,\gamma)$ to
determine the conductance $\gamma$ from the data $\LG$.

%%%%%%%%%%%%%%%%%%%%%%%%%%%%%%%%%%%%%%%%%%%%%%%%%%%%%%%%%%%%%%%%%%%%%%%
\subsection{Solving the 2D problem with optimal grids}
\label{sect:fmog}
To approximate $\s(\bx)$ from the network conductance $\gamma$ we
modify the reconstruction mapping introduced in
section~\ref{sect:Lay_rat} for layered media.  The approximation is
obtained by interpolating the output of the reconstruction mapping on
the optimal grid computed for the reference $\sr \equiv 1$. This grid
is described in sections \ref{sect:FBgrids} and \ref{sect:RAT_interp}.
But which interpolation should we take?  If we could have grids with
as many points as we wish, the choice of the interpolation would not
matter.  This was the case in section \ref{sect:Lay_ISP}, where we
studied the continuum limit $n \to \infty$ for the inverse spectral
problem. The EIT problem is exponentially unstable and the whole idea
of our approach is to have a sparse parametrization of the unknown
$\s$.  Thus, $n$ is typically small, and the approximation of $\s$
should go beyond ad-hoc interpolations of the parameters returned by
the reconstruction mapping.  We show in section~\ref{sect:fbgn} how to
approximate $\s$ with a Gauss-Newton iteration preconditioned with the
reconstruction mapping.  We also explain briefly how one can introduce
prior information about $\s$ in the inversion method.

%%%%%%%%%%%%%%%%%%%%%%%%%%%%%%%%%%%%%%%%%%%%%%%%%%%%%%%%%%%%%%%%%%%%%%%
\subsubsection{The reconstruction mapping}
\label{sect:2Drecmapping}
The idea behind the reconstruction mapping is to interpret the
resistor network $(\Gamma,\gamma)$ determined from the measured $\LG =
\BM_n(\Lambda_\s)$ as a finite volumes discretization of the equation
(\ref{eq:CONDeq}) on the optimal grid computed for $\sr \equiv 1$.
This is what we did in section~\ref{sect:Lay_rat} for the layered
case, and the approach extends to the two dimensional problem.

The conductivity is related to the conductances $\gamma(E)$, for $E
\in \E$, via quadrature rules that approximate the current fluxes
\eqref{eq:RAlex} through the dual edges. We could use for example
the quadrature in \cite{BDM-10,BDMG-10,MamonovPhD}, where the
conductances are 
\begin{equation}
 \gamma_{a,b} = \sigma(P_{a,b}) \frac{
 L(\Sigma_{a,b})}{L(E_{a,b})},
 \label{eq:quad}
\end{equation}
$(a, b) \in \left\{ \left(i,j \pm \frac{1}{2}\right), \; \left(i \pm
    \frac{1}{2}, j\right) \right\}$ and $L$ denotes the arc length of
the primary and dual edges $E$ and $\Sigma$ (see section
\ref{sect:FVol} for the indexing and edge notation). Another example
of quadrature is given in \cite{BorDruGue}. It is specialized to
tensor product grids in a disk, and it coincides with the quadrature
\eqref{eq:alphas} in the case of layered media.  For inversion
purposes, the difference introduced by different quadrature rules
is small (see \cite[Section 2.4]{BDM-10}).

To define the reconstruction mapping $\Qc_n$, we solve two inverse problems
for resistor networks. One with the measured data $\LG =
\BM_n(\Lambda_\s)$, to determine the conductance $\gamma$, and one with
the computed data $\LGr = \BM_n(\Lambda_\sr)$, for the reference $\sr
\equiv 1$. The latter gives the reference conductance $\gamma^{(o)}$ 
which we associate with the geometrical factor in (\ref{eq:quad})
\begin{equation}
\gamma^{(o)}_{a,b} \approx \frac{
 L(\Sigma_{a,b})}{L(E_{a,b})},
\end{equation}
so that we can write
\begin{equation}
 \s(P_{a,b}) \approx \sigma_{a,b} =
 \frac{\gamma_{a,b}}{\gamma_{a,b}^{(o)}}.
 \label{eq:2Dratios}
\end{equation}
Note that \eqref{eq:2Dratios} becomes \eqref{eq:LQn} in the layered
case, where \eqref{eq:alphas} gives $\alpha_j =
h_\theta/\gamma_{j+\frac{1}{2},q}$ and $\hal_j = h_\theta
\gamma_{j,q+\frac{1}{2}}$. The factors $h_\theta$ cancel out.

Let us call $\D_n$ the set in $\R^{e}$ of $e = n(n-1)/2$ independent
measurements in $\BM_n(\Lambda_\s)$, obtained by removing the
redundant entries. Note that there are $e$ edges in the network, as
many as the number of the data points in $\D_n$, given for
example by the entries in the upper triangular part of
$\BM_n(\Lambda_\s)$, stacked column by column in a vector in
$\R^{e}$.  By the consistency of the measurements
(section~\ref{sect:consistency}), $\D_n$ coincides with the set of the
strictly upper triangular parts of the DtN maps of circular planar
resistor networks with $n$ boundary nodes. The mapping $\Qc_n: \D_n
\to \R^{e}_+$ associates to the measurements in $\D_n$ the $e$
positive values $\s_{a,b}$ in \eqref{eq:2Dratios}.

\begin{figure}[t]
 \begin{center}
 \includegraphics[width=0.9\textwidth]{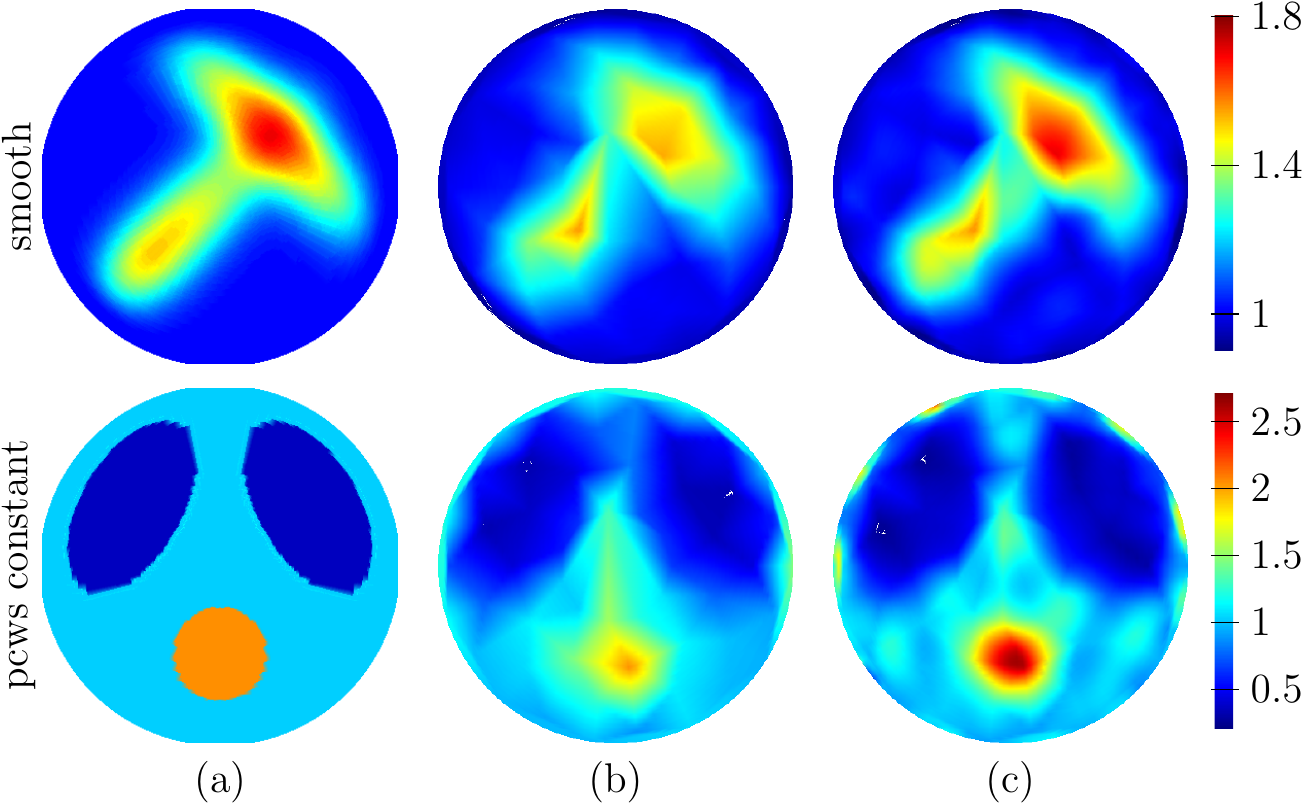}
 \end{center}
 \caption{(a) True conductivity phantoms. (b) The output
 of the reconstruction mapping $\Qc_n$, linearly interpolated on a grid
 obtained for layered media as in section~\ref{sect:RAT_interp}.
 (c) One step of Gauss-Newton improves the reconstructions.}
 \label{fig:fbrec}
\end{figure}

We illustrate in Figure~\ref{fig:fbrec}(b) the output of the mapping
$\Qc_n$, linearly interpolated on the optimal grid. It gives a good
approximation of the conductivity that is improved further in
Figure~\ref{fig:fbrec}(c) with the Gauss-Newton iteration described
below. The results in Figure~\ref{fig:fbrec} are obtained by
solving the inverse problem for the networks with a fast layer peeling
algorithm \cite{CurtMooMor}.  Optimization can also be used for this
purpose, at some additional computational cost. In any case, because
we have relatively few $n(n-1)/2$ parameters, the cost is negligible
compared to that of solving the forward problem on a fine grid.

%%%%%%%%%%%%%%%%%%%%%%%%%%%%%%%%%%%%%%%%%%%%%%%%%%%%%%%%%%%%%%%%%%%%%%%
\subsubsection{The optimal grids and sensitivity functions}
\label{sect:2Doptgrid}
The definition of the tensor product optimal grids considered in
sections \ref{sect:FBgrids} and \ref{sect:Layered} does not extend to
partial boundary measurement setups or to non-layered reference
conductivity functions. We present here an alternative approach to
determining the location of the points $P_{a,b}$ at which we approximate
the conductivity in the output (\ref{eq:2Dratios}) of the
reconstruction mapping. This approach extends to arbitrary setups, and
it is based on the sensitivity analysis of the conductance function
$\gamma$ to changes in the conductivity \cite{BDMG-10}.

The {\em sensitivity grid} points are defined as the maxima of the
sensitivity functions $D_\s\gamma_{a,b}(\bx)$. They are the points at
which the conductances $\gamma_{a,b}$ are most sensitive to changes in
the conductivity. The sensitivity functions $D_\s \gamma (\bx)$ are
obtained by differentiating the identity $\Lambda_{\gamma(\s)} =
\BM_n(\Lambda_\s)$ with respect to $\s$,
\begin{equation}
\left( D_\sigma \gamma \right) (\bx) = \left( \left. D_\gamma
\LG \right|_{\LG =
\BM_n(\Lambda_\sigma)} \right)^{-1} \mathrm{vec} \left(
\BM_n (D\K_\sigma) (\bx) \right) , \qquad \bx \in \Omega.
\label{eqn:sensdef}
\end{equation}
The left hand side is a vector in $\R^{e}$. Its
$k-$th entry is the Fr\'echet derivative of conductance
$\gamma_{k}$ with respect to changes in the conductivity $\s$. The
entries of the Jacobian $D_\gamma \LG \in \R^{e \times e}$
are 
\begin{equation}
\left( D_\gamma \LG \right)_{jk} = \left( \mbox{vec} \left(
\frac{\partial \LG}{\partial \gamma_k} \right) \right)_j,
\label{eq:jacgamma}
\end{equation}
where $\mbox{vec}(A)$ denotes the operation of stacking in a vector in
$\R^e$ the entries in the strict upper triangular part of a matrix $A
\in \mathbb{R}^{n \times n}$.  The last factor in (\ref{eqn:sensdef})
is the sensitivity of the measurements to changes of the conductivity, 
given by 
\begin{equation}
  \left( \BM_n ( D \K_\sigma ) \right)_{ij}(\bx) = 
  \left\{ \begin{tabular}{ll}
      $\displaystyle\int\limits_{\B \times \B} 
      \chi_i(x) D \K_\sigma(\bx;x,y) \chi_j(y) dx dy$, & $i \neq j$, \\ \\
      $- \displaystyle\sum\limits_{k \neq i} \int\limits_{\B \times \B} 
      \chi_i(x) D \K_\sigma(\bx;x,y) \chi_k(y) dx dy$, & $i = j$.
\end{tabular} \right. 
\end{equation}
Here  $\K_\sigma(x,y)$ is the kernel of the DtN map evaluated at points
$x$ and $y \in \B$. Its Jacobian to changes in the conductivity is
\begin{equation}
 D \K_\sigma(\bx; x, y) = \sigma(x) \sigma(y) 
\left\{ \nabla_\bx (\bn(x) \cdot \nabla_x  G(x,\bx)) \right\} \cdot 
\left\{ \nabla_\bx (\bn(y) \cdot \nabla_y  G(\bx,y)) \right\}, 
\label{eqn:jacobiankernel}
\end{equation}
where $G$ is the Green's function of the differential operator $u \to \nabla \cdot (\sigma
\nabla u)$ with Dirichlet boundary conditions, and $\bn(x)$ is the outer
unit normal at $x \in \B$. For more details on the calculation of the
sensitivity functions see \cite[Section 4]{BDMG-10}.

%%%%%%%%%%%%%%%%%%%%%%%%%%%%%%%%%%%%%%%%%%%%%%%%%%%%%%%%%%%%%%%%%%%%%
\begin{figure}[t]
\begin{center}
  \begin{tabular}{ccc}
  $D_\s \gamma_{1,1/2} / \gamma_{1,1/2}^{(0)}$ & 
  $D_\s \gamma_{3/2,0} / \gamma_{3/2,0}^{(0)}$ & 
  $D_\s \gamma_{2,1/2} / \gamma_{2,1/2}^{(0)}$\\
  \includegraphics[width=0.25\textwidth]{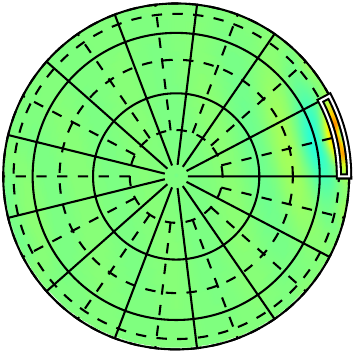} &
  \includegraphics[width=0.25\textwidth]{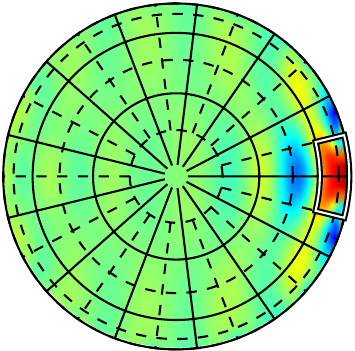} & 
  \includegraphics[width=0.25\textwidth]{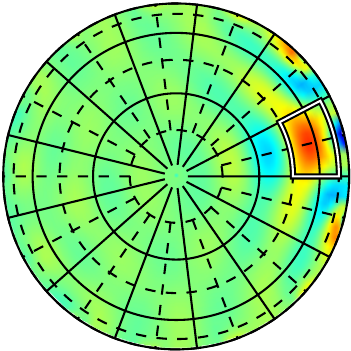}\\
  \includegraphics[width=0.25\textwidth]{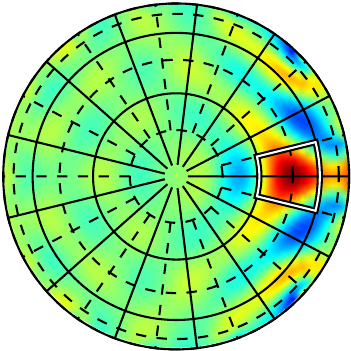} & 
  \includegraphics[width=0.25\textwidth]{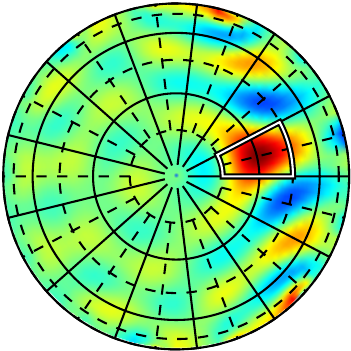} & 
  \includegraphics[width=0.25\textwidth]{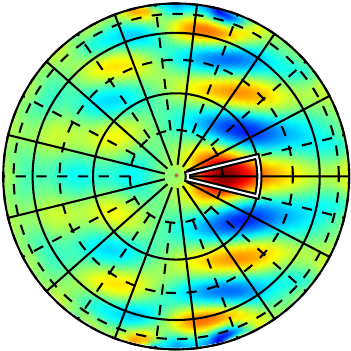}\\
  $D_\s \gamma_{5/2,0} / \gamma_{5/2,0}^{(0)}$ & 
  $D_\s \gamma_{3,1/2} / \gamma_{3,1/2}^{(0)}$ & 
  $D_\s \gamma_{7/2,0} / \gamma_{7/2,0}^{(0)}$
 \end{tabular}
\end{center}
 \caption{Sensitivity functions $\diag(1/\gamma^{(0)}) D_\s \gamma$
 computed around the conductivity $\s=1$ for $n=13$. The images
 have a linear scale from dark blue to dark red spanning $\pm$ their
 maximum in absolute value. Light green corresponds to zero. We only
 display 6 sensitivity functions, the other ones can be obtained by
 integer multiple of $2\pi/13$ rotations. The primary grid is
 displayed in solid lines and the dual grid in dotted lines. The maxima
 of the sensitivity functions are very close to those of the optimal
 grid (intersection of solid and dotted lines).}
 \label{fig:idsig}
\end{figure}

The definition of the sensitivity grid points is
\begin{equation}
P_{a,b} = \arg\max_{\bx \in \Omega}\, (D_\sigma
\gamma_{_{a,b}})(\bx), \quad \mbox{evaluated at} ~ ~
\s = \sr \equiv 1.
\label{eqn:sensgriddef}
\end{equation}
We display in Figure~\ref{fig:idsig} the sensitivity functions with
the superposed optimal grid obtained as in
section~\ref{sect:RAT_interp}. Note that the maxima of the sensitivity
functions are very close to the optimal grid points in the full
measurements case.

%%%%%%%%%%%%%%%%%%%%%%%%%%%%%%%%%%%%%%%%%%%%%%%%%%%%%%%%%%%%%%%%%%%%%%%
\subsubsection{The preconditioned Gauss-Newton iteration}
\label{sect:fbgn}
Since the reconstruction mapping $\Qc_n$ gives good reconstructions
when properly interpolated, we can think of it as an approximate
inverse of the forward map $\BM_n(\Lambda_\s)$ and use it as a {\em
  non-linear} preconditioner. Instead of minimizing the misfit in the
data, we solve the optimization problem
\begin{equation}
 \min_{\s>0} \| \Qc_n(\text{vec}\,(\BM_n(\Lambda_\s))) -
 \Qc_n(\text{vec}\,(\BM_n(\Lambda_{\s_*}))) \|^2_2.
 \label{eq:minimization}
\end{equation}
Here  $\s_*$
is the conductivity that we would like to recover. For simplicity the
minimization \eqref{eq:minimization} is
formulated with noiseless data and no regularization. We refer to
\cite{BGM-11} for a study of the effect of noise and regularization on
the minimization \eqref{eq:minimization}. 

The positivity constraints in \eqref{eq:minimization} can be dealt
with by solving for the log-conductivity $\kappa = \ln (\s)$ instead
of the conductivity $\s$. With this change of variable, the residual
in \eqref{eq:minimization} can be minimized with the standard
Gauss-Newton iteration, which we write in terms of the sensitivity
functions \eqref{eqn:sensdef} evaluated at $\s^{(j)} = \exp
\kappa^{(j)}$:
\begin{equation}
 \kappa^{(j+1)} = \kappa^{(j)} - \left(
 \diag(1/\gamma^{(0)}) \; D_\s \gamma \; \diag(\exp\kappa^{(j)})
 \right)^\dagger
 \left[\Qc_n(\text{vec}\,(\BM_n(\Lambda_{\exp \kappa^{(j)}}))) -
 \Qc_n(\text{vec}\,(\BM_n(\Lambda_{\s_*})))\right].
 \label{eq:gniter}
\end{equation}
The superscript $\dagger$ denotes the Moore-Penrose pseudoinverse and
the division is understood componentwise.  We take as initial guess
the log-conductivity $\kappa^{(0)} = \ln \s^{(0)}$, where $\s^{(0)}$
is given by the linear interpolation of
$\Qc_n(\text{vec}\,(\BM_n(\Lambda_{\s_*}))$ on the optimal grid (i.e.
the reconstruction from section~\ref{sect:2Drecmapping}). Having such
a good initial guess helps with the convergence of the Gauss-Newton
iteration.  Our numerical experiments indicate that the residual in
\eqref{eq:minimization} is mostly reduced in the first iteration
\cite{BorDruGue}.  Subsequent iterations do not change significantly
the reconstructions and result in negligible reductions of the
residual in \eqref{eq:minimization}. Thus, for all practical purposes,
the preconditioned problem is linear. We have also observed in
\cite{BorDruGue,BGM-11} that the conditioning of the linearized
problem is significantly reduced by the preconditioner $\Qc_n$.

\begin{remark}
  The conductivity obtained after one step of the Gauss-Newton
  iteration is in the span of the sensitivity functions
  \eqref{eqn:sensdef}. The use of the sensitivity functions as an
  optimal parametrization of the unknown conductivity was studied in
  \cite{BGM-11}. Moreover, the same preconditioned Gauss-Newton idea
  was used in \cite{GuevaraPhD} for the inverse spectral problem of
  section~\ref{sect:Lay_rat}.
\end{remark}

We illustrate the improvement of the reconstructions with one
Gauss-Newton step in Figure~\ref{fig:fbrec} (c). If prior information
about the conductivity is available, it can be added in the form of a
regularization term in \eqref{eq:minimization}. An example using total
variation regularization is given in \cite{BorDruGue}.

\section{Two dimensional media and partial boundary measurements}
\label{sect:2DPB}
\setcounter{equation}{0}

In this section we consider the two dimensional EIT problem with
partial boundary measurements.  As mentioned in section
\ref{sect:intro}, the boundary $\mathcal{B}$ is the union of the
accessible subset $\mathcal{B}_A$ and the inaccessible subset
$\mathcal{B}_I$. The accessible boundary $\mathcal{B}_A$ may consist
of one or multiple connected components. We assume that the
inaccessible boundary is grounded, so the partial boundary
measurements are a set of Cauchy data $\left\{
  \left.u\right|_{{\mathcal B}_A}, \left. \left( \sigma \bn \cdot
      \nabla u \right) \right|_{{\mathcal B}_A} \right\}$, where $u$
satisfies (\ref{eq:CONDeq}) and $\left. u \right|_{{\mathcal B}_I} =
0$. The inverse problem is to determine $\sigma$ from these Cauchy
data.

Our inversion method described in the previous sections extends to the
partial boundary measurement setup. But there is a significant
difference concerning the definition of the optimal grids. The tensor
product grids considered so far are essentially one dimensional, and
they rely on the rotational invariance of the problem for $\sr \equiv
1$. This invariance does not hold for the partial boundary
measurements, so new ideas are needed to define the optimal grids. We
present two approaches in sections \ref{sect:CT} and \ref{sect:ST}.
The first one uses circular planar networks with the same topology as
before, and mappings that take uniformly distributed points on $\B$ to
points on the accessible boundary $\B_A$. The second one uses networks
with topologies designed specifically for the partial boundary
measurement setups. The underlying two dimensional optimal grids are
defined with sensitivity functions.

\subsection{Coordinate transformations for the partial data problem}
\label{sect:CT}

The idea of the approach described in this section is to map the
partial data problem to one with full measurements at equidistant
points, where we know from section \ref{sect:2DFB} how to define the
optimal grids. Since $\Omega$ is a unit disk, we can do this with 
diffeomorphisms of the unit disk to itself.

Let us denote such a diffeomorphism by $F$ and its inverse $F^{-1}$ by
$G$.  If the potential $u$ satisfies (\ref{eq:CONDeq}), then the transformed
potential $\widetilde{u}(x) = u(F(x))$ solves the same equation with
conductivity $\widetilde{\sigma}$ defined by
\begin{equation}
\widetilde{\sigma}(x) = \left. 
\frac{G' (y) \sigma(y) \left(G' (y) \right)^T}{\left| \det G' (y) \right|} 
\right|_{y = F(x)},
\label{eqn:diffeo}
\end{equation}
where $G'$ denotes the Jacobian of $G$. The conductivity
$\widetilde{\sigma}$ is the \emph{push forward} of $\sigma$ by $G$,
and it is denoted by $G_* \sigma$.  Note that if $G' (y) \left(G' (y)
\right)^T \neq I$ and $\det G' (y) \neq 0$, then $\widetilde{\sigma}$
is a symmetric positive definite tensor.  If its eigenvalues are
distinct, then the push forward of an isotropic conductivity is
anisotropic.

The push forward $g_*\Lambda_{\sigma}$ of the DtN map is written in
terms of the restrictions of diffeomorphisms $G$ and $F$ to the
boundary. We call these restrictions $g = \left. G
\right|_{\mathcal{B}}$ and $f = \left. F \right|_{\mathcal{B}}$ and
write
\begin{equation}
((g_* \Lambda_{\sigma}) \uB)(\theta) = 
\left. (\Lambda_\sigma (\uB \circ g) )(\tau) \right|_{\tau=f(\theta)}, 
\qquad \theta \in [0, 2\pi),
\label{eqn:DtNpushfwd}
\end{equation}
for $\uB \in H^{1/2}(\mathcal{B})$. It is shown in
\cite{sylvester1990aib} that the DtN map is \emph{invariant} under the
push forward in the following sense
\begin{equation}
g_* \Lambda_\sigma = \Lambda_{G_* \sigma}.
\label{eqn:DtNinvar}
\end{equation}
Therefore, given (\ref{eqn:DtNinvar}) we can compute the push forward
of the DtN map, solve the inverse problem with data
$g_*\Lambda_\sigma$ to obtain $G_* \sigma$, and then map it back using
the inverse of (\ref{eqn:DtNpushfwd}). This requires the full
knowledge of the DtN map. However, if we use the discrete analogue of
the above procedure, we can transform the discrete measurements of
$\Lambda_\sigma$ on $\mathcal{B}_A$ to discrete measurements at 
equidistant points on $\B$, from which we can estimate
$\widetilde{\sigma}$ as described in section \ref{sect:2DFB}.

There is a major obstacle to this procedure: The EIT problem is
uniquely solvable just for isotropic conductivities. Anisotropic
conductivities are determined by the DtN map only up to a
boundary-preserving diffeomorphism \cite{sylvester1990aib}. Two
distinct approaches to overcome this obstacle are described in
sections \ref{sect:conformal} and \ref{sect:quasiconformal}. The first
one uses conformal mappings $F$ and $G$, which preserve the isotropy
of the conductivity, at the expense of rigid placement of the
measurement points.  The second approach uses extremal quasiconformal
mappings that minimize the artificial anisotropy of $\widetilde \s$
introduced by the placement at our will of the measurement points in
$\B_A$.

\subsubsection{Conformal mappings}
\label{sect:conformal}

The push forward $G_* \sigma$ of an isotropic $\sigma$ is isotropic if
$G$ and $F$ satisfy $G' \left( (G')^T \right) = I$ and $F'
\left((F')^T \right) = I$. This means that the diffeomorphism is
\emph{conformal} and the push forward is simply
\begin{equation}
G_* \sigma = \sigma \circ F.
\end{equation}
Since all conformal mappings of the unit disk to itself belong to the
family of M\"{o}bius transforms \cite{lavrentiev1987methods}, $F$ must
be of the form
\begin{equation}
  F(z) = e^{i \omega} \frac{ z - a }{ 1 - \overline{a} z}, \qquad 
  z \in \mathbb{C}, ~~ |z| \leq 1, ~~ 
\omega \in [0, 2\pi), ~~ a \in \mathbb{C}, ~~ |a|<1,
\label{eqn:moebius}
\end{equation}
where we associate $\mathbb{R}^2$ with the complex plane $\mathbb{C}$.
Note that the group of transformations (\ref{eqn:moebius}) is
extremely rigid, its only degrees of freedom being the numerical
parameters $a$ and $\omega$.

\begin{figure}[t!]
 \begin{center}
   \includegraphics[height=0.35\textwidth]{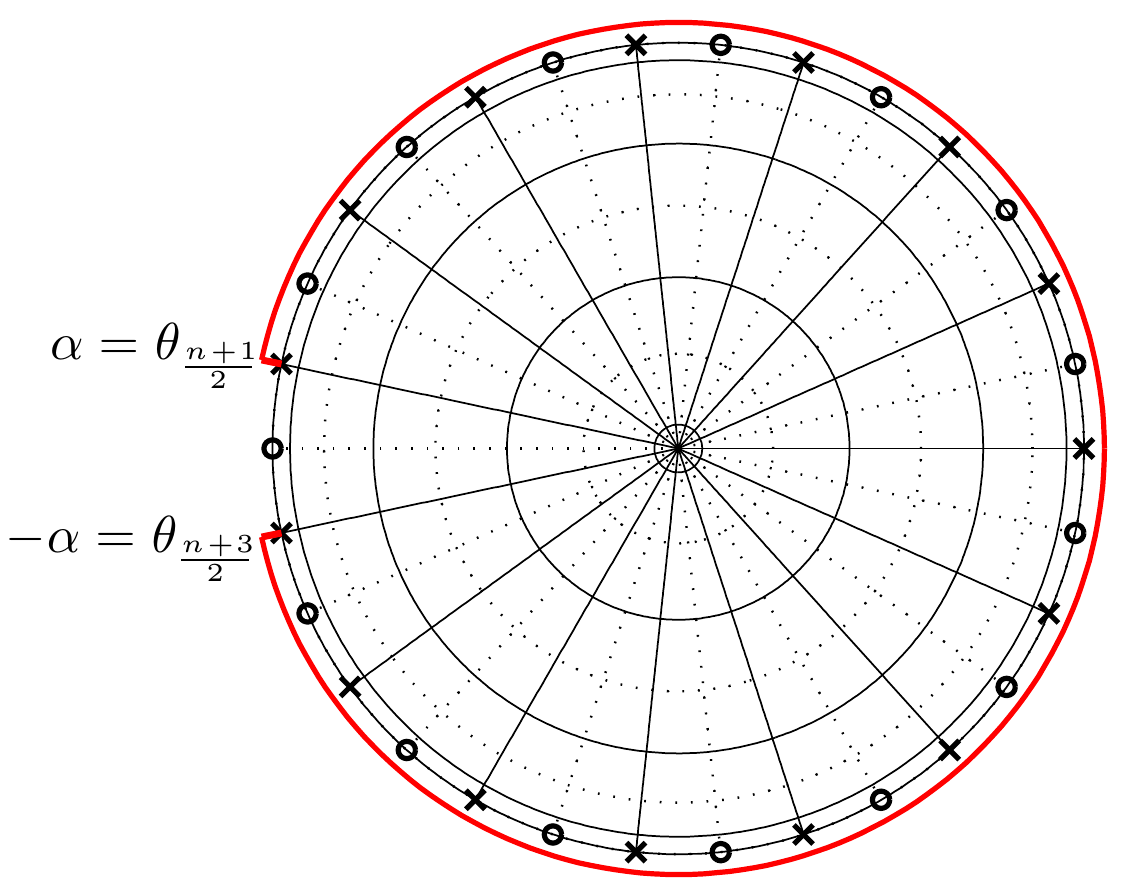}
   \includegraphics[height=0.35\textwidth]{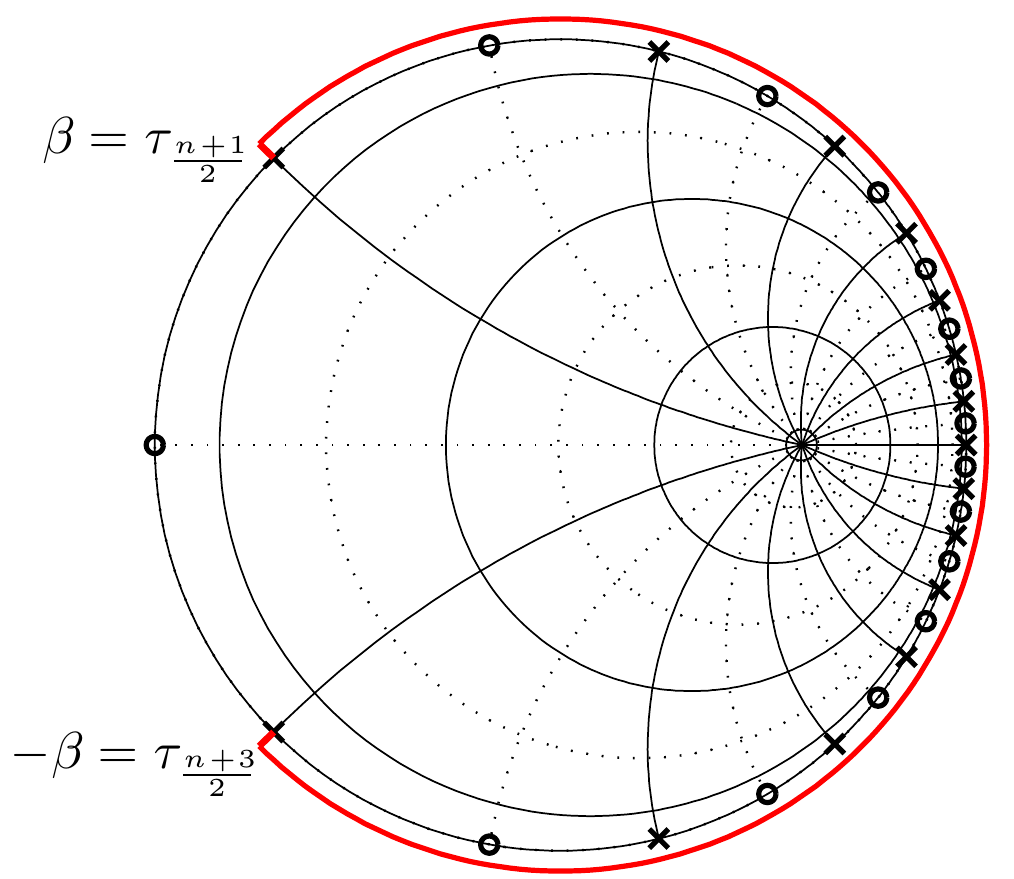}
 \end{center}
 \caption[The optimal grid under conformal
 mapping]{\label{fig:confgrid} \renewcommand{\baselinestretch}{1}
   \small\normalsize The optimal grid in the unit disk (left) and its
   image under the conformal mapping $F$ (right).  Primary grid lines
   are solid black, dual grid lines are dotted black.  Boundary grid
   nodes: primary $\times$, dual $\circ$.  The accessible boundary
   segment $\mathcal{B}_A$ is shown in solid red.  }
\end{figure}

To use the full data discrete inversion procedure from section
\ref{sect:2DFB} we require that $G$ maps the accessible boundary
segment $\mathcal{B}_A = \left\{ e^{i \tau} \; | \; \tau \in [-\beta,
  \beta] \right\}$ to the whole boundary with the exception of one
segment between the equidistant measurement points $\theta_k$,
$k=(n+1)/2,\; (n+3)/2$ as shown in Figure \ref{fig:confgrid}. This
determines completely the values of the parameters $a$ and $\omega$ in
(\ref{eqn:moebius}) which in turn determine the mapping $f$ on the
boundary. Thus, we have no further control over the positioning of the
measurement points $\tau_k = f(\theta_k)$, $k=1,\ldots,n$.

As shown in Figure \ref{fig:confgrid} the lack of control over
$\tau_k$ leads to a grid that is highly non-uniform in angle. In fact
it is demonstrated in \cite{BDM-10} that as $n$ increases there is no
asymptotic refinement of the grid away from the center of $\B_A$,
where the points accumulate.  However, since the limit $n \to \infty$
is unattainable in practice due to the severe ill-conditioning of the
problem, the grids obtained by conformal mapping can still be useful
in practical inversion. We show reconstructions with these grids in
section \ref{sect:numericalpartial}.

\subsubsection{Extremal quasiconformal mappings}
\label{sect:quasiconformal}

To overcome the issues with conformal mappings that arise due to the
inherent rigidity of the group of conformal automorphisms of the unit
disk, we use here quasiconformal mappings.  A quasiconformal mapping
$F$ obeys a Beltrami equation in $\Omega$
\begin{equation}
  \frac{\partial F}{\partial \overline{z}} = \mu(z) 
\frac{\partial F}{\partial z}, \quad 
  \| \mu \|_{\infty} < 1,
\label{eqn:beltrami}
\end{equation}
with a Beltrami coefficient $\mu(z)$ that measures how much $F$
differs from a conformal mapping. If $\mu \equiv 0$, then
(\ref{eqn:beltrami}) reduces to the Cauchy-Riemann equation and $F$ is
conformal. The magnitude of $\mu$ also provides a measure of the 
anisotropy $\kappa$ of the push forward of $\sigma$ by $F$.  The
definition of the anisotropy is
\begin{equation}
  \kappa(F_* \sigma, z) = 
  \frac{\sqrt{\lambda_1(z) / \lambda_2(z)} - 1}{\sqrt{\lambda_1(z) / 
\lambda_2(z)} + 1},
\label{eqn:aniso}
\end{equation} 
where $\lambda_1(z)$, $\lambda_2(z)$ are the largest and the smallest
eigenvalues of $F_* \sigma$ respectively. The connection between $\mu$
and $\kappa$ is given by
\begin{equation}
\kappa \left( F_* \sigma, z \right) = |\mu(z)|,
\label{eqn:pushfwd}
\end{equation}
and the maximum anisotropy is
\begin{equation}
\kappa(F_* \sigma) = \sup_z \kappa(F_* \sigma, z) = \| \mu \|_{\infty}.
\end{equation}

Since the unknown conductivity is isotropic, we would like to minimize
the amount of artificial anisotropy that we introduce into the
reconstruction by using $F$. This can be done with extremal
quasiconformal mappings, which minimize $\|\mu \|_\infty$ under
constraints that fix $f = \left.  F \right|_{\mathcal{B}}$, thus
allowing us to control the positioning of the measurement points
$\tau_k = f(\theta_k)$, for $k=1,\ldots,n$.

For sufficiently regular boundary values $f$ there exists a unique
extremal quasiconformal mapping that is known to be of a
Teichm\"{u}ller type \cite{strebel1976eet}. Its Beltrami coefficient
satisfies
\begin{equation}
\mu(z) = \| \mu \|_{\infty} \frac{ \overline{\phi(z)} }{ |\phi(z)| },
\label{eqn:teichphi}
\end{equation}
for some holomorphic function $\phi(z)$ in $\Omega$. Similarly, we can
define the Beltrami coefficient for $G$, using a holomorphic function
$\psi$. It is established in \cite{reich1976quasiconformal} that $F$
admits a decomposition
\begin{equation}
F = \Psi^{-1} \circ A_K \circ \Phi,
\label{eqn:teichdecomp}
\end{equation}
where
\begin{equation}
  \Phi(z) = \int \sqrt{\phi(z)} dz, \qquad \Psi(\zeta) = 
\int \sqrt{\psi(\zeta)} d\zeta,
\label{eqn:defphipsi}
\end{equation}
are conformal away from the zeros of $\phi$ and $\psi$, and 
\begin{equation}
A_K(x + iy) = Kx + iy 
\end{equation}
is an affine stretch, the only source of anisotropy in
(\ref{eqn:teichdecomp}):
\begin{equation}
  \kappa \left( F_* \sigma \right) = \| \mu \|_\infty = 
\left|\frac{K-1}{K+1} \right|.
\label{eqn:quasianiso}
\end{equation}

\begin{figure}[t!]
 \begin{center}
   \includegraphics[height=0.17\textwidth]{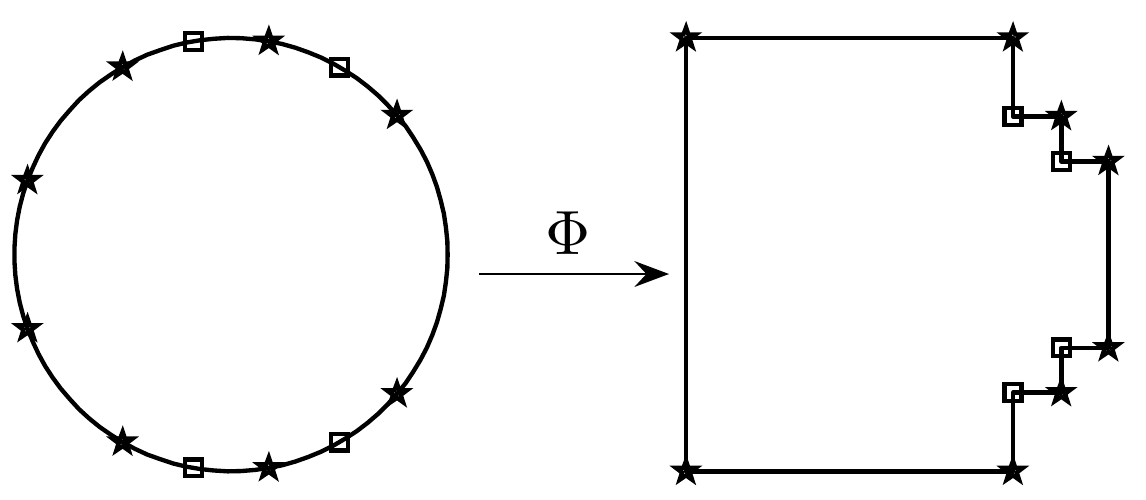}
   \includegraphics[height=0.17\textwidth]{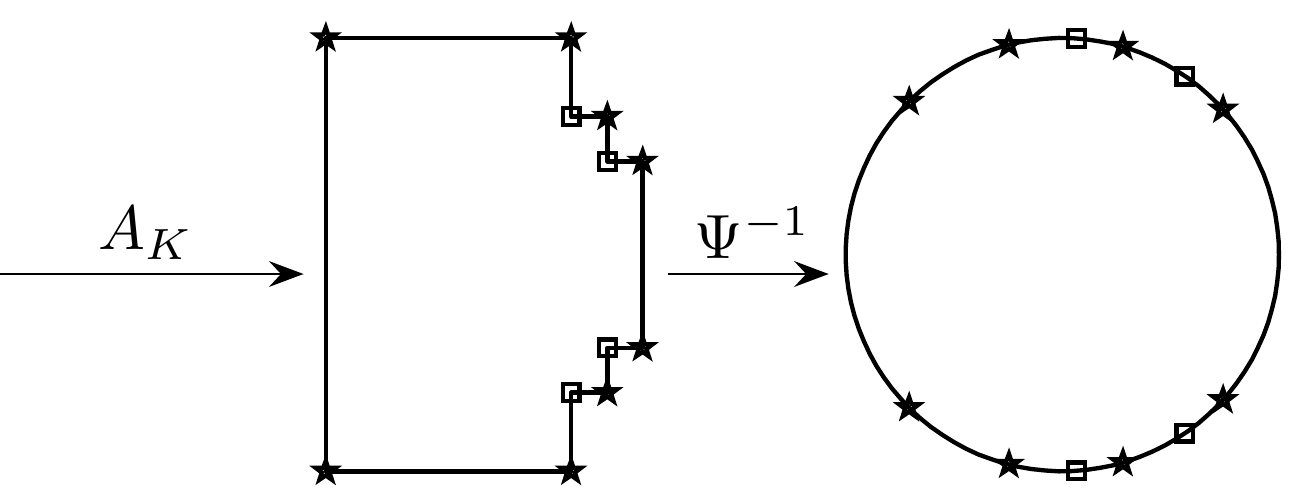}
 \end{center}
 
 \caption[Decomposition of a Teichm\"{u}ller
 mapping]{\label{fig:teichdecomp} \renewcommand{\baselinestretch}{1}
   \small\normalsize Teichm\"{u}ller mapping decomposed into conformal
   mappings $\Phi$ and $\Psi$, and an affine transform $A_K$. The
   poles of $\phi$ and $\psi$ and their images under $\Phi$ and $\Psi$
   are $\bigstar$, the zeros of $\phi$ and $\psi$ and their images
   under $\Phi$ and $\Psi$ are $\square$.  }
\end{figure}

Since only the behavior of $f$ at the measurement points $\theta_k$ is
of interest to us, it is possible to construct explicitly the mappings
$\Phi$ and $\Psi$ \cite{BDM-10}.  They are Schwartz-Christoffel
conformal mappings of the unit disk to polygons of special form, as
shown in Figure \ref{fig:teichdecomp}. See \cite[Section 3.4]{BDM-10}
for more details.

\begin{figure}[t!]
 \begin{center}
   \includegraphics[height=0.35\textwidth]{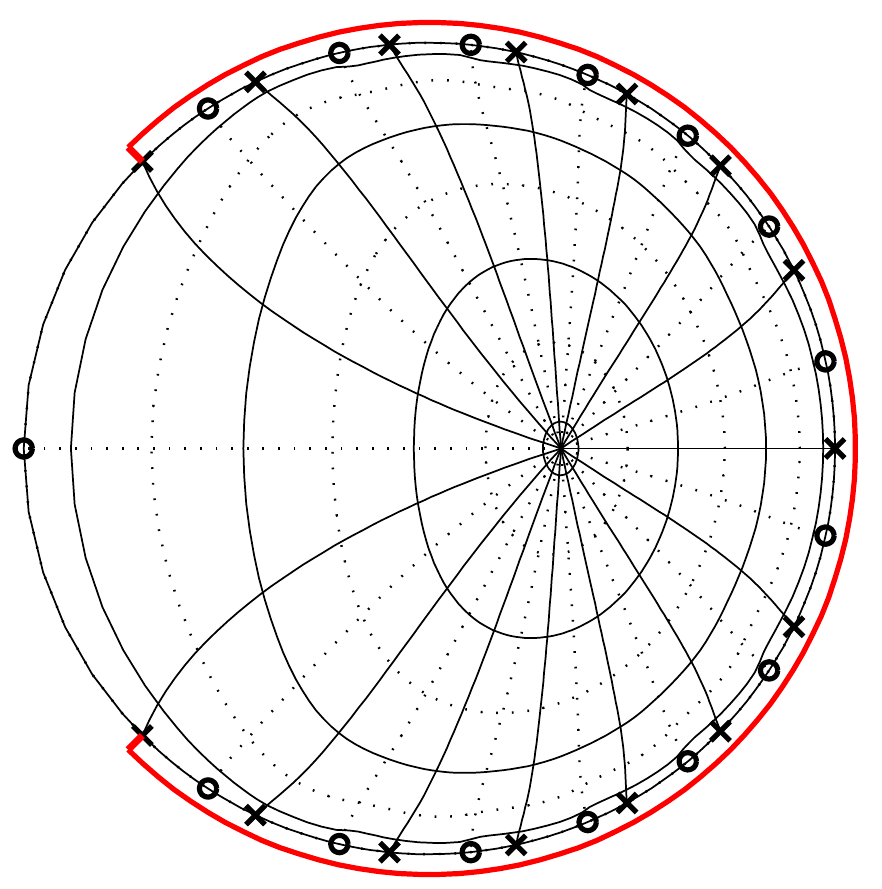}
   \hskip0.3in
   \includegraphics[height=0.35\textwidth]{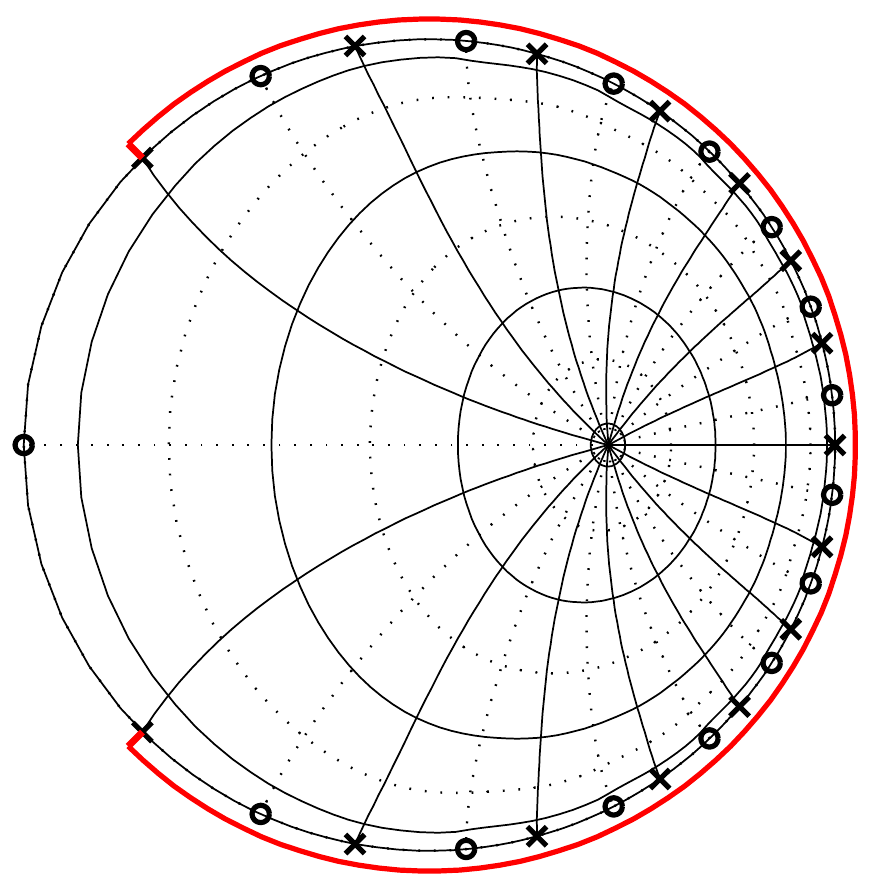}
 \end{center}
 
 \caption[The optimal grid under Teichm\"{u}ller
 mapping]{\label{fig:teichgrid} \renewcommand{\baselinestretch}{1}
   \small\normalsize The optimal grid under the quasiconformal
   Teichm\"{u}ller mappings $F$ with different $K$.  Left: $K=0.8$
   (smaller anisotropy); right: $K=0.66$ (higher anisotropy).  Primary
   grid lines are solid black, dual grid lines are dotted black.
   Boundary grid nodes: primary $\times$, dual $\circ$.  The accessible
   boundary segment $\mathcal{B}_A$ is shown in solid red.  }
\end{figure}

We demonstrate the behavior of the optimal grids under the extremal
quasiconformal mappings in Figure \ref{fig:teichgrid}. We present the
results for two different values of the affine stretching constant
$K$.  As we increase the amount of anisotropy from $K=0.8$ to
$K=0.66$, the distribution of the grid nodes becomes more uniform. The
price to pay for this more uniform grid is an increased amount of
artificial anisotropy, which may detriment the quality of the
reconstruction, as shown in the numerical examples in section
\ref{sect:numericalpartial}.

\subsection{Special network topologies for the partial data problem}
\label{sect:ST}

The limitations of the construction of the optimal grids with
coordinate transformations can be attributed to the fact that there is
no non-singular mapping between the full boundary $\mathcal{B}$ and
its proper subset $\mathcal{B}_A$.  Here we describe an alternative
approach, that avoids these limitations by considering networks with
different topologies, constructed specifically for the partial
measurement setups. The one-sided case, with the accessible boundary
$\mathcal{B}_A$ consisting of one connected segment, is in section
\ref{sect:piramidal}. The two sided case, with $\mathcal{B}_A$ the
union of two disjoint segments, is in section \ref{sect:twosided}. The
optimal grids are constructed using the sensitivity analysis of the
discrete and continuum problems, as explained in sections
\ref{sect:2Doptgrid} and \ref{sect:sensgrid}.

\subsubsection{Pyramidal networks for the one-sided problem}
\label{sect:piramidal}

We consider here the case of $\B_A$ consisting of one connected
segment of the boundary.  The goal is to choose a topology of the
resistor network based on the flow properties of the continuum partial
data problem. Explicitly, we observe that since the potential
excitation is supported on $\mathcal{B}_A$, the resulting currents
should not penetrate deep into $\Omega$, away from $\mathcal{B}_A$.
The currents are so small sufficiently far away from $\mathcal{B}_A$
that in the discrete (network) setting we can ask that there is no
flow escaping the associated nodes. Therefore, these nodes are
interior ones.  A suitable choice of networks that satisfy such
conditions was proposed in \cite{BDMG-10}.  We call them
\emph{pyramidal} and denote their graphs by $\Gamma_n$, with $n$ the
number of boundary nodes.

\begin{figure}[t!]
 \begin{center}
   \includegraphics[height=0.178\textwidth]{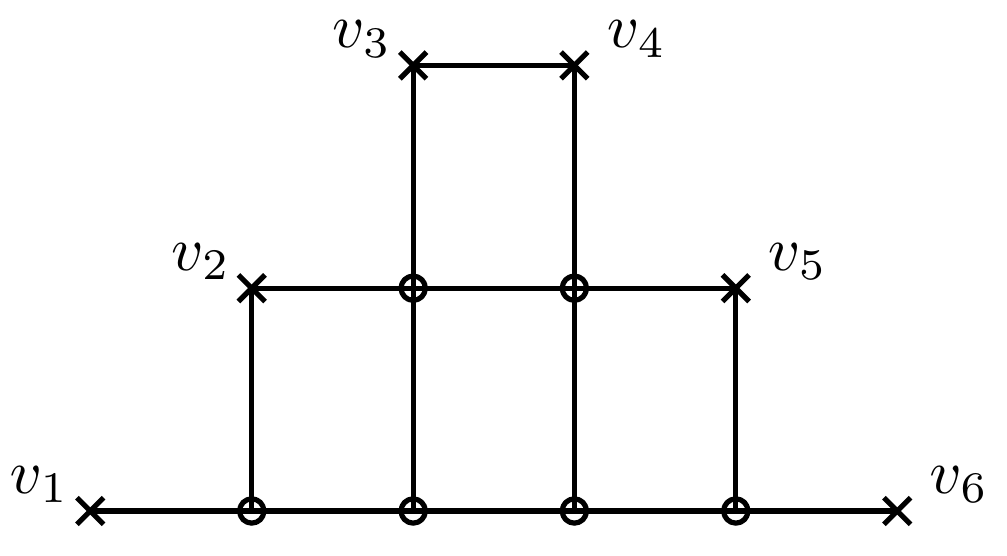}
   \hskip0.05\textwidth
   \includegraphics[height=0.250\textwidth]{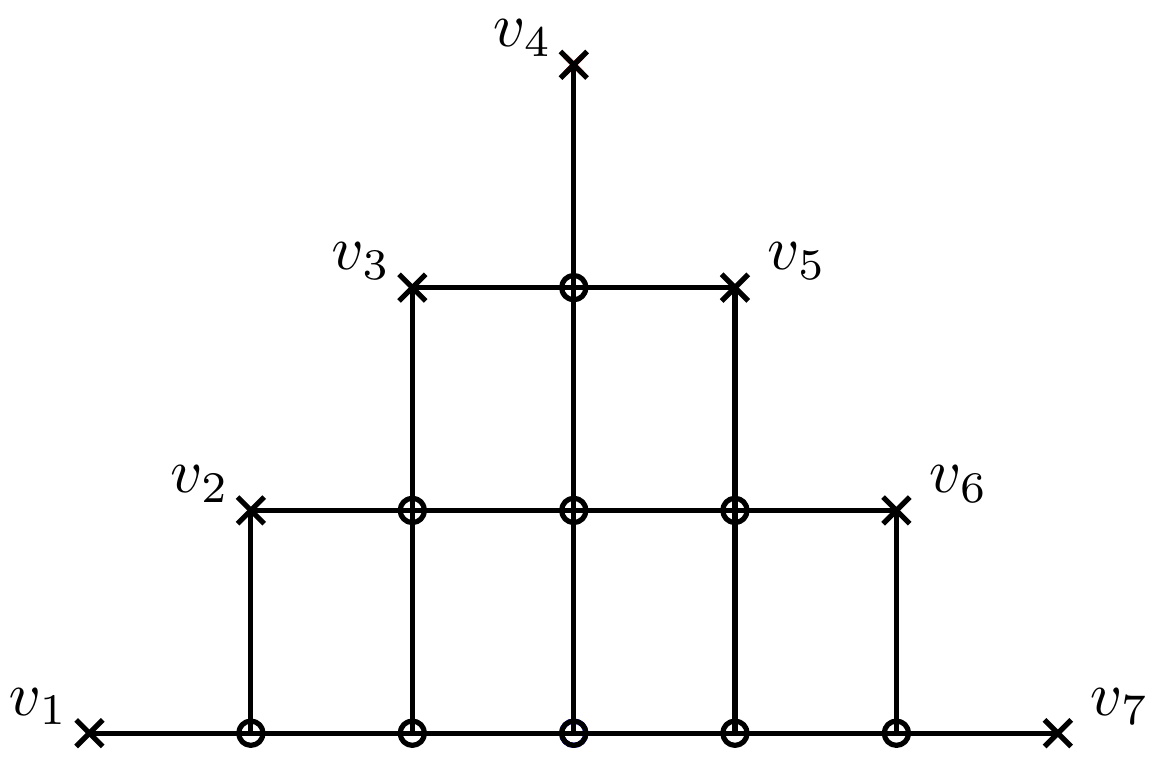}
 \end{center}
 \caption[Pyramidal networks]{\label{fig:pyramid67}
   \renewcommand{\baselinestretch}{1} \small\normalsize Pyramidal
   networks $\Gamma_n$ for $n=6$, $7$.  The boundary nodes $v_j$,
   $j=1,\ldots,n$ are indicated with $\times$ and the interior nodes
   with $\circ$.  }
\end{figure}

We illustrate two pyramidal graphs in Figure \ref{fig:pyramid67}, for
$n = 6$ and $7$. Note that it is not necessary that $n$ be odd for the
pyramidal graphs $\Gamma_n$ to be critical, as was the case in the
previous sections. In what follows we refer to the edges of $\Gamma_n$
as vertical or horizontal according to their orientation in Figure
\ref{fig:pyramid67}. Unlike the circular networks in which all the
boundary nodes are in a sense adjacent, there is a gap between the
boundary nodes $v_1$ and $v_n$ of a pyramidal network. This gap is
formed by the bottommost $n-2$ interior nodes that enforce the
condition of zero normal flux, the approximation of the lack
of penetration of currents away from $\mathcal{B}_A$.

It is known from \cite{CurtIngMor, BDMG-10} that the pyramidal
networks are critical and thus uniquely recoverable from the DtN map.
Similar to the circular network case, pyramidal networks can be
recovered using a layer peeling algorithm in a finite number of
algebraic operations.  We recall such an algorithm below, from
\cite{BDMG-10}, in the case of even $n=2m$.  A similar procedure can
also be used for odd $n$.

\begin{alg}
\label{alg:pyramidal}
To determine the conductance $\gamma$ of the pyramidal network $\left(\Gamma_n, \gamma\right)$ 
from the DtN map $\bm{\Lambda}^{(n)}$, perform the following steps:
\begin{enumerate}
\item[(1)] To compute the conductances of horizontal and vertical
  edges emanating from the boundary node $v_p$, for each $p=1,\ldots,2m$,
  define the following sets:\\
$Z = \{ v_1, \ldots, v_{p-1}, v_{p+1}, \ldots, v_m\}$, 
$C = \{ v_{m+2}, \ldots, v_{2m} \}$, \\
$H = \{ v_1, \ldots, v_{p} \}$ and 
$V = \{ v_p, \ldots, v_{m+1} \}$, in the case  $p \leq m$.\\
$Z = \{ v_{m+1}, \ldots, v_{p-1}, v_{p+1}, \ldots, v_{2m}\}$, 
$C = \{ v_1, \ldots, v_{m-1} \}$, \\
$H = \{ v_p, \ldots, v_{2m} \}$ and 
$V = \{ v_m, \ldots, v_p \}$, for  $m+1 \leq p \leq 2m$.
\item[(2)] Compute the conductance $\gamma(E_{p,h})$ of the horizontal 
edge emanating from $v_p$ using
\begin{equation}
  \gamma(E_{p,h}) =  \left( \bm{\Lambda}^{(n)}_{p, H} -
    \bm{\Lambda}^{(n)}_{p, C} \; \left( \bm{\Lambda}^{(n)}_{Z, C} 
    \right)^{-1} \;
    \bm{\Lambda}^{(n)}_{Z, H} \right) {\bf 1}_{H},
\label{eqn:alggammah}
\end{equation}
compute the conductance $\gamma(E_{p,v})$ of the vertical edge
emanating from $v_p$ using
\begin{equation}
  \gamma(E_{p,v}) =  \left( \bm{\Lambda}^{(n)}_{p, V} -
    \bm{\Lambda}^{(n)}_{p, C} \; \left( \bm{\Lambda}^{(n)}_{Z, C} 
    \right)^{-1} \;
    \bm{\Lambda}^{(n)}_{Z, V} \right) {\bf 1}_{V},
\label{eqn:alggammav}
\end{equation}
where ${\bf 1}_V$ and ${\bf 1}_H$ are column vectors of all ones.
\item[(3)] Once $\gamma(E_{p,h})$, $\gamma(E_{p,v})$ are known, peel
  the outer layer from $\Gamma_n$ to obtain the subgraph
  $\Gamma_{n-2}$ with the set $\mathcal{S} = \{ w_1, \ldots,
  w_{2m-2}\}$ of boundary nodes.  Assemble the blocks ${\bf
    K}_{\mathcal{SS}}$, ${\bf K}_{\mathcal{SB}}$, ${\bf
    K}_{\mathcal{BS}}$, ${\bf K}_{\mathcal{BB}}$ of the Kirchhoff
  matrix of $(\Gamma_n, \gamma)$, and compute the updated DtN map
  $\bm{\Lambda}^{(n-2)}$ of the smaller network
  $(\Gamma_{n-2},\gamma)$, as follows
\begin{equation}
  \bm{\Lambda}^{(n-2)} = - {\bf K}_{\mathcal{SS}}^{\prime} - 
  {\bf K}_{\mathcal{SB}} \; {\bf P}^T \; 
  \left( {\bf P} \; (\bm{\Lambda}^{(n)} - {\bf K}_{\mathcal{BB}}) \; 
    {\bf P}^T \right)^{-1} \; {\bf P} \; {\bf K}_{\mathcal{BS}}.
\label{eqn:algupdate}
\end{equation}
Here ${\bf P} \in \mathbb{R}^{(n-2) \times n}$ is a projection
operator: ${\bf P} {\bf P}^T = {\bf I}_{n-2}$, and ${\bf
  K}_{\mathcal{SS}}^{\prime}$ is a part of ${\bf K}_{\mathcal{SS}}$
that only includes the contributions from the edges connecting
$\mathcal{S}$ to $\mathcal{B}$.
\item[(4)] If $m=1$ terminate. Otherwise, decrease $m$ by 1, update
  $n=2m$ and go back to step 1.
\end{enumerate}
\end{alg}

Similar to the layer peeling method in \cite{CurtMooMor}, Algorithm
\ref{alg:pyramidal} is based on the construction of special solutions.
In steps 1 and 2 the special solutions are constructed implicitly, to
enforce a unit potential drop on edges $E_{p,h}$ and $E_{p,v}$
emanating from the boundary node $v_p$.  Since the DtN map is known,
so is the current at $v_p$, which equals to the conductance of an edge
due to a unit potential drop on that edge. Once the conductances are
determined for all the edges adjacent to the boundary, the layer of
edges is peeled off and the DtN map of a smaller network
$\Gamma_{n-2}$ is computed in step 3. After $m$ layers have been
peeled off, the network is completely recovered. The algorithm is
studied in detail in \cite{BDMG-10}, where it is also shown that all
matrices that are inverted in (\ref{eqn:alggammah}),
(\ref{eqn:alggammav}) and (\ref{eqn:algupdate}) are non-singular.

\begin{remark}
The DtN update formula (\ref{eqn:algupdate}) provides an
interesting connection to the layered case. It can be viewed as a
matrix generalization of the continued fraction representation
(\ref{eq:CFr}).  The difference between the two formulas is that
(\ref{eq:CFr}) expresses the eigenvalues of the DtN map, while
(\ref{eqn:algupdate}) gives an expression for the DtN map itself.
\end{remark}
\subsubsection{The two-sided problem}
\label{sect:twosided}

\begin{figure}[t!]
 \begin{center}
   \includegraphics[height=0.35\textwidth]{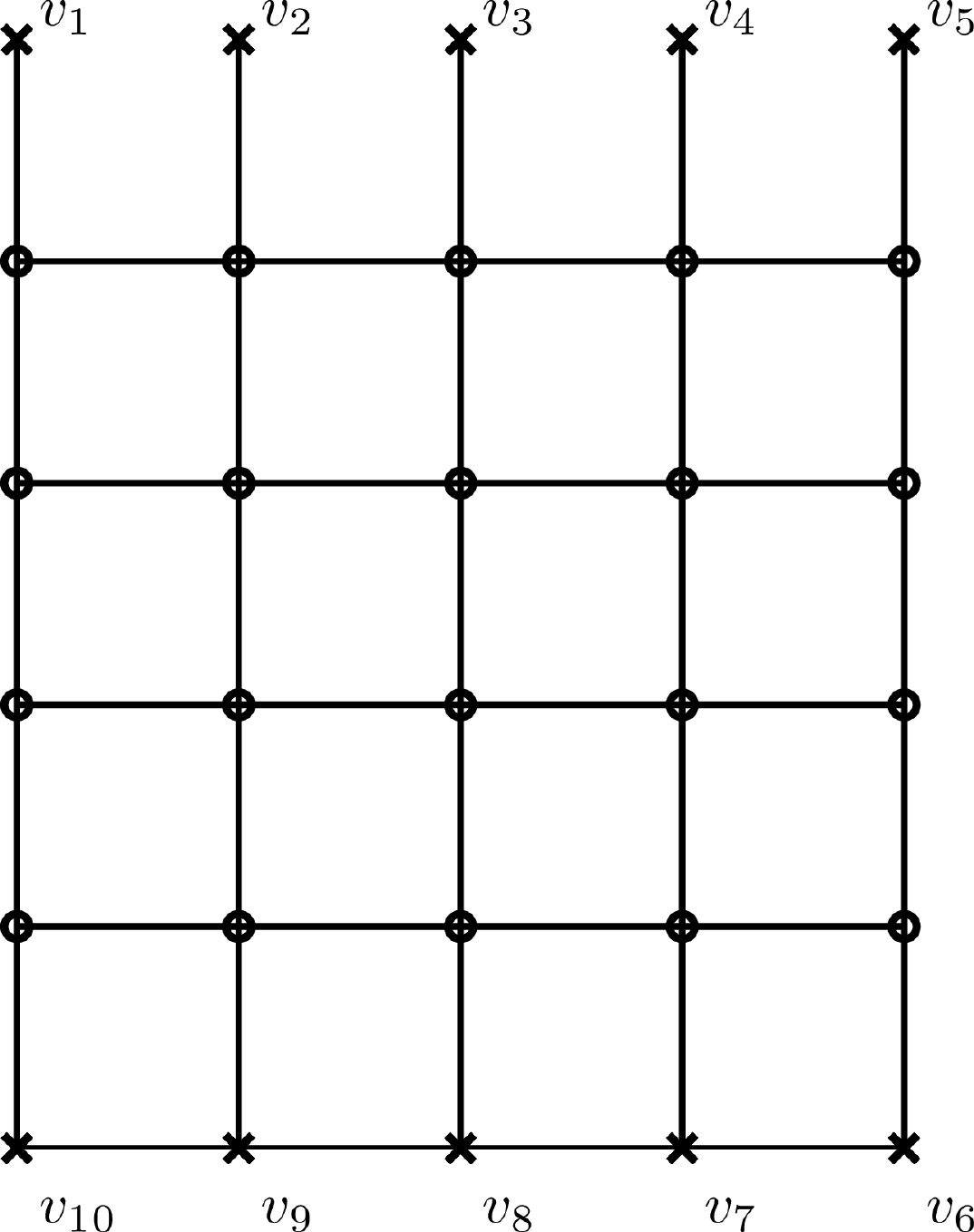}
 \end{center}
 \caption[Two-sided network]{\label{fig:twosidenet}
   \renewcommand{\baselinestretch}{1} \small\normalsize Two-sided
   network $T_n$ for $n=10$.  Boundary nodes $v_j$, $j=1,\ldots,n$ are
   $\times$, interior nodes are $\circ$.  }
\end{figure}

We call the problem two-sided when the accessible boundary
$\mathcal{B}_A$ consists of two disjoint segments of $\mathcal{B}$. A
suitable network topology for this setting was introduced in
\cite{BGM-11}.  We call these networks \emph{two-sided} and denote
their graphs by $T_n$, with $n$ the number of boundary nodes assumed
even $n=2m$. Half of the nodes are on one segment of the boundary and
the other half on the other, as illustrated in Figure
\ref{fig:twosidenet}.  Similar to the one-sided case, the two groups
of $m$ boundary nodes are separated by the outermost interior nodes,
which model the lack of penetration of currents away from the
accessible boundary segments.  One can verify that the two-sided
network is critical, and thus it can be uniquely recovered from the
DtN map by the Algorithm \ref{alg:twosided} introduced in
\cite{BGM-11}.

When referring to either the horizontal or vertical edges of a two
sided network, we use their orientation in Figure
\ref{fig:twosidenet}.
\begin{alg}
\label{alg:twosided}
To determine the conductance $\gamma$ of the two-sided network $(T_n,
\gamma)$ from the DtN map $\bm{\Lambda}_\gamma$, perform the following
steps:
\begin{enumerate}
\item[(1)] Peel the lower layer of horizontal resistors:\\ For $p =
  m+2, m+3, \ldots, 2m$ define the sets $Z = \{ p+1,
  p+2,\ldots,p+m-1\}$ and $C = \{ p-2, p-3, \ldots, p-m\}$.  The
  conductance of the edge $E_{p,q,h}$ between $v_p$ and $v_q$, $q=p-1$
  is given by
\begin{equation} 
\label{eqn:gammapq}
\gamma(E_{p,q,h}) = - \bm{\Lambda}_{p, q} +
\bm{\Lambda}_{p, C} (\bm{\Lambda}_{Z,C})^{-1} \bm{\Lambda}_{Z, q}.
\end{equation}
Assemble a symmetric tridiagonal matrix $\bm{A}$ with off-diagonal entries
$ - \gamma(E_{p,p-1,h}) $ and rows summing to zero.  Update the lower
right $m$-by-$m$ block of the DtN map by subtracting $\bm{A}$ from it.
\item[(2)] Let $s = m-1$.
\item[(3)] Peel the top and bottom layers of vertical resistors:\\ For
  $p=1,2,\ldots,2m$ define the sets $L = \{p-1,p-2,\ldots,p-s\}$ and
  $R = \{p+1,p+2,\ldots,p+s\}$. If $p<m/2$ for the top layer, or
  $p>3m/2$ for the bottom layer, set $Z=L$, $C=R$.  Otherwise let
  $Z=R$, $C=L$. The conductance of the vertical edge emanating from
  $v_p$ is given by
\begin{equation} 
\label{eqn:gammapp}
\gamma(E_{p,v}) = \bm{\Lambda}_{p, p} - 
\bm{\Lambda}_{p, C} (\bm{\Lambda}_{Z,C})^{-1} \bm{\Lambda}_{Z, p}.
\end{equation}
Let $\bm{D} = \mbox{diag} \left( \gamma(E_{p,v}) \right)$ and update
the DtN map
\begin{equation}
\label{eqn:dtnupdate}
\bm{\Lambda}_\gamma = - \bm{D} - \bm{D} \left( 
\bm{\Lambda}_\gamma + \bm{D} \right)^{-1} \bm{D}.
\end{equation}
\item[(4)] If $s=1$ go to step (7). Otherwise decrease $s$ by $2$.
\item[(5)] Peel the top and bottom layers of horizontal
  resistors:\\ For $p=1,2,\ldots,2m$ define the sets $L = \{
  p-1,p-2,\ldots,p-s\}$ and $R = \{ p+2,p+3,\ldots,p+s+1\}$. If
  $p<m/2$ for the top layer, or $p<3m/2$ for the bottom layer, set
  $Z=L$, $C=R$, $q=p+1$. Otherwise let $Z=R$, $C=L$, $q=p-1$.  The
  conductance of the edge connecting $v_p$ and $v_q$ is given by
  (\ref{eqn:gammapq}). Update the upper left and lower right blocks of
  the DtN map as in step (1).
\item[(6)] If $s=0$ go to step (7), otherwise go to (3).
\item[(7)] Determine the last layer of resistors. If $m$ is odd the
  remaining vertical resistors are the diagonal entries of the DtN
  map. If $m$ is even, the remaining resistors are horizontal. The
  leftmost of the remaining horizontal resistors $\gamma(E_{1,2,h})$ is
  determined from (\ref{eqn:gammapq}) with $p=1$, $q=m+1$, $C=\{1,2\}$,
  $Z=\{m+1,m+2\}$ and a change of sign. The rest are determined by
\begin{equation}
  \gamma(E_{p,p+1,h}) = \left( \bm{\Lambda}_{p, H} -
    \bm{\Lambda}_{p, C} (\bm{\Lambda}_{Z,C})^{-1} \bm{\Lambda}_{Z, H} \right) \mathbf{1},
\end{equation}
where $p=2,3,\ldots,m-1$, $C = \{p-1, p, p+1\}$, $Z = \{p+m-1, p+m,
p+m+1\}$, $H = \{ p+m-1, p+m\}$, and $\mathbf{1}$ is a vector
$(1,1)^T$.
\end{enumerate}
\end{alg}

Similar to Algorithm \ref{alg:pyramidal}, Algorithm \ref{alg:twosided}
is based on the construction of special solutions examined in
\cite{CurtMooMor, curtMorBook}.  These solutions are designed to
localize the flow on the outermost edges, whose conductance we
determine first. In particular, formulas (\ref{eqn:gammapq}) and
(\ref{eqn:gammapp}) are known as the ``boundary edge'' and ``boundary
spike'' formulas \cite[Corollaries 3.15 and 3.16]{curtMorBook}.

\subsubsection{Sensitivity grids for pyramidal and two-sided networks}
\label{sect:sensgrid}

\begin{figure}[t!]
 \begin{center}
   \includegraphics[height=0.37\textwidth]{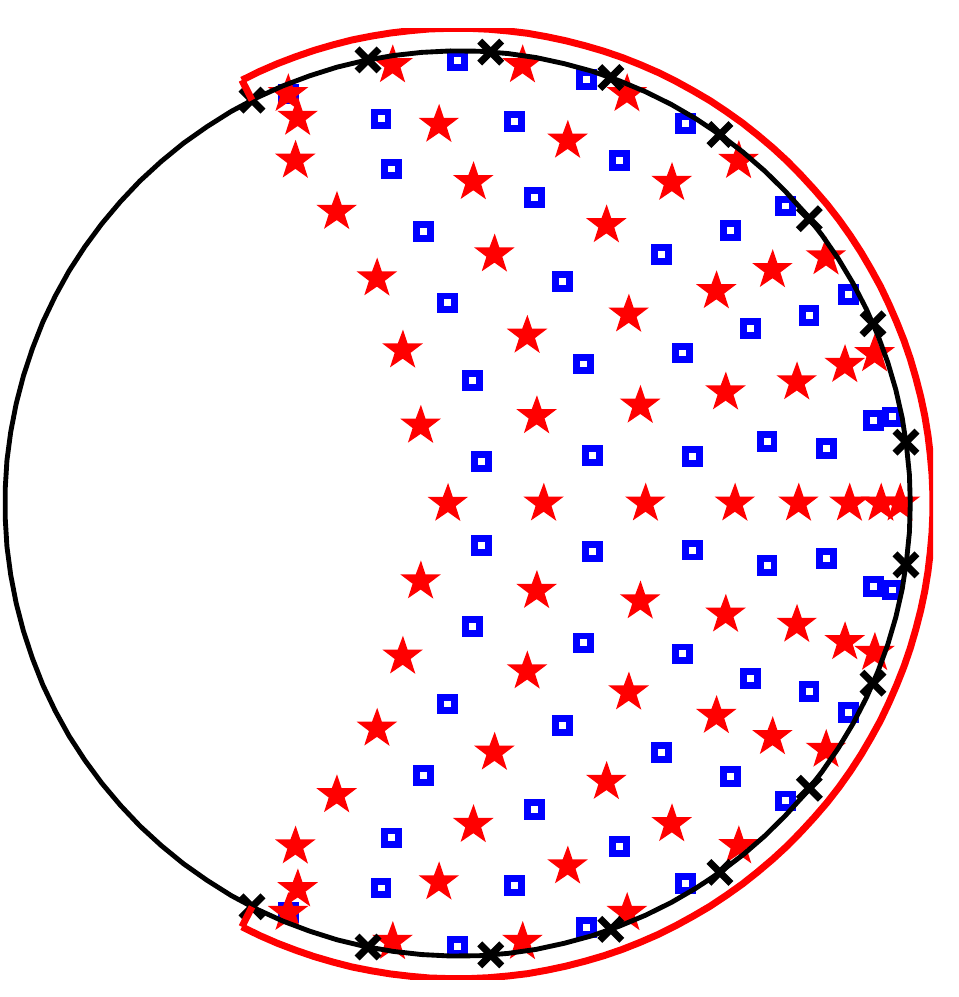}
   \hskip0.05\textwidth
   \includegraphics[height=0.37\textwidth]{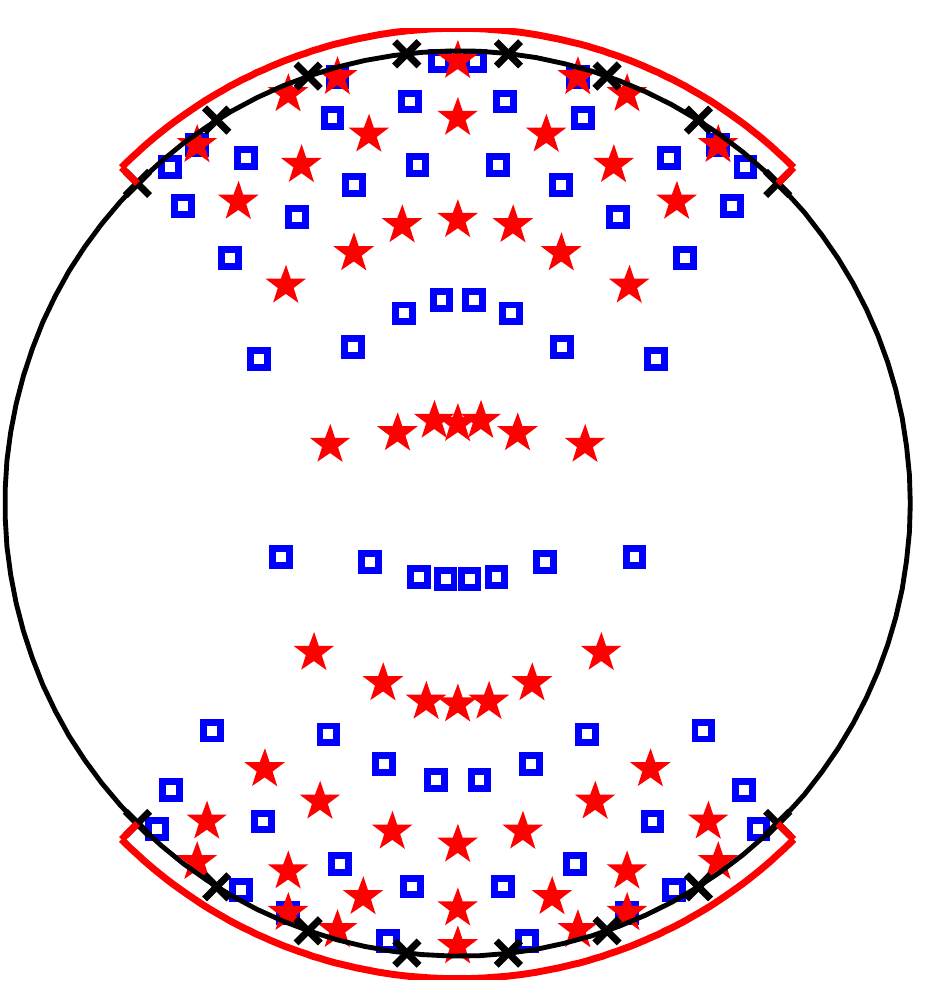}
 \end{center}
 \caption[Sensitivity grids]{\label{fig:sensgrids}
   \renewcommand{\baselinestretch}{1} \small\normalsize Sensitivity
   optimal grids in the unit disk for the pyramidal network $\Gamma_n$
   (left) and the two-sided network $T_n$ (right) with $n=16$.  The
   accessible boundary segments $\mathcal{B}_A$ are solid red.  Blue
   $\square$ correspond to vertical edges, red $\bigstar$ correspond
   to horizontal edges, measurement points are black $\times$.  }
\end{figure}

The underlying grids of the pyramidal and two-sided networks are truly
two dimensional, and they cannot be constructed explicitly as in
section \ref{sect:Layered} by reducing the problem to a one
dimensional one.  We define the grids with the sensitivity function
approach described in section \ref{sect:2Doptgrid}.  The computed
sensitivity grid points are presented in Figure \ref{fig:sensgrids},
and we observe a few important properties.  First, the neighboring
points corresponding to the same type of resistors (vertical or
horizontal) form rather regular virtual quadrilaterals. Second, the
points corresponding to different types of resistors interlace in
the sense of lying inside the virtual quadrilaterals formed by the
neighboring points of the other type.  Finally, while there is some
refinement near the accessible boundary (more pronounced in the
two-sided case), the grids remain quite uniform throughout the covered
portion of the domain.
% Such behavior is typical of grids with good approximation
% properties.

Note from Figure \ref{fig:twosidenet} that the graph $T_n$ lacks the
upside-down symmetry. Thus, it is possible to come up with two sets of
optimal grid nodes, by fitting the measured DtN map
$\BM_n(\Lambda_\s)$ once with a two-sided network and the second time
with the network turned upside-down. This way the number of nodes in
the grid is essentially doubled, thus doubling the resolution of the
reconstruction. However, this approach can only improve resolution in
the direction transversal to the depth, as shown in \cite[Section
2.5]{BGM-11}.

\subsection{Numerical results}
\label{sect:numericalpartial}

\begin{figure}[t!]
 \begin{center}
   \includegraphics[width=\textwidth]{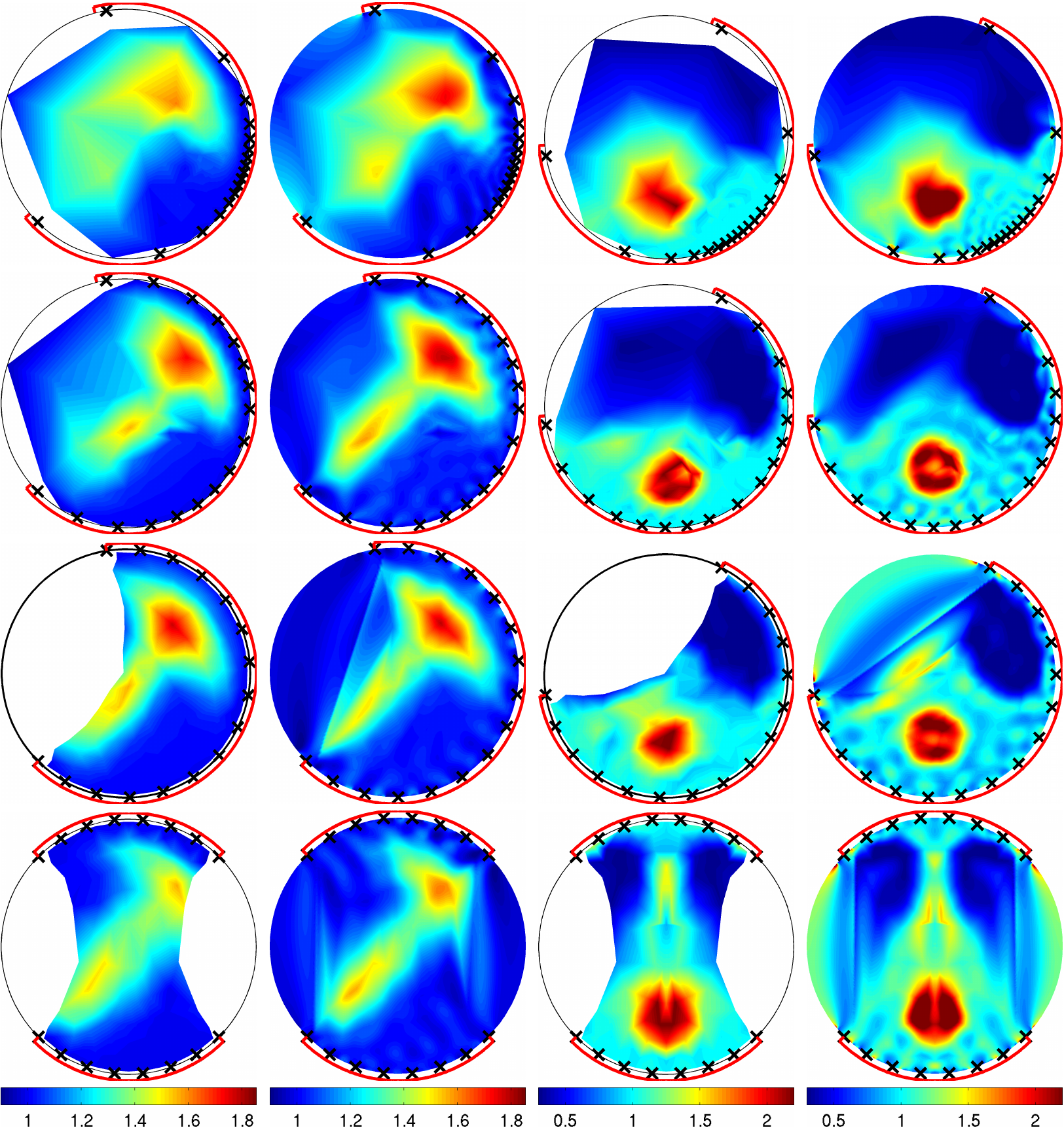}
 \end{center}
 \caption[Partial data reconstructions]{\label{fig:partialrec}
   \renewcommand{\baselinestretch}{1} \small\normalsize
   Reconstructions with partial data. Same conductivities are used as
   in figure \ref{fig:fbrec}.  Two leftmost columns: smooth
   conductivity. Two rightmost columns: piecewise constant chest
   phantom.  Columns 1 and 3: piecewise linear reconstructions.
   Columns 2 and 4: reconstructions after one step of Gauss-Newton
   iteration (\ref{eq:gniter}).  Rows from top to bottom: conformal
   mapping, quasiconformal mapping, pyramidal network, two-sided
   network.  Accessible boundary $\mathcal{B}_A$ is solid red. Centers
   of supports of measurement (electrode) functions $\chi_q$ are
   $\times$.  }
\end{figure}

We present in this section numerical reconstructions with partial
boundary measurements.  The reconstructions with the four methods from
sections \ref{sect:conformal}, \ref{sect:quasiconformal},
\ref{sect:piramidal} and \ref{sect:twosided} are compared row by row
in Figure \ref{fig:partialrec}. We use the same two test
conductivities as in Figure \ref{fig:fbrec}(a). Each row in Figure
\ref{fig:partialrec} corresponds to one method. For each test
conductivity, we show first the piecewise linear interpolation of the
entries returned by the reconstruction mapping $\Qc_n$, on the optimal
grids (first and third column in Figure \ref{fig:partialrec}). Since
these grids do not cover the entire $\Omega$, we display the results
only in the subset of $\Omega$ populated by the grid points.  We also
show the reconstructions after one-step of the Gauss-Newton iteration
(\ref{eq:gniter}) (second and fourth columns in Figure
\ref{fig:partialrec}).

As expected, the reconstructions with the conformal mapping grids are
the worst. The highly non-uniform conformal mapping grids cannot
capture the details of the conductivities away from the middle of the
accessible boundary. The reconstructions with quasiconformal grids
perform much better, capturing the details of the conductivities much
more uniformly throughout the domain. Although the piecewise linear
reconstructions $\Qc_n$ have slight distortions in the geometry, these
distortions are later removed by the first step of the Gauss-Newton
iteration. The piecewise linear reconstructions with pyramidal and
two-sided networks avoid the geometrical distortions of the
quasiconformal case, but they are also improved after one step of the
Gauss-Newton iteration.

Note that while the Gauss-Newton step improves the geometry of the
reconstructions, it also introduces some spurious oscillations.  This
is more pronounced for the piecewise constant conductivity phantom
(fourth column in Figure \ref{fig:partialrec}).  To overcome this
problem one may consider regularizing the Gauss-Newton iteration
(\ref{eq:gniter}) by adding a penalty term of some sort. For example,
for the piecewise constant phantom, we could penalize the total
variation of the reconstruction, as was done in \cite{BorDruGue}.

\section{Summary}
We presented a discrete approach to the numerical solution of the
inverse problem of electrical impedance tomography (EIT) in two
dimensions.  Due to the severe ill-posedness of the problem, it is
desirable to parametrize the unknown conductivity $\s(\bx)$ with as
few parameters as possible, while still capturing the best attainable
resolution of the reconstruction. To obtain such a parametrization, we
used a discrete, model reduction formulation of the problem.  The
discrete models are resistor networks with special graphs.  

We described in detail the solvability of the model reduction problem.
First, we showed that boundary measurements of the continuum Dirichlet
to Neumann (DtN) map $\Lambda_\s$ for the unknown $\s(\bx)$ define
matrices that belong to the set of discrete DtN maps for resistor
networks.  Second, we described the types of network graphs
appropriate for different measurement setups. By appropriate we mean
those graphs that ensure unique recoverability of the network from its
DtN map. Third, we showed how to determine the networks.

We established that the key ingredient in the connection between the
discrete model reduction problem (inverse problem for the network) and
the continuum EIT problem is the optimal grid. The name optimal refers
to the fact that finite volumes discretizations on these grids give
spectrally accurate approximations of the DtN map, the data in EIT. We
defined reconstructions of the conductivity using the optimal grids,
and studied them in detail in three cases: (1) The case of layered
media and full boundary measurements, where the problem can be reduced
to one dimension via Fourier transforms. (2) The case of two
dimensional media with measurement access to the entire boundary.  (3)
The case of two dimensional media with access to a subset of the
boundary.

We presented the available theory behind our inversion approach and
illustrated its performance with numerical simulations.

\section*{Acknowledgements}
The work of L. Borcea was partially supported by the National Science
Foundation grants DMS-0934594, DMS-0907746 and by the Office of Naval
Research grant N000140910290. The work of F. Guevara Vasquez was
partially supported by the National Science Foundation grant
DMS-0934664. The work of A.V. Mamonov was partially supported by the
National Science Foundation grants DMS-0914465 and DMS-0914840.  LB, FGV and
AVM were also partially supported by the National Science Foundation and the
National Security Agency, during the Fall 2010 special semester on Inverse
Problems at MSRI, Berkeley, CA.

\appendix
\section{The quadrature formulas}
\label{ap:Quadrature}
\setcounter{equation}{0} 

To understand definitions (\ref{eq:alphas}), recall Figure
\ref{fig:FinVol}. Take for example the dual edge
\[
\Sigma_{i-\frac{1}{2},j} =
(P_{i-\frac{1}{2},j-\frac{1}{2}},P_{i-\frac{1}{2},j-\frac{1}{2}}),\]
where $P_{i-\frac{1}{2},j-\frac{1}{2}} = \hr_i(\cos \hT_j,\sin
\hT_j)$. We have from (\ref{eq:RAlex}), and the change of variables to
$z(r)$ that
\begin{equation*}
\int_{\Sigma_{i - \frac{1}{2}, j}} \hspace{-0.2in}
\sigma(\bx)\bn(\bx) \cdot \nabla u(\bx) d s (\bx) =
\int_{\hT_j}^{\hT_{j+1}} \hspace{-0.05in} \hr_{i-1} \s (\hr_{i-1})
\frac{\partial u(\hr_{i-1},\T)}{\partial r} d \T \approx - h_\T
\frac{\partial u(\hr_{i-1},\T_j)}{\partial z} \approx  \frac{h_\T
\left(U_{i-1,j}-U_{i,j}\right)}{z(r_{i}) - z(r_{i-1})},
\end{equation*}
which gives the first equation in (\ref{eq:alphas}). Similarly, the
flux across \[\Sigma_{i, j+\frac{1}{2}} =
(P_{i-\frac{1}{2},j+\frac{1}{2}},P_{i+\frac{1}{2},j+\frac{1}{2}}),\]
is given by 
\begin{eqnarray*}
\int_{\Sigma_{i,j+ \frac{1}{2}}} \hspace{-0.2in}
\sigma(\bx)\bn(\bx) \cdot \nabla u(\bx) d s (\bx) &=&
\int_{\hr_{i}}^{\hr_{i-1}} \hspace{-0.05in} \frac{\s (r)}{r}
\frac{\partial u(r,\hT_{j+1})}{\partial \T} dr \approx
\frac{\partial
u(r_i,\hT_{j+1})}{\partial \T} \int_{\hr_{i}}^{\hr_{i-1}} \hspace{-0.05in} \frac{\s (r)}{r} dr 
\\ &\approx&
\left(\hz(\hr_{i})-\hz(\hr_{i-1}) \right) \frac{U_{i,j+1}-U(i,j)}{h_\T},
\end{eqnarray*}
which gives the second equation in (\ref{eq:alphas}). 

\section{Continued fraction representation}
\label{ap:CF}
\setcounter{equation}{0} 
Let us begin with the system of equations satisfied by the potential $V_j$, which we 
rewrite as 
\begin{eqnarray}
b_j &=& b_{j+1} + \hal_{j+1} \la V_{j+1}, \qquad j = 0, 1, \ldots
\ell, \nonumber \\ b_0 &=& \Phi_{_\B}, \label{eq:B1} \\ V_{\ell+1} &=&
0, \nonumber
\end{eqnarray}
where we let 
\begin{equation}
V_j = V_{j+1} + \al_j b_j.
\label{eq:B2}
\end{equation}
Combining the first equation in (\ref{eq:B2}) with (\ref{eq:B2}), we
obtain the recursive relation
\begin{equation}
\frac{b_j}{V_j} = \cfrac{1}{\al_j + \cfrac{1}{\hal_{j+1} \la +
\cfrac{b_{j+1}}{V_{j+1}}}}, \qquad j = 1, 2, \ldots, \ell, \label{eq:B3}
\end{equation}
which we iterate for $j$ decreasing from $j = \ell-1$ to $1$, and starting 
with 
\begin{equation}
\frac{b_\ell}{V_\ell} = \frac{1}{\al_\ell}.
\end{equation}
The latter follows from the first equation in (\ref{eq:B1}) evaluated
at $j = \ell$, and boundary condition $V_{\ell+1} = 0$.  We obtain
that
\begin{equation}
\Fd(\la) = V_1/\Phi_{_\B} = \frac{V_1}{b_0} = \cfrac{V_1}{b_1 + \hal_1 \la V_1} = 
\cfrac{1}{\hal_1 \la + \cfrac{b_1}{V_1}}
\label{eq:B4}
\end{equation}
has the continued fraction representation (\ref{eq:CFr}).

\section{Derivation of results (\ref{eq:RR}-\ref{eq:RR1})}
\label{ap:DRR}
To derive equation (\ref{eq:RR}) let us begin with the Fourier series of the 
electrode functions
\begin{equation}
\chi_q(\T) = \sum_{k \in \mathbb{Z}} C_q(k) e^{i k \T} = \sum_{k \in
  \mathbb{Z}} \overline{C_q(k)} e^{-i k \T},
\label{eq:C1}
\end{equation}
where the bar denotes complex conjugate and the coefficients are 
\begin{equation}
C_q(\T) = \frac{1}{2\pi} \int_0^{2\pi} \chi_q(\T) e^{i k \T} d \T =
\frac{e^{i k \T_q}}{2 \pi} \mbox{sinc} \left( \frac{k h_\T}{2}
\right).
\label{eq:C3}
\end{equation}
Then, we have 
\begin{eqnarray}
\left(\LG\right)_{p,q} &=& \int_0^{2 \pi} \chi_p(\T) \Lambda_\s
\chi_q(\T) d \T = \sum_{k,k' \in \mathbb{Z}} C_p(k) \overline{C_q(k')}
\int_0^{2 \pi} e^{i k \T} \Lambda_\s e^{-i k' \T} d \T \nonumber
\\ &=& \frac{1}{2 \pi} \sum_{k \in \mathbb{Z}} e^{i k (\T_p-\T_q)}
f(k^2) \left[\mbox{sinc}\left( \frac{k h_\T}{2} \right)\right]^2, \qquad p \ne q.
\label{eq:C4}
\end{eqnarray}
The diagonal entries are 
\begin{equation}
\left(\LG\right)_{p,p} = - \sum_{q \ne p} \left(\LG\right)_{p,q} = -
\frac{1}{2 \pi} \sum_{k \in \mathbb{Z}} e^{i k \T_p} f(k^2) \left[
\mbox{sinc} \left( \frac{k h_\T}{2} \right)\right]^2 \sum_{q \ne p}
e^{- i k \T_q}.
\label{eq:C5}
\end{equation}
But 
\begin{equation}
\sum_{q \ne p}
e^{- i k \T_q} = \sum_{q = 1}^n e^{-i \frac{2 \pi k}{n} (q-1)} - e^{-i k \T_p} = 
e^{i \pi k (1-1/n)} \frac{\sin(\pi k)}{\sin(\pi k/n)} - e^{-i k \T_p} = 
n \delta_{k,0} - e^{i k \T_p}.
\label{eq:C6}
\end{equation}
Since $f(0) = 0$, we obtain from (\ref{eq:C5}-\ref{eq:C6}) that
(\ref{eq:C4}) holds for $p = q$, as well. This is the result
(\ref{eq:RR}).  Moreover, (\ref{eq:RR1}) follows from
\begin{eqnarray} 
\left( \LG \left[ e^{i k \T} \right] \right)_p = \sum_{q=1}^n
\left(\LG\right)_{p,q} e^{i k \T_q} = \frac{1}{2 \pi} \sum_{k_1 \in
  \mathbb{Z}} e^{i k_1 \T_p} f(k_1^2) \left[ \mbox{sinc} \left(
\frac{k_1 h_\T}{2} \right)\right]^2 \sum_{q=1}^n e^{i (k-k_1) \T_q},
\end{eqnarray}
and the identity
\begin{equation}
\sum_{q = 1}^n e^{i (k-k_1) \T_q} = \sum_{q = 1}^n e^{i \frac{2 \pi
    (k-k_1)}{n} (q-1)}= n \delta_{k,k_1}.
\label{eq:C6}
\end{equation}

\section{Rational interpolation and Euclidean division}
\label{ap:Euclid}
Consider the case $\mhf = 1$, where $F(\la) = 1/\Fd(\la)$ follows from
(\ref{eq:CFr}). We rename the coefficients as
\begin{equation}
\kappa_{2j-1} = \hal_j, \quad \kappa_{2j} = \al_j, \quad j = 1, \ldots \ell,
\label{eq:D.1}
\end{equation}
and let $\la = x^2$ to obtain 
\begin{equation}
\frac{F(x^2)}{x} = \kappa_1 x + \cfrac{1}{\kappa_2 x+\dots
 \cfrac{1}{\kappa_{2\ell-1}x + \cfrac{1}{\kappa_{2\ell}x}}}.
\label{eq:D.2}
\end{equation}
To determine $\kappa_j$, for $j = 1, \ldots, 2 \ell$, we write first
(\ref{eq:D.2}) as the ratio of two polynomials of $x$, $P_{2 \ell}(x)$
and $Q_{2 \ell-1}(x)$ of degrees $2\ell$ and $2\ell-1$ respectively,
and seek their coefficients $c_j$,
\begin{equation}
\frac{F(x^2)}{x} = \frac{P_{2 \ell}(x)}{Q_{2 \ell-1}(x)} = \frac{c_{2
    \ell} x^{2 \ell} + c_{2(\ell-1)} x^{2(\ell-1)} + \ldots + c_2 x^2
  + c_0}{ c_{2 \ell -1}x^{2 \ell -1} + c_{2 \ell -3} x^{2 \ell -3} +
  \ldots + c_1 x}.
\label{eq:D.3}
\end{equation}
We normalize the ratio by setting $c_0 = -1$. 

Now suppose that we have measurements of $F$ at $\la_k = x_k^2$, for 
$k = 1, \ldots, 2 \ell$, and introduce the notation
\begin{equation}
\frac{F(x_k^2)}{x_k} = D_k.
\label{eq:D.4}
\end{equation}
We obtain from (\ref{eq:D.3}) the following linear system of equations
for the coefficients 
\begin{equation}
P_{2 \ell}(x_k) - D_k Q_{2 \ell-1}(x_k) = 0, \qquad k = 1, \ldots, 2 \ell,
\label{eq:D.5}
\end{equation}
or in matrix form
\begin{equation} 
\left( \begin{array}{cccccc} -D_1 x_1 & x_1^2 & - D_1 x_1^3 & \ldots &
  - D_1 x_1^{2 \ell-1} & x_1^{2 \ell} \\ -D_2 x_2 & x_2^2 & - D_2
  x_2^3 & \ldots & - D_2 x_2^{2 \ell-1} & x_2^{2 \ell} \\ &&& \vdots
  \\ -D_{2 \ell} x_{2 \ell} & x_{2 \ell}^2 & - D_{2 \ell} x_{2 \ell}^3
  & \ldots & - D_{2 \ell} x_{2 \ell}^{2 \ell-1} & x_{2 \ell}^{2 \ell}
\end{array} \right) 
\left( \begin{array}{c} c_1 \\ c_2 \\ \vdots \\ c_{2
    \ell} \end{array}\right) = {\bf 1},
\label{eq:D.6}
\end{equation}
with right hand side a vector of all ones. The coefficients are
obtained by inverting the Vandermonde-like matrix in
(\ref{eq:D.6}). In the special case of the rational interpolation
(\ref{eq:Rat.5}), it is precisely a Vandermonde matrix.  Since the
condition number of such matrices grows exponentially with their size
\cite{gautschi1987lower}, the determination of $\{c_{j}\}_{j = 1,
  \ldots, 2 \ell}$ is an ill-posed problem, as stated in Remark
\ref{rem.1}.

Once we have determined the polynomials $P_{2 \ell}(x)$ and $Q_{2 \ell
  -1}(x)$, we can obtain $\{\kappa_j\}_{j=1, \ldots 2 \ell}$ by
Euclidean polynomial division. Explicitly, let us introduce a new
polynomial $ {P}_{2 \ell-2}(x) = \widetilde{c}_{2 \ell -2} x^{2 \ell
  -2} + \ldots \widetilde{c}_0, $ so that
\begin{equation}
\kappa_2 x+\dots \cfrac{1}{\kappa_3 x + \ldots
  \cfrac{1}{\kappa_{2\ell-1}x + \cfrac{1}{\kappa_{2\ell}x}}} =
\frac{{Q}_{2 \ell -1}(x)}{{P}_{2 \ell-2}(x)},
\qquad \kappa_1 x + \frac{{P}_{2
    \ell-2}(x)}{{Q}_{2 \ell -1}(x)} = \frac{P_{2
    \ell}(x)}{Q_{2 \ell -1}(x)}.
\label{eq:D8}
\end{equation}
Equating powers of $x$ we get 
\begin{equation}
\kappa_1 = \frac{c_{2 \ell}}{c_{2 \ell-1}}, 
\label{eq:D7}
\end{equation}
and the coefficients of the polynomial ${P}_{2 \ell-2}(x)$ are
determined by
\begin{eqnarray}
\widetilde{c}_{2 j} &=& c_{2 j} - \kappa_1 c_{2 j - 1}, \qquad j = 1,
\ldots, \ell -1, \\\widetilde{c}_0 &=& c_0.
\label{eq:D9}
\end{eqnarray}
Then, we proceed similarly to get $\kappa_2$, and introduce a new
polynomial $Q_{2 \ell -3}(x)$ so that
\begin{equation}
\kappa_3 x+\dots \cfrac{1}{\kappa_{4}x + \ldots \cfrac{1}{\kappa_{2
      \ell -1} x + \cfrac{1}{\kappa_{2\ell}x}}} = \frac{P_{2 \ell
    -2}(x)}{Q_{2 \ell-3}(x)}, \qquad \kappa_2 x + \frac{Q_{2
    \ell-3}(x)}{P_{2 \ell -2}(x)} = \frac{Q_{2 \ell-1}(x)}{P_{2 \ell
    -2}(x)}.
\label{eq:D10}
\end{equation}
Equating powers of $x$ we get $\kappa_2 = {c_{2 \ell
    -1}}/{\widetilde{c}_{2 \ell -2}}$ and the polynomial $Q_{2\ell
  -3}(x)$ and so on.

\section{The Lanczos iteration}
\label{ap:Lanczos}
Let us write the Jacobi matrix (\ref{eq:tA}) as
\begin{equation}
\tA = \left(
\begin{array}{ccccccc}
-a_1 & b_1 & 0 & \ldots & \ldots & 0 & 0 \\ b_1 & -a_2 & b_2 & 0 &
\ldots & 0 & 0 \\ \ddots & \ddots & \ddots & \ddots & \ddots & \ddots
& \ddots \\ 0 & 0 & \ldots & \ldots & 0 & b_{\ell-1} & -a_\ell
\end{array} \right),
\label{eq:La1}
\end{equation}
where $-a_j$ are the negative diagonal entries and $b_j$ the positive
off-diagonal ones. Let also 
\begin{equation}
-\Delta = \mbox{diag}(-\delta_1^2,\ldots, -\delta_\ell^2)
\label{eq:La2} 
\end{equation}
be the diagonal matrix of the eigenvalues and 
\begin{equation}
\tY_j = \mbox{diag}(\sqrt{\hal_1}, \ldots , \sqrt{\hal_\ell}) {\bf Y}_j
\label{eq:La3}
\end{equation}
the eigenvectors. They are orthonormal and the matrix $\tY = (\tY_1,
\ldots, \tY_\ell)$ is orthogonal
\begin{equation}
\tY \tY^T = \tY^T \tY = {\bf I}.  
\label{eq:La4}
\end{equation}
The spectral theorem gives that $\tA = - \tY \Delta \tY^T$ or,
equivalently,
\begin{equation}
\tA \tY = - \tY \Delta.  
\label{eq:La5} 
\end{equation}
The Lanczos iteration \cite{trefethen1997numerical,chu2002structured}
determines the entries $a_j$ and $b_j$ in $\tA$ by taking equations
(\ref{eq:La5}) row by row.

Let us denote the rows of $\tY$ by
\begin{equation}
\bW_j = {\bf e}_j^T \tY, \quad j = 1, \ldots, \ell,
\label{eq:La6}
\end{equation}
and observe from (\ref{eq:La5}) that they are orthonormal 
\begin{equation}
\bW_j \bW_q =
\delta_{j,q} 
\label{eq:La7}
\end{equation}
We get for $j = 1$ that
\begin{equation}
\|\bW_1\|^2 = \sum_{j=1}^{\ell} \hal_1 Y^2_{1,j} = \hal_1
\sum_{j=1}^\ell \xi_j = 1,  
\label{eq:La8} 
\end{equation}
which determines $\hal_1$, and  we can set 
\begin{equation}
\bW_1 = \sqrt{\hal_1} \left( \sqrt{\xi_1}, \ldots, \sqrt{\xi_\ell}
\right).
\label{eq:La8p}
\end{equation}

The first row in equation (\ref{eq:La5}) gives
\begin{equation}
-a_1 \bW_1 + b_1 \bW_2 = - \bW_1 \Delta, 
\label{eq:La9} 
\end{equation}
and using the orthogonality in (\ref{eq:La7}), we obtain 
\begin{eqnarray}
a_1 = \bW_1 \Delta \bW_1^T = \sum_{j=1}^\ell \delta_j^2 \xi_j,
\label{eq:La10} \qquad  b_1 = \| a_1 \bW_1 - \bW_1 \Delta \|,
\end{eqnarray}
and 
\begin{eqnarray} \label{eq:La11} 
W_2 = b_1^{-1} \left( a_1 \bW_1 - \bW_1 \Delta\right).
\end{eqnarray}

The second row in equation (\ref{eq:La5}) gives
\begin{equation}
b_1 \bW_1 -a_2 \bW_2 +  b_2 \bW_3 = - \bW_2 \Delta, 
\label{eq:La12} 
\end{equation}
and we can compute $a_2$ and $b_2$ as follows,
\begin{eqnarray}
\label{eq:La13}
a_2 = \bW_2 \Delta \bW_2^T, 
\qquad  b_2 = \| a_2 \bW_2 - \bW_2 \Delta - b_1 \bW_1\|.
\end{eqnarray}
Moreover, 
\begin{equation}
 \bW_3 = b_2^{-1} \left(a_2 \bW_2 - \bW_2 \Delta - b_1 \bW_1\right),
\end{equation}
and the equation continues to the next row.

Once we have determined $\{a_j\}_{j=1, \ldots, \ell}$ and
$\{b_j\}_{1,\ldots, \ell-1}$ with the Lanczos iteration described
above, we can compute $\{\al_j,\hal_j\}_{j=1,\ldots, \ell}$. We
already have from (\ref{eq:La8}) that
\begin{equation}
\hal_1 = 1/\sum_{j=1}^\ell \xi_j.  
\label{eq:La14} 
\end{equation}
The remaining parameters are determined from the identities 
\begin{equation}
a_j = \frac{1}{\hal_1 \al_1} \delta_{j,1} + (1-\delta_{j,1})
\frac{1}{\hal_j} \left( \frac{1}{\al_j} + \frac{1}{\al_{j-1}} \right),
\qquad b_j = \frac{1}{\al_j \sqrt{\hal_j \hal_{j+1}}}.
\end{equation}

\section{Proofs of Lemma \ref{lem.3} and Corollary \ref{cor.1}}
\label{ap:ProofLem.3}
To prove Lemma \ref{lem.3}, let $\bAq$ be the tridiagonal matrix with
entries defined by $\{\aq_j,\haq_j\}_{j=1, \ldots, \ell}$, like in
(\ref{eq:JacA}).  It is the discretization of the operator in
(\ref{eq:TT}) with $\s \leadsto \sq$.  Similarly, let $\Ar$ be the
matrix defined by $\{\ar_j,\har_j\}_{j=1, \ldots, \ell}$, the
discretization of the second derivative operator for conductivity
$\sr$.  By the uniqueness of solution of the inverse spectral problem
and (\ref{eq:CL12}-\ref{eq:CL14}), the matrices $\bAq$ and $\Ar$ are
related by
\begin{equation}
\mbox{diag} \left(\sqrt{\frac{\haq_1}{\har_1}}, \ldots,
\sqrt{\frac{\haq_\ell}{\har_\ell}}\right) \bAq \mbox{diag}
\left(\sqrt{\frac{\har_1}{\haq_1}}, \ldots,
\sqrt{\frac{\har_\ell}{\haq_\ell}}\right) = \Ar - \bq \, {\bf I}.
\label{eq:CL15}
\end{equation}
They have eigenvectors $\bYq_j$ and $\Yr_j$ respectively, related by 
\begin{equation}
\mbox{diag}
\left(\sqrt{\haq_1}, \ldots,
\sqrt{\haq_\ell}\right) \bYq_j = \mbox{diag}
\left(\sqrt{\har_1}, \ldots,
\sqrt{\har_\ell}\right) \Yr_j, \qquad j = 1, \ldots, \ell, 
\label{eq:CL15p}
\end{equation}
and the matrix $\tY$ with columns (\ref{eq:CL15p}) is orthogonal.
Thus, we have the identity
\begin{equation}
\left( \tY \tY^T \right)_{11}= \haq_1 \sum_{j=1}^{\ell} \xiq = 
\har_1 \sum_{j=1}^{\ell} \xir = 1,
\label{eq:CL17}
\end{equation}
which gives $\haq_1 = \har_1$ by (\ref{eq:CL12}) or, equivalently
\begin{equation}
\sq_1 = \frac{\haq_1}{\har_1} = 1 = \sq(0).
\label{eq:CL18}
\end{equation}
Moreover, straightforward algebraic manipulations of the equations in
(\ref{eq:CL15}) and definitions (\ref{eq:Q0}) give the finite
difference equations (\ref{eq:Q1}).  $\Box$.

To prove Corollary \ref{cor.1}, recall the definitions (\ref{eq:Q0})
and (\ref{eq:GQ}) to write
\begin{equation}
\sum_{p=1}^j \haq_j = \int_0^{\hzeq_{j+1}} \sq(\zeta) d \zeta =
\sum_{p=1}^j \har_p \sq_{p} = \sum_{p=1}^j \har_p \sq(\zer_{p}) + o(1).
\label{eq:F1}
\end{equation}
Here we used the convergence result in Theorem \ref{thm.2} and denote
by $o(1)$ a negligible residual in the limit $\ell \to \infty$.
We have 
\begin{equation}
\int_{\hzer_{j+1}}^{\hzeq_{j+1}} \sq(\zeta) d \zeta = \sum_{p=1}^j
\har_p \sq(\zer_{p}) - \int_0^{\hzer_{j+1}} \sq(\zeta) d \zeta +
o(1),
\end{equation}
and therefore
\begin{equation}
\left| \hzeq_{j+1}-\hzer_{j+1} \right| \le C \left|
\int_0^{\hzer_{j+1}} \sq(\zeta) d \zeta - \sum_{p=1}^j \har_p
\sq(\zer_{p}) \right| + o(1), \qquad C = 1/\min_\zeta \sq(\zeta).
\end{equation}
But the first term in the bound is just the error of the quadrature on
the optimal grid, with nodes at $\zer_j$ and weights $\har_j =
\hzer_{{j+1}}-\hzer_{j}$, and it converges to zero by the properties
of the optimal grid stated in Lemma \ref{lem.2} and the smoothness of
$\sq(\zeta)$. Thus, we have shown that
\begin{equation}
  \left| \hzeq_{j+1}-\hzer_{j+1} \right| \to 0, \quad \mbox{as} ~ 
\ell \to \infty, 
\end{equation}
uniformly in $j$. The proof for the primary nodes $\zeq_j$ is
similar. $\Box$.

\section{Perturbation analysis}
\label{ap:perturbation}
It is shown in \cite[Appendix B]{BorDruKni} that the skew-symmetric
matrix $\bB$ given in (\ref{eq:PC2}) has eigenvalues $\pm i \de_j$ and
eigenvectors
\begin{equation} {\mathcal Y}(\pm \de_j) = \frac{1}{\sqrt{2}} \left(
    {\mathcal Y}_1(\de_j),\pm i \widehat{{\mathcal Y}}_1(\de_j),
    \ldots, {\mathcal Y}_\ell(\de_j),\pm i \widehat{{\mathcal
        Y}}_\ell(\de_j) \right)^T,
\label{eq:PA1}
\end{equation}
where 
\begin{equation}
\left({\mathcal Y}_1(\de_j), \ldots, {\mathcal Y}_\ell(\de_j)\right)^T =
\mbox{diag}\left(\hal_1^{\frac{1}{2}}, \ldots,
\hal_\ell^{\frac{1}{2}}\right) \bY_j, \qquad
\left(\widehat{{\mathcal Y}}_1(\de_j), \ldots,
\widehat{{\mathcal Y}}_\ell(\de_j)\right)^T =
\mbox{diag}\left(\al_1^{\frac{1}{2}}, \ldots,
\al_\ell^{\frac{1}{2}}\right) \widehat{\bY}_j,
\label{eq:PA2}
\end{equation}
$\bY_j = \left(Y_{1,j},\ldots, Y_{\ell,j}\right)^T$ are the
eigenvectors of matrix $\bA$ for eigenvalues $-\de_j^2$ and $
\widehat{\bY}_j = \left(\widehat{Y}_{1,j},\ldots,
\widehat{Y}_{\ell,j}\right)^T$ is the vector with entries
\begin{equation}
\widehat{Y}_{p,j} = \frac{Y_{p+1,j}-Y_{p,j}}{\de_j \al_j}.
\label{eq:PA3}
\end{equation}

\subsection{Discrete Gel'fand--Levitan formulation}
It is difficult to carry a precise perturbation analysis of the
recursive Lanczos iteration that gives $\bB$ from the spectral data.
We use instead the following discrete Gel'fand--Levitan formulation
due to Natterer \cite{natterer1994dgl}. 

Consider the ``reference'' matrix $\bB^r$,
for an arbitrary, but fixed $r \in [0,1]$, and define the lower
triangular, transmutation matrix $\bGG$, satisfying
\begin{equation}
\bE \bGG \bB = \bE \bB^r \bGG, \quad {\bf e}_1^T \bGG = {\bf e}_1^T, 
\label{res.4}
\end{equation}
where $\bE = {\bf I} - {\bf e}_{2\ell} {\bf e}_{2\ell}^T$. 
Clearly, if $\bB = \bB^r$, then $\bGG = \bGG^r = \mbox{I}$, the
identity.  In general $\bGG$ is lower triangular and it is uniquely
defined as shown with a Lanczos iteration argument in \cite[Section
6.2]{BorDruKni}.

Next, consider the initial value problem
\begin{equation}
  \bE \bB \bPhi(\la) = i \la \bE \bPhi(\la), \quad {\bf e}_1^T \bPhi(\la) = 1,
\label{res.5}
\end{equation}
which has a unique solution $\bPhi(\la) \in \mathbb{C}^{2\ell}$, as
shown in \cite[Section 6.2]{BorDruKni}.  When $\la = \pm \de_j$, one
of the eigenvalues of $\bB$, we have
\begin{equation}
  \bPhi(\pm \de_j )= \frac{\sqrt{2}}{{\mathcal Y}_{1}(\de_{j})} 
{\mathcal Y}(\pm \de_j) =
  \sqrt{\frac{2}{\hal_1 \xi_j}}\, {\mathcal Y}(\pm \de_j), \label{res.8}
\end{equation}
and (\ref{res.5}) holds even for $\bE$ replaced by the identity
matrix. The analogue of (\ref{res.5}) for $\bB^r$ is
\begin{equation}
\bE \bB^r \bPhi^r(\la) = i \la \bE \bPhi^r(\la), \quad 
{\bf e}_1^T \bPhi^r(\la) = 1,
\label{res.6}
\end{equation}
and, using (\ref{res.4}) and the lower triangular structure of $\bGG$,
we obtain
\begin{equation}
\bPhi^r(\pm \de_j) = \bGG \bPhi(\pm \de_j), \quad 1 \le j \le \ell.
\label{res.7}
\end{equation}
Equivalently, in matrix form (\ref{res.7}) and (\ref{res.8}) give 
\begin{equation}
\bmP^r = \bGG \bmP = \bGG \bTTe \bS,
\end{equation}
where $\bmP$ is the matrix with columns (\ref{res.8}), $\bTTe$ is the
orthogonal matrix of eigenvectors of $\bB$ with columns
(\ref{eq:PA1}), and $\bS$ is the diagonal scaling matrix
\begin{equation}
  \bS = \sqrt{\frac{2}{\hal_1}} \mbox{diag}\left(\xi_1^{-1/2}, 
    \xi_1^{-1/2},\ldots, 
    \xi_\ell^{-1/2},  \xi_\ell^{-1/2}\right).
\label{eq:ScalingM}
\end{equation}
Then, letting 
\begin{equation}
\bFF = \bmP^r \bS^{-1} = \bGG \bTTe
\end{equation}
and using the orthogonality of $\bTTe$ we get 
\begin{equation}
\bFF \overline{\bFF}^T = \bGG \bGG^T,
\label{res.12}
\end{equation}
where the bar denotes complex conjugate.  Moreover, equation
(\ref{res.4}) gives
\begin{equation}
\bE \bB^r \bFF = \bE \bB^r \bGG \bTTe = \bE \bGG \bB \bTTe = i \bE
\bGG \bTTe \bDe = i \bE \bFF \bDe,
\label{eq:FinMM}
\end{equation}
where $i \bDe = i \mbox{diag}\left(\de_1,-\de_1,\ldots,
\de_\ell,-\de_\ell\right)$ is the matrix of the eigenvalues of $\bB$.

The discrete Gel'fand-Levitan's inversion method proceeds as follows:
Start with a known reference matrix $\bB^r$, for some $r\in[0,1]$. The
usual choice is $\bB^0 = \bB^{(o)}$, the matrix corresponding to the
constant coefficient $\sr \equiv 1$.  Determine $\bmP^r$ from
(\ref{res.6}), with a Lanczos iteration as explained in \cite[Section
  6.2]{BorDruKni}. Then, $\bFF = \bmP^r \bS^{-1}$ is determined by the
spectral data $\de_j^r$ and $\xi_j^r$, for $1 \le j \le \ell$. The
matrix $\bGG$ is obtained from (\ref{res.12}) by a Cholesky
factorization, and $\bB$ follows by solving (\ref{res.4}), using a
Lanczos iteration.

\subsection{Perturbation estimate}
Consider the perturbations $d \de_j = \Delta \de_j dr$ and $d \xi_j =
\Delta \xi_j dr$ of the spectral data of reference matrix $\bB^r$. 
We denote the perturbed quantities with a tilde  as in 
\begin{equation}
\widetilde{\bDe} = \bDe^r + d \bDe, \quad 
\widetilde{\bS} = \bS^r + d \bS, \quad 
\widetilde{\bTTe} = \bTTe^r + d \bTTe, \quad 
\widetilde{\bFF} = \bTTe^r + d \bFF,
\label{eq:PERT1}
\end{equation}
with $\bDe$, $\bS$, $\bTTe$ and $\bFF$ defined above. Note that
$\bFF^r = \bTTe^r$, because $\bGG^r = {\bf I}$.  Substituting
(\ref{eq:PERT1}) in (\ref{eq:FinMM}) and using (\ref{res.6}), we get 
\begin{equation}
\bE \bB^r d \bFF = i \bE \,\bTTe^r d \bDe + i \bE\, d \bFF \,\bDe^r.
\label{eq:PERT2}
\end{equation}
Now multiply by $\overline{\bTTe^r}^T$ on the right and use that
$\bDe^r \overline{\bTTe^r}^T = - i\overline{\bTTe^r}^T \bB^r$ to
obtain that $d \bW = d \bFF \overline{\bTTe^r}^T $ satisfies
\begin{equation}
\bE \bB^r d \bW - \bE \, d \bW \, \bB^r = i \bE \,\bTTe^r d \bDe \,
\overline{\bTTe^r}^T,
\label{eq:PERT4}
\end{equation}
with initial condition
\begin{equation}
{\bf e}_1^T d \bW = {\bf e}_1^T d \bFF \, \overline{\bTTe^r}^T =
\left( d \sqrt{\frac{\hal_1 \xi_1}{2}}, d \sqrt{\frac{\hal_1
    \xi_1}{2}}, \ldots, d \sqrt{\frac{\hal_\ell \xi_\ell}{2}}, d
\sqrt{\frac{\hal_\ell \xi_\ell}{2}}\right) \overline{\bTTe^r}^T.
\label{eq:PERT5}
\end{equation}
Similarly, we get from (\ref{res.4}) and $\bGG^r = {\bf I}$ that
\begin{equation}
E\, d \bB + \bE\, d \bGG \, \bB^r = \bE \bB^r d \bGG, \quad 
{\bf e}_1^T d \bGG = {\bf 0}.
\label{eq:PERT6}
\end{equation}
Furthermore, equation (\ref{res.12}) and $\bFF^r = \bTTe^r$ give 
\begin{equation}
d \bFF \, \overline{\bTTe^r}^T + \bTTe^r \, d \bFF = d \bW +
\overline{d \bW}^T = d \bGG + d \bGG^T.
\label{eq:PERT7}
\end{equation}

Equations (\ref{eq:PERT4}-(\ref{eq:PERT7}) allow us to estimate
$d\beta_j/\beta_j^r$. Indeed, consider the $j,j+1$ component in
(\ref{eq:PERT6}) and use (\ref{eq:PERT7}) and the structure of $\bGG$,
$d \bGG$ and $\bB^r$ to get
\begin{equation}
\frac{d \beta_j}{\beta_j^r} = d G_{j+1,j+1}-d G_{j,j} = d W_{j+1,j+1}-
d W_{j,j} , \quad j = 1, \ldots, 2 \ell -1.
\label{eq:PERT8}
\end{equation}
The right hand side is given by the components of $d \bW$ satisfying
(\ref{eq:PERT4}-\ref{eq:PERT5}) and calculated explicitly in
\cite[Appendix C]{BorDruKni} in terms of the eigenvalues and
eigenvectors of $\bB^r$. Then, the estimate 
\begin{equation}
\sum_{j=1}^{2 \ell -1} \left| \frac{d \beta_j}{\beta_j^r} \right| 
\le C_1 dr 
\end{equation}
which is equivalent to (\ref{eq:PC7}) follows after some calculation
given in \cite[Section 6.3]{BorDruKni}, using the assumptions
(\ref{eq:CL6}) on the asymptotic behavior of $\Delta \de_j$ and
$\Delta \xi_j$, i.e., of $\de_j^r - \der_j = r \Delta \de_j$ and
$\xi_j^r - \xir_j = r \Delta \xi_j$.

\bibliography{BIBLIO} \bibliographystyle{plain}

\begin{thebibliography}{10}

\bibitem{alessandrini1988sdc}
G.~Alessandrini.
\newblock {Stable determination of conductivity by boundary measurements}.
\newblock {\em Applicable Analysis}, 27(1):153--172, 1988.

\bibitem{alessandrini2005lipschitz}
G.~Alessandrini and S.~Vessella.
\newblock {Lipschitz stability for the inverse conductivity problem}.
\newblock {\em Advances in Applied Mathematics}, 35(2):207--241, 2005.

\bibitem{ameur2002refinement}
H.B. Ameur, G.~Chavent, and J.~Jaffr{\'e}.
\newblock {Refinement and coarsening indicators for adaptive parametrization:
  application to the estimation of hydraulic transmissivities}.
\newblock {\em Inverse Problems}, 18:775, 2002.

\bibitem{ameur2002regularization}
H.B. Ameur and B.~Kaltenbacher.
\newblock {Regularization of parameter estimation by adaptive discretization
  using refinement and coarsening indicators}.
\newblock {\em JOURNAL OF INVERSE AND ILL POSED PROBLEMS}, 10(6):561--584,
  2002.

\bibitem{astala2005cip}
K.~Astala, L.~P{\"a}iv{\"a}rinta, and M.~Lassas.
\newblock {Calder{\'o}n's Inverse Problem for Anisotropic Conductivity in the
  Plane}.
\newblock {\em Communications in Partial Differential Equations},
  30(1):207--224, 2005.

\bibitem{asvadurov2004ofd}
S.~Asvadurov, V.~Druskin, M.N. Guddati, and L.~Knizhnerman.
\newblock {On optimal finite-difference approximation of PML}.
\newblock {\em SIAM Journal on Numerical Analysis}, 41(1):287--305, 2004.

\bibitem{asvadurov2000adg}
S.~Asvadurov, V.~Druskin, and L.~Knizhnerman.
\newblock {Application of the difference Gaussian rules to solution of
  hyperbolic problems}.
\newblock {\em Journal of Computational Physics}, 158(1):116--135, 2000.

\bibitem{asvadurov2007optimal}
S.~Asvadurov, V.~Druskin, and S.~Moskow.
\newblock Optimal grids for anisotropic problems.
\newblock {\em Electronic Transactions on Numerical Analysis}, 26:55--81, 2007.

\bibitem{barcelo2001sic}
J.A. Barcelo, T.~Barcelo, and A.~Ruiz.
\newblock {Stability of the inverse conductivity problem in the plane for less
  regular conductivities}.
\newblock {\em Journal of Differential Equations}, 173(2):231--270, 2001.

\bibitem{biesel}
O.D. Biesel, D.V. Ingerman, J.A. Morrow, and W.T. Shore.
\newblock {Layered Networks, the Discrete Laplacian, and a Continued Fraction
  Identity}.
\newblock http://www.math.washington.edu/\textasciitilde
  reu/papers/current/william/layered.pdf.

\bibitem{borcea2002electrical}
L.~Borcea.
\newblock {Electrical impedance tomography. Topical review.}
\newblock {\em Inverse Problems}, 18(6):99--136, 2002.

\bibitem{BorDru}
L.~Borcea and V.~Druskin.
\newblock {Optimal finite difference grids for direct and inverse
  Sturm-Liouville problems}.
\newblock {\em Inverse Problems}, 18(4):979--1002, 2002.

\bibitem{BorDruGue}
L.~Borcea, V.~Druskin, and F.~Guevara~Vasquez.
\newblock {Electrical impedance tomography with resistor networks}.
\newblock {\em Inverse Problems}, 24(3):035013 (31pp), 2008.

\bibitem{BorDruKni}
L.~Borcea, V.~Druskin, and L.~Knizhnerman.
\newblock {On the Continuum Limit of a Discrete Inverse Spectral Problem on
  Optimal Finite Difference Grids}.
\newblock {\em Communications on Pure and Applied Mathematics}, 58(9):1231,
  2005.

\bibitem{BDM-10}
L.~Borcea, V.~Druskin, and A.V. Mamonov.
\newblock Circular resistor networks for electrical impedance tomography with
  partial boundary measurements.
\newblock {\em Inverse Problems}, 26(4):045010, 2010.

\bibitem{BDMG-10}
L.~Borcea, V.~Druskin, A.V. Mamonov, and F.~Guevara~Vasquez.
\newblock Pyramidal resistor networks for electrical impedance tomography with
  partial boundary measurements.
\newblock {\em Inverse Problems}, 26(10):105009, 2010.

\bibitem{BGM-11}
L.~Borcea, F.~Guevara~Vasquez, and A.~V. Mamonov.
\newblock Uncertainty quantification for electrical impedance tomography with
  resistor networks.
\newblock submitted to Inverse Problems.
  ArXiv:\href{http://arxiv.org/abs/1105.1183}{1105.1183v1 [math-ph].}

\bibitem{brown1997uniqueness}
R.M. Brown and G.~Uhlmann.
\newblock {Uniqueness in the inverse conductivity problem for nonsmooth
  conductivities in two dimensions }.
\newblock {\em Commun. Partial Diff. Eqns}, 22:1009--27, 1997.

\bibitem{chadan1997introduction}
K.~Chadan.
\newblock {\em {An introduction to inverse scattering and inverse spectral
  problems}}.
\newblock Society for Industrial Mathematics, 1997.

\bibitem{chu2002structured}
M.T. Chu and G.H. Golub.
\newblock {Structured inverse eigenvalue problems}.
\newblock {\em Acta Numerica}, 11(-1):1--71, 2002.

\bibitem{ColM}
C.~F. Coleman and J.~R. McLaughlin.
\newblock Solution of the inverse spectral problem for an impedance with
  integrable derivative, i, ii.
\newblock {\em Comm. Pure Appl. Math.}, 46(2):145--212, 1993.

\bibitem{CurtMooMor}
E.~Curtis, E.~Mooers, and J.A. Morrow.
\newblock Finding the conductors in circular networks from boundary
  measurements.
\newblock {\em RAIRO - Mathematical Modelling and Numerical Analysis},
  28:781--814, 1994.

\bibitem{CurtIngMor}
E.B. Curtis, D.~Ingerman, and J.A. Morrow.
\newblock Circular planar graphs and resistor networks.
\newblock {\em Linear Algebra and its Applications}, 23:115--150, 1998.

\bibitem{curtMorBook}
E.B. Curtis and J.A. Morrow.
\newblock {\em {Inverse problems for electrical networks}}.
\newblock World Scientific, 2000.

\bibitem{deverdiere1994rep}
Y.C. de~Verdi{\`e}re.
\newblock {Reseaux electriques planaires I}.
\newblock {\em Commentarii Mathematici Helvetici}, 69(1):351--374, 1994.

\bibitem{deverdiere1996rep}
Y.C. de~Verdi{\`e}re, I.~Gitler, and D.~Vertigan.
\newblock {Reseaux electriques planaires II}.
\newblock {\em Commentarii Mathematici Helvetici}, 71(1):144--167, 1996.

\bibitem{druskin1982usi}
V.~Druskin.
\newblock {The unique solution of the inverse problem of electrical surveying
  and electrical well-logging for piecewise-continuous conductivity}.
\newblock {\em Izv. Earth Physics}, 18:51--3, 1982.

\bibitem{druskin1985udt}
V.~Druskin.
\newblock {On uniqueness of the determination of the three-dimensional
  underground structures from surface measurements with variously positioned
  steady-state or monochromatic field sources}.
\newblock {\em Sov. Phys.--Solid Earth}, 21:210--4, 1985.

\bibitem{DruKni}
V.~Druskin and L.~Knizhnerman.
\newblock {Gaussian spectral rules for second order finite-difference schemes}.
\newblock {\em Numerical Algorithms}, 25(1):139--159, 2000.

\bibitem{druskin2000gsr}
V.~Druskin and L.~Knizhnerman.
\newblock {Gaussian spectral rules for the three-point second differences: I. A
  two-point positive definite problem in a semi-infinite domain}.
\newblock {\em SIAM Journal on Numerical Analysis}, 37(2):403--422, 2000.

\bibitem{DruMos}
V.~Druskin and S.~Moskow.
\newblock {Three-point finite-difference schemes, Pade and the spectral
  Galerkin method. I. One-sided impedance approximation}.
\newblock {\em Mathematics of Computation}, 71(239):995--1020, 2002.

\bibitem{druskin2002three}
V.~Druskin and S.~Moskow.
\newblock {Three-point finite-difference schemes, Pad{\'e} and the spectral
  Galerkin method. I. One-sided impedance approximation}.
\newblock {\em Mathematics of computation}, 71(239):995--1020, 2002.

\bibitem{gautschi1987lower}
W.~Gautschi and G.~Inglese.
\newblock Lower bounds for the condition number of vandermonde matrices.
\newblock {\em Numerische Mathematik}, 52(3):241--250, 1987.

\bibitem{gel1951determination}
I.M. Gel'fand and B.M. Levitan.
\newblock {On the determination of a differential equation from its spectral
  function}.
\newblock {\em Izvestiya Rossiiskoi Akademii Nauk. Seriya Matematicheskaya},
  15(4):309--360, 1951.

\bibitem{gelfand1951dde}
I.M. Gel'fand and B.M. Levitan.
\newblock {On the determination of a differential equation from its spectral
  function}.
\newblock {\em Izvestiya Rossiiskoi Akademii Nauk. Seriya Matematicheskaya},
  15(4):309--360, 1951.

\bibitem{Godunov}
S.~K. Godunov and V.~S. Ryabenkii.
\newblock {\em The theory of difference schemes --- An introduction}.
\newblock North Holland, Amsterdam, 1964.

\bibitem{GuevaraPhD}
F.~Guevara~Vasquez.
\newblock {\em {On the Parametrization of Ill-posed Inverse Problems Arising
  from Elliptic Partial Differential Equations}}.
\newblock PhD thesis, Rice University, Houston, TX, USA, 2006.

\bibitem{hochstadt1973inverse}
H.~Hochstadt.
\newblock {The inverse sturm-liouville problem}.
\newblock {\em Communications on Pure and Applied Mathematics},
  26(5-6):715--729, 1973.

\bibitem{imanuvilov2008gup}
O.Y. Imanuvilov, G.~Uhlmann, and M.~Yamamoto.
\newblock {Global uniqueness from partial Cauchy data in two dimensions}.
\newblock {\em Arxiv preprint arXiv:0810.2286}, 2008.

\bibitem{IngerLayer}
D.~Ingerman.
\newblock Discrete and continuous {D}irichlet-to-{N}eumann maps in the layered
  case.
\newblock {\em SIAM Journal on Mathematical Analysis}, 31:1214--1234, 2000.

\bibitem{IngDruKni}
D.~Ingerman, V.~Druskin, and L.~Knizhnerman.
\newblock {Optimal finite difference grids and rational approximations of the
  square root I. Elliptic problems}.
\newblock {\em Communications on Pure and Applied Mathematics},
  53(8):1039--1066, 2000.

\bibitem{MorInger}
D.~Ingerman and J.~A. Morrow.
\newblock {On a characterization of the kernel of the Dirichlet-to-Neumann map
  for a planar region}.
\newblock {\em SIAM Journal on Applied Mathematics}, 29:106--115, 1998.

\bibitem{isaacson}
D.~Isaacson.
\newblock {Distinguishability of conductivities by electric current computed
  tomography}.
\newblock {\em IEEE transactions on medical imaging}, 5(2):91--95, 1986.

\bibitem{kac1974spectral}
IS~Kac and MG~Krein.
\newblock {On the spectral functions of the string}.
\newblock {\em Amer. Math. Soc. Transl}, 103(2):19--102, 1974.

\bibitem{kohn1984dcb}
R.~Kohn and M.~Vogelius.
\newblock {Determining conductivity by boundary measurements}.
\newblock {\em Communications on Pure and Applied Mathematics}, 37:289--98,
  1984.

\bibitem{kohn1985dcb}
R.~Kohn and M.~Vogelius.
\newblock {Determining conductivity by boundary measurements II. Interior
  results}.
\newblock {\em Communications on Pure and Applied Mathematics}, 38(5), 1985.

\bibitem{lang2005undergraduate}
S.~Lang.
\newblock {\em Undergraduate algebra}.
\newblock Springer Verlag, 2005.

\bibitem{lavrentiev1987methods}
M.A. Lavrentiev and B.V. Shabat.
\newblock {\em {Methods of the complex variable function theory (in Russian)}}.
\newblock Nauka, Moscow, 1987.

\bibitem{levitan1987isl}
B.M. Levitan.
\newblock {\em {Inverse Sturm-Liouville Problems}}.
\newblock VSP, 1987.

\bibitem{macmillan2004first}
H.R. MacMillan, T.A. Manteuffel, and S.F. McCormick.
\newblock {First-order system least squares and electrical impedance
  tomography}.
\newblock {\em SIAM Journal on Numerical Analysis}, 42(2):461--483, 2004.

\bibitem{MamonovMasters}
A.V. Mamonov.
\newblock {Resistor Network Approaches to the Numerical Solution of Electrical
  Impedance Tomography with Partial Boundary Measurements}.
\newblock Master's thesis, Rice University, Houston, TX, USA, 2009.

\bibitem{MamonovPhD}
A.V. Mamonov.
\newblock {\em {Resistor Networks and Optimal Grids for the Numerical Solution
  of Electrical Impedance Tomography with Partial Boundary Measurements}}.
\newblock PhD thesis, Rice University, Houston, TX, USA, 2010.

\bibitem{mandache2001eii}
N.~Mandache.
\newblock {Exponential instability in an inverse problem for the Schrodinger
  equation}.
\newblock {\em Inverse Problems}, 17(5):1435--1444, 2001.

\bibitem{marchenko2011sturm}
V.A. Marchenko.
\newblock {\em {Sturm-Liouville operators and applications}}.
\newblock Chelsea Pub Co, 2011.

\bibitem{mclaughlin1987uniqueness}
J.R. McLaughlin and W.~Rundell.
\newblock {A uniqueness theorem for an inverse Sturm--Liouville problem}.
\newblock {\em Journal of mathematical physics}, 28:1471, 1987.

\bibitem{nachman1996gut}
A.I. Nachman.
\newblock {Global uniqueness for a two-dimensional inverse boundary value
  problem}.
\newblock {\em Annals of Mathematics}, pages 71--96, 1996.

\bibitem{Natanson}
I.~Natanson.
\newblock {\em Theory of functions of a real variable}, volume~1.
\newblock Ungar Pub Co, New York, 1961.

\bibitem{natterer1994dgl}
F.~Natterer.
\newblock {A discrete Gelfand-Levitan theory}.
\newblock Technical report, Technical report, Institut fuer Numerische und
  instrumentelle Mathematik, 1994.

\bibitem{nikishin1991rational}
E.M. Nikishin and V.N. Sorokin.
\newblock {\em {Rational approximations and orthogonality}}.
\newblock Amer Mathematical Society, 1991.

\bibitem{Trub}
J.~P\"oschel and E.~Trubowitz.
\newblock {\em Inverse spectral theory. Pure and Applied Mathematics}, volume
  130.
\newblock Academic Press, Inc., Boston, MA, 1987.

\bibitem{quarteroni1999domain}
A.~Quarteroni and A.~Valli.
\newblock {\em Domain decomposition methods for partial differential
  equations}.
\newblock Oxford University Press, USA, 1999.

\bibitem{reich1976quasiconformal}
E.~Reich.
\newblock {Quasiconformal mappings of the disk with given boundary values}.
\newblock {\em Lecture Notes in Mathematics}, 505:101--137, 1976.

\bibitem{strebel1976eet}
K.~Strebel.
\newblock {On the existence of extremal Teichm{\"u}ller mappings}.
\newblock {\em Journal d'Analyse Math{\'e}matique}, 30(1):464--480, 1976.

\bibitem{sylvester1990aib}
J.~Sylvester.
\newblock {An anisotropic inverse boundary value problem}.
\newblock {\em Communications on Pure and Applied Mathematics}, 43(2):201--232,
  1990.

\bibitem{trefethen1997numerical}
L.N. Trefethen and D.~Bau.
\newblock {\em Numerical linear algebra}.
\newblock Number~50. Society for Industrial Mathematics, 1997.

\end{thebibliography}
\end{document}